\documentclass[iop,revtex4]{emulateapj}
\usepackage{natbib}
\usepackage{color}
\usepackage{amsmath}
\usepackage{graphicx,setspace,amssymb}
\usepackage{ccaption}
\usepackage{lscape}

\shorttitle{Radio Continuum Measurements of local U/LIRGs}
\shortauthors{Barcos-Mu\~noz et al.}

\begin{document}
\title{A 33 GHz Survey of Local Major Mergers: Estimating the Sizes of the Energetically Dominant Regions from High Resolution Measurements of the Radio Continuum}

\author{L. Barcos-Mu\~noz\altaffilmark{1,2,3}}
\author{A. K. Leroy\altaffilmark{4}}
\author{A. S. Evans\altaffilmark{2,3}}
\author{J. Condon\altaffilmark{2}}
\author{G. C. Privon\altaffilmark{5,6}}
\author{T. A. Thompson\altaffilmark{4,7}}
\author{L. Armus\altaffilmark{8}}
\author{T. D\'{i}az-Santos\altaffilmark{9}}
\author{J. M. Mazzarella\altaffilmark{10}}
\author{D. S. Meier\altaffilmark{11,12}}
\author{E. Momjian\altaffilmark{12}}
\author{E. J. Murphy\altaffilmark{2}}
\author{J. Ott\altaffilmark{12}}
\author{D. B. Sanders\altaffilmark{13}}
\author{E. Schinnerer\altaffilmark{14}}
\author{S. Stierwalt\altaffilmark{2}}
\author{J. A. Surace\altaffilmark{15}}
\author{F. Walter\altaffilmark{14}}
\email{loreto.barcos@alma.cl}
\altaffiltext{1}{Joint ALMA Observatory, Alonso de C\'{o}rdova 3107, Vitacura, Santiago, Chile}
\altaffiltext{2}{National Radio Astronomy Observatory, 520 Edgemont Road, Charlottesville, VA 22903, USA}
\altaffiltext{3}{Department of Astronomy, University of Virginia, 530 McCormick Road, Charlottesville, VA 22904, USA}
\altaffiltext{4}{Astronomy Department, The Ohio State University, 140 W 18th St, Columbus, OH 43210, USA}
\altaffiltext{5}{Instituto de Astrof\'{i}sica, Facultad de F\'{i}sica, Pontificia universidad Cat\'{o}lica de Chile, Casilla 306, Santiago, Chile}
\altaffiltext{6}{Departamento de Astronom\'{i}a, Universidad de Concepci\'{o}n, Casilla 160-C, Concepci\'{o}n, Chile}
\altaffiltext{7}{Department of Astronomy and Center for Cosmology \& Astro-Particle Physics, The Ohio State University, Columbus, OH 43210, USA}
\altaffiltext{8}{Spitzer Science Center, California Institute of Technology, MC 220-6, 1200 East California Boulevard, Pasadena, CA 91125, USA}
\altaffiltext{9}{N\'{u}cleo de Astronom\'{i}a de la Facultad de Ingenier\'{i}a, Universidad Diego Portales, Av. Ej\'{e}rcito Libertador 441, Santiago, Chile}
\altaffiltext{10}{Infrared Processing and Analysis Center, MS 100-22, California Institute of Technology, Pasadena, CA 91125, USA}
\altaffiltext{11}{New Mexico Institute of Mining and Technology, 801 Leroy Place, Socorro, NM 87801, USA}
\altaffiltext{12}{National Radio Astronomy Observatory, P.O. Box O, 1003 Lopezville Road, Socorro, NM 87801, USA}
\altaffiltext{13}{Institute for Astronomy, University of Hawaii, 2680 Woodlawn Dr., Honolulu, HI 96816, USA}
\altaffiltext{14}{Max-Planck-Institut f\"{u}r Astronomie, K\"{o}nigstuhl 17, D-69117 Heidelberg, Germany}
\altaffiltext{15}{Spitzer Science Center, MS 314-6, California Institute of Technology, Pasadena, CA 91125, USA}

\begin{abstract}
We present Very Large Array observations of the 33 GHz radio continuum emission from 22 local ultraluminous and luminous infrared (IR) galaxies (U/LIRGs). These observations have spatial (angular) resolutions of 30--720~pc (0$\,\farcs$07-0$\,\farcs$67) in a part of the spectrum that is likely to be optically thin. This allows us to estimate the size of the energetically dominant regions. We find half-light radii from 30~pc to 1.7~kpc. The 33 GHz flux density correlates well with the IR emission, and we take these sizes as indicative of the size of the region that produces most of the energy. Combining our 33 GHz sizes with unresolved measurements, we estimate the IR luminosity and star formation rate per area, and the molecular gas surface and volume densities. These quantities span a wide range (4 dex) and include some of the highest values measured for any galaxy (e.g., $\mathrm{\Sigma_{SFR}^{33GHz} \leq 10^{4.1}~M_{\odot}~yr^{-1}~kpc^{-2}}$). At least $13$ sources appear Compton thick ($\mathrm{N_{H}^{33GHz} \geq 10^{24}~cm^{-2}}$). Consistent with previous work, contrasting these data with observations of normal disk galaxies suggests a nonlinear and likely multi-valued relation between SFR and molecular gas surface density, though this result depends on the adopted CO-to-H$_{2}$ conversion factor and the assumption that our 33 GHz sizes apply to the gas. 11 sources appear to exceed the luminosity surface density predicted for starbursts supported by radiation pressure and supernovae feedback, however we note the need for more detailed observations of the inner disk structure. U/LIRGs with higher surface brightness exhibit stronger [{\sc Cii}] 158$\mu$m deficits, consistent with the suggestion that high energy densities drive this phenomenon.
\end{abstract} 


\section{Introduction}

Luminous and ultraluminous infrared (IR) galaxies (LIRGs: $10^{11}~L_{\odot} \leq L_{IR}$ [8-1000$\mu$m] $< 10^{12}~L_{\odot}$, ULIRGs: $L_{IR} \geq 10^{12}~L_{\odot}$) host some of the most extreme environments in the local universe. Local U/LIRGs are primarily triggered by galaxy interactions and mergers \citep[e.g.,][and references therein]{S&M96}. During this process, large amounts of gas are funneled into the central few kpc. There, the gas fuels prodigious star formation and/or AGN activity. This activity is heavily embedded in dust and gas, which reprocesses the emergent light into the IR, giving rise to the high IR luminosities of U/LIRGs.

Their enormous gas surface densities, gas volume densities, energy densities, and high star formation rates \citep[SFRs; up to a few times 100~M$_{\odot}~{\rm yr}^{-1}$ based on L$_{IR}$, e.g.,][]{Solomon97,D&S98,Evans02} make the local U/LIRGs crucial laboratories to understand the physics of star formation and feedback in an extreme regime. Indeed, these systems have among the highest SFR and gas surface densities measured for any galaxy population in the local universe \citep[e.g.,][]{D&S98,Liu15,Lutz16}. These extreme conditions may lead U/LIRGs to convert gas into stars in a mode that is distinct from what we find in main-sequence galaxies like the Milky Way, and more similar to extreme starbursts observed at high redshift. In this scenario, U/LIRGs and their high redshift counterparts produce a higher rate of star formation per unit gas mass compared to ``main sequence galaxies'' at both low and high redshift \citep[e.g.,][]{Daddi10,Genzel10}.

The combination of high opacity, high gas surface density, and on-going star formation also makes these galaxies key testbeds for theories exploring the balance between feedback and gravity \citep[e.g.,][]{Murray05,Shetty11}. For example, \citet{Thompson05} have argued that the most extreme local U/LIRGs may represent ``Eddington limited'' star-forming systems or ``maximal starbursts'', which produce stars at the maximum capacity allowed for the considered feedback mechanism, i.e., radiation pressure on dust.

Exploring the physics of U/LIRGs requires knowing their intensive properties, i.e., the luminosity, or mass, per unit area or volume. The extreme nature of these systems is most evident when the high luminosity is viewed in the context of the very small area from which it emerges. In turn, measuring these intensive quantities requires knowing the size of the region where star formation is on-going. This is a challenging measurement. Even the nearest U/LIRGs are quite distant ($50$--$150$~Mpc) compared to prototypes of more quiescent main-sequence galaxies. Thus very high angular resolution is required to study them. Compounding the challenge, U/LIRGs host enormous amounts of dust (e.g., A$_{V}$ $\sim$ 1000 for Arp 220 \citealt{Lutz96}), rendering them optically thick at optical and even infrared wavelengths. They are also opaque at very long radio wavelengths due to free-free absorption \citep[e.g.,][]{Condon90}, leaving them transparent only over a limited regime, from radio to sub-millimeter wavelengths (for the extreme case of Arp 220, see \citealt{BM15}).

Interferometric radio imaging is the ideal, and almost only, way to measure the sizes of the energetically dominant regions in the centers of local U/LIRGs. Radio interferometers make it possible to achieve the high angular resolution required to resolve the compact central starbursts, while cm-wave photons penetrate the high dust columns that prevent measurements of the inner regions at optical wavelengths. The two dominant radio continuum emission mechanisms at cm wavelengths, free-free (``thermal'') and synchrotron (``nonthermal'') emission, both trace the distribution of recent star formation and can indicate AGN activity, if present.

Following this logic, \citet{Condon90} and \citet{Condon91} used the old (pre-upgrade) Very Large Array (VLA) to study the energetically dominant regions in U/LIRGs at 1.49 GHz \citep[angular resolution $\geq$1$\,\farcs$5,][]{Condon90} and 8.44 GHz \citep[angular resolution $\geq0\,\farcs25$,][]{Condon91}. Their constraints on the sizes of the star-forming/AGN dominated regions in these systems are still some of the strongest measurements twenty five years later. 

Because the VLA has fixed antenna array configurations, higher frequency observations provide the logical pathway to better angular resolution, and so better size constraints for the local U/LIRGs. However, the spectral index of radio emission from galaxies is negative over the range $\nu \sim 1{-}50$~GHz, so that galaxies are fainter at higher frequencies. The sensitivity of the historic VLA receivers was also lower at high frequency. As a result, efforts to imaging these systems using the historic VLA at $\nu \gtrsim 10$~GHz were limited.

With the upgrade from the old VLA to the Karl G. Jansky Very Large Array (VLA), this situation changed. Both the bandwidth and receiver sensitivity improved, thereby improving the ability of the VLA to image the radio continuum from U/LIRGs at high frequency (and thus high angular resolution). Given the current VLA capabilities, the Ka band (26.5 $-$ 40 GHz) offers the ideal balance between low opacity in the source, high angular resolution, and good sensitivity. We demonstrated this capability in \citet{BM15}, where we used the VLA at Ka band to make the sharpest-ever image that recovered all of the flux density of the nuclear disks of Arp 220.

Here we extend the work of \citet{BM15} to a sample of 22 of the most luminous northern U/LIRGs. This is the first high resolution, high sensitivity, 33 GHz continuum survey of local U/LIRGs. The angular resolution (beam size) of the VLA at $\nu = 33$~GHz improves compared to the $8.44$~GHz of \citet{Condon91} by at least a factor of two.

The paper proceeds as follows. In Section \ref{sec:red}, we describe the survey and the data reduction process. In Section \ref{sec:results}, we present the measurements. We explore the physical implications of these measurements in Section \ref{sec:impl_radio}. In Section \ref{sec:discussion}, we discuss the nature of the energy emission at 33 GHz, the implied physical conditions in these systems, the implications of our measurements for star formation scaling relations, and whether the systems are maximal starbursts. We summarize our conclusions in Section \ref{sec:conclusions}, and the Appendix presents detailed notes on individual systems.

Throughout this paper, intrinsic quantities are derived by adopting the cosmology H$_{0}$ = 73 km s$^{-1}$ Mpc$^{-1}$, $\mathrm{\Omega_{vacuum}=0.73}$ and $\mathrm{\Omega_{matter}=0.27}$, with recessional velocities corrected to the frame of the cosmic microwave background.

\section{Sample, Observations, and Data Reduction} 
\label{sec:red}

\subsection{Observations}

We used the Karl G. Jansky Very Large Array (VLA) to observe radio continuum emission from the most luminous nearby LIRGs and ULIRGs. Our sample (see Table \ref{table:tbl-1}) consists of 22 sources from the IRAS Revised Bright Galaxy Sample \citep[RBGS;][]{Sanders03}. These galaxies have infrared luminosities $\mathrm{L_{IR}[8-1000 \mu m] = 10^{11.6} - 10^{12.6}~L_{\odot}}$ and were selected to be northern enough to be observed by the VLA, i.e., $\delta > -15^{\circ}$. These systems are also a subset of the Great Observatories All-sky LIRG Survey \citep[GOALS;][]{Armus09}, for which multiwavelength data are available.

As part of the resident shared risk project AL746, we observed the radio continuum emission from each source at C band (4--8 GHz) and Ka band (26.5--40 GHz). For each observation we used dual polarization mode with two 1 GHz-wide basebands. Each band was split into eight 128 MHz spectral windows (spw's) with 64 channels each. We centered the 1 GHz basebands at $\sim 4.7$ and $7.2$ GHz in C band and $\sim 29$ and $36$ GHz in Ka band.

In order to recover emission across a wide range of angular scales, we observed our sample in each frequency range in separate sessions using each of the four VLA configurations (A, B, C and D, from highest to lowest angular resolution). Observations spanned the period 2010 August 2 to 2011 August 16. In the D and C configurations, we observed each source for five minutes. In the B configuration, we observed each source for ten minutes split between two five-minute scans. In the A configuration we observed most sources for 20 minutes, split into four five-minute scans. Due to scheduling constraints, eight sources were not observed in the A configuration at Ka band; these are identified with an asterisk in Table \ref{table:tbl-3}. Thus the total time on source for most targets was $\sim$ 40 minutes per band.

At the beginning of each session, we observed either 3C~48 or 3C~286, which was used to set the flux density scale and calibrate the bandpass. Through the rest of the session we alternated between observations of science targets and a secondary calibrator within a few degrees of each science target. We used observations of these secondary calibrators to measure phase and gain variations due to atmospheric/ionospheric and instrumental fluctuations. Table \ref{table:tbl-2} summarizes the calibrators used for each science target.

These data have also appeared in \citet{Leroy11} and \citet{BM15}. \citet{Leroy11} presented first results from our observations at both C and Ka band but used only observations from the two most compact VLA configurations. \citet{BM15} presented C and Ka band observations using all four configurations for the specific case of Arp 220. In this paper, we report on the full survey, emphasizing the Ka band observations and the combination of all four array configurations. These represent the highest resolution, highest sensitivity radio observations for these galaxies published to date. The C band observations combining all four array configurations will be reported in an upcoming paper focused on the resolved spectral energy distribution, i.e., across the disks of the systems in our sample (Barcos-Mu\~{n}oz et al. in preparation).

\subsection{Data Processing}

We used the Common Astronomy Software Application \citep[CASA,][]{McM07} to calibrate, inspect, and analyze the data. We followed a standard VLA reduction procedure, including calibrating the bandpass, phase, and amplitude response of each antenna. We set the overall flux density scale using ``Perley-Butler 2010" models for the primary calibrators and assuming that the Ka band emission shares the same structure as the VLA-provided Q-band model.

Once the data were calibrated, we imaged each science target. To do this, we used the task \texttt{CLEAN} in mode \texttt{mfs} (multi-frequency synthesis) \citep{SaW94}, with \textit{Briggs} weighting setting \texttt{robust=0.5}. For each array configuration, we imaged each baseband independently. Whenever possible, we iterated this imaging with phase and amplitude self calibration. The number of self calibration iterations varied from zero to eight based on the signal to noise of the data, with four iterations typical. After several iterations of phase-only self calibration, when possible, we also performed amplitude self calibration. We always solved for only relative variations in the amplitudes gains among the antennas (\texttt{solnorm=True} in CASA's \texttt{gaincal}), and so avoided forcing the flux of the source to some value.

After self calibrating the two basebands independently, we combined both into a single image using \texttt{clean}'s multifrequency synthesis mode and \texttt{nterms=2}. The latter allows us to model the frequency dependence of the sky emission with two Taylor coefficients. After the described combination we ended up with four images per source (one per array configuration). Finally, we jointly imaged all self-calibrated data, combining all eight measurement sets (four configurations and two basebands). This combined image represents our best data product, using all of our observations with sensitivity to a wide range of angular scales. In the highest signal-to-noise cases, for example UGC 08058 (Mrk 231) and UGC 09913 (Arp 220), we performed further self calibration during this final imaging step. 

These final images have a nominal frequency $\nu = 32.5$~GHz\footnote{Throughout the paper we use 33 and 32.5 GHz interchangeably, however for calculation/estimation purposes we use $\nu = 32.5$~GHz.} and a typical rms noise $\mathrm{26~\mu Jy~beam^{-1}}$. Table \ref{table:tbl-3} reports the exact beam size and rms noise for the combined image for each target.

\subsection{Additional Data}
\label{sec:add_data}

We combine our survey with previous observations of our sample at 1.49 GHz (beam FWHM $\sim$ 15$\arcsec$) from \citet{Condon90}. We also use the 5.95 GHz flux densities (beam FWHM $\sim$ 0.4$\arcsec$) from \citet{Leroy11} and CO (1$-$0) flux densities, obtained using the ARO 12-m antenna (FWHM = 1'), the latter of which will be reported in Privon et al. (in preparation). We present a compilation of the flux densities at these different frequencies, along with 32.5 GHz flux densities measured from our new images, in Table \ref{table:tbl-3}. The uncertainty in the 1.49 GHz flux density values are assumed to be dominated by flux density calibration errors ($\sim$ 5\%, see Section III in \citealt{Condon90}).

Five of our sources lack flux density measurements at 1.49 GHz. For three of these --- VII Zw 031, CGCG 448-020 and IRAS F23365+3608 --- we use the 1.4 GHz flux density from the NVSS catalog \citep{Condon98}. For IRAS 19542+1110 and IRAS 21101+5810, we take the values at 1.425 GHz measured by \citet{Condon96}. We use these flux densities interchangeably with the $1.49$~GHz fluxes, but assign them a larger ($10\%$) uncertainty in these cases to reflect some uncertainty in the spectral index.

\section{Results}
\label{sec:results}

In Figure \ref{fig:fig1}, contour and color maps show new VLA $\nu = 32.5$~GHz images for our sample of 22 local U/LIRGs. These are the first $33$~GHz images of these systems that have both high resolution and sensitivity to a wide range of angular scales. We use them to measure: (1) the area of the energetically dominant region in each galaxy, (2) the integrated flux density of each target at 33~GHz, and (3) the contribution (by area and flux density) of compact regions to the integrated properties of each system. In Tables \ref{table:tbl-3} and \ref{table:tbl-4} we report the measured areas and integrated flux densities at 33 GHz, along with the integrated flux densities from the literature that we use to study the spectral index, and so the nature of the radio emission.

\begin{figure*}[tbh]
\centering
\includegraphics[scale=0.65]{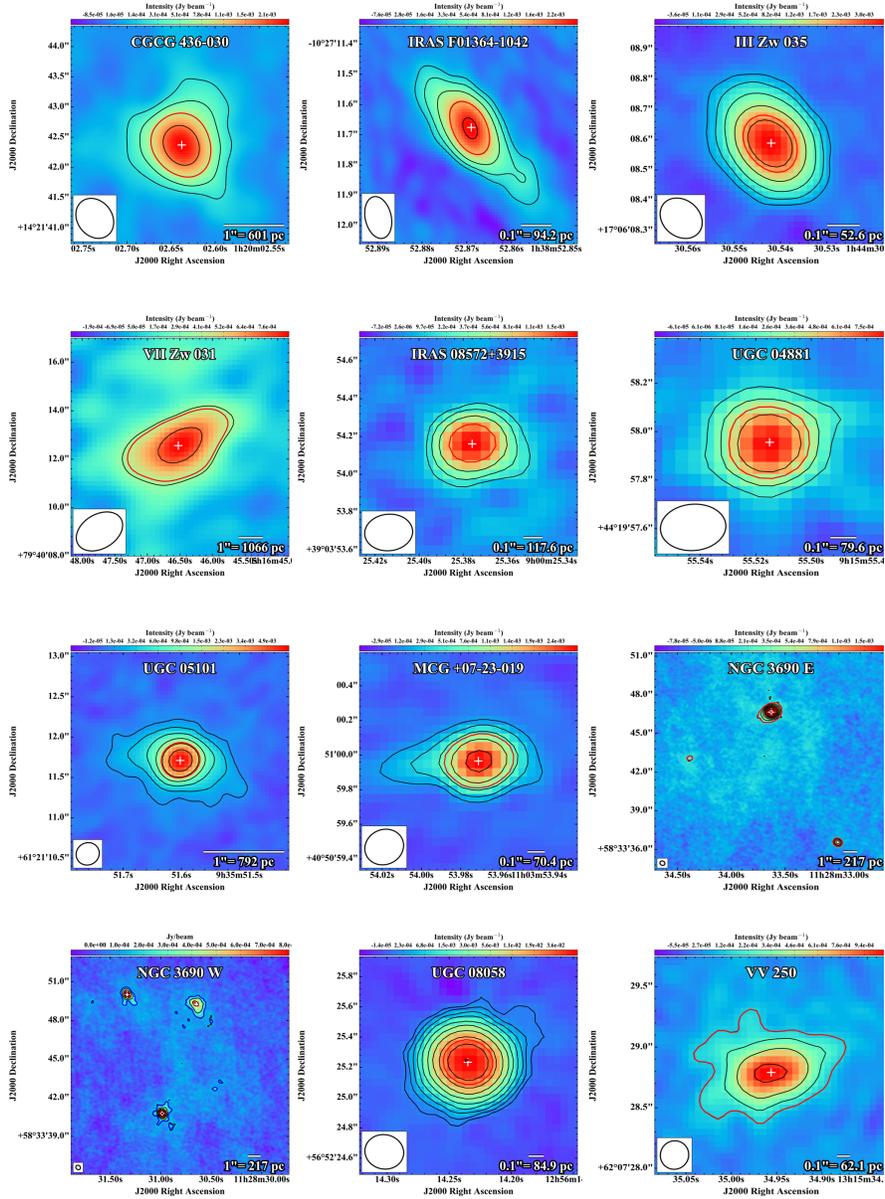}
\vspace*{-10mm}
\caption{Contour and color maps of $\sim$33 GHz continuum emission of each galaxy in our sample. The contours are spaced by in factor of two in intensity, with the outermost contour set to 5 times the rms noise in the map. The beam for each map appears as a boxed black ellipse in the bottom left corner. The scale bar for each map appears in the bottom right corner. The white crosses indicate the location of compact sources whose properties were derived from Gaussian fits (see Section \ref{sec:comp}). The red contour encloses 50\% of the total flux density at 33~GHz; we use the area inside this contour, A$_{50}$, as a characteristic size for energetically dominant part of the galaxy. Most of the emission in our sample is compact, with only a few systems showing considerable extended emission (e.g., VV 340a) and others showing a combination of compact and extended emission (e.g., UGC 08387).\label{fig:fig1}}
\end{figure*}

\begin{figure*}
\centering
\includegraphics[scale=0.65]{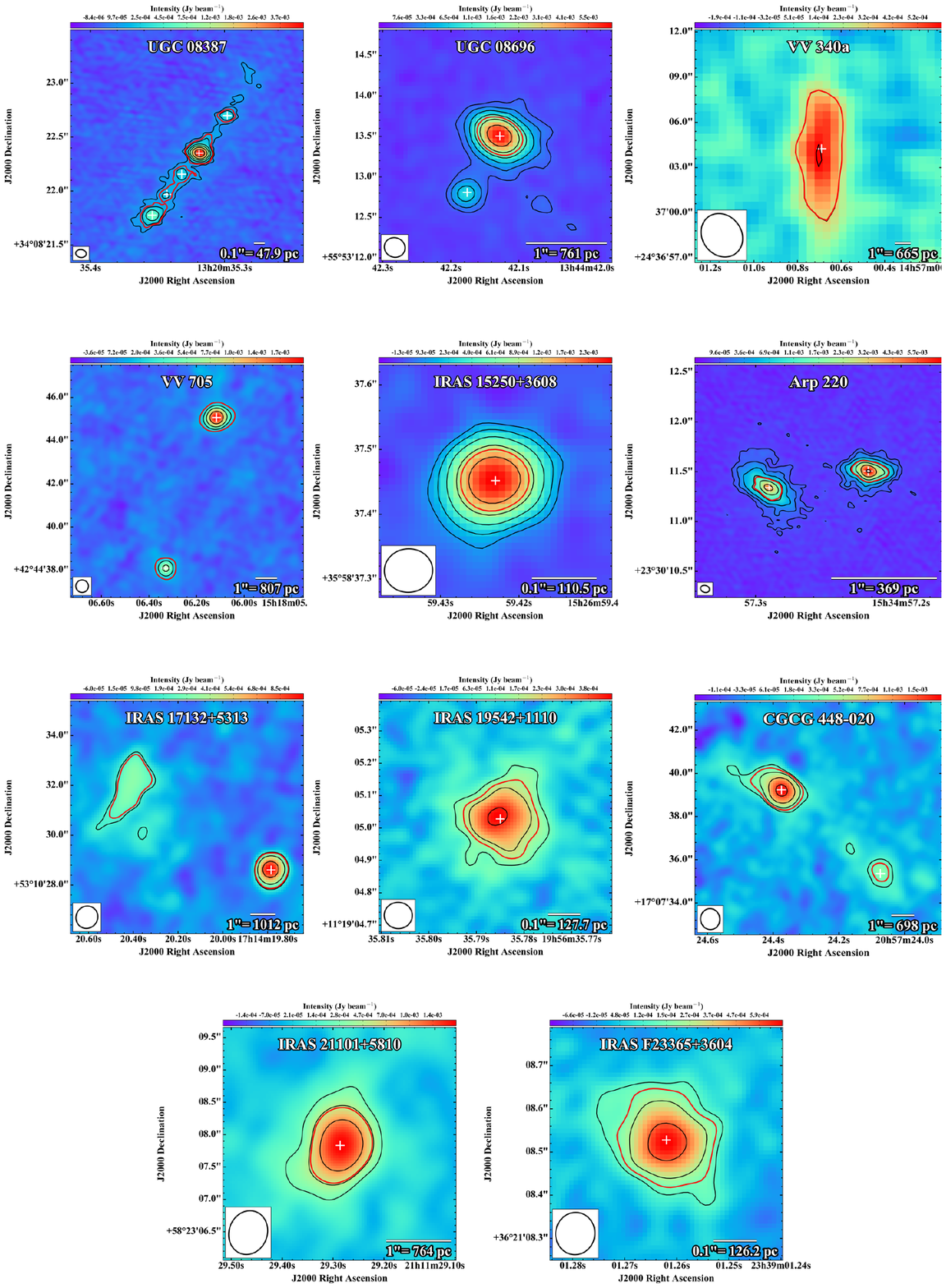}
\vspace*{-10mm}
\contcaption{\it Continued.}
\end{figure*}

\subsection{Flux Densities at $\nu = 32.5$~GHz}
\label{sec:Iflux}

We measure integrated flux densities for each source from the lowest angular resolution observations, which were obtained using the VLA in its D configuration. The maximum recoverable scale for the D configuration, $\approx 22''$, corresponds to $\sim$ 16 kpc at the 165 Mpc median distance of our sample. This would cover most of the star forming activity in a local disk galaxy \citep[e.g.,][]{Schruba11}. U/LIRGs are observed to be much more compact, with sizes less than a few kpc based on previous radio \citep[][]{Liu15}, near-IR \citep{Haan11}, mid-IR \citep{DS10}, and far-IR observations \citep{Lutz16}. Therefore, we expect negligible missing flux in the D configuration-based flux densities.

Confirming this expectation, most of our targets appear unresolved in the D configuration-only images, which have beam sizes $\approx2\,\farcs7$. We obtained the flux densities reported in Table \ref{table:tbl-3} using CASA task \texttt{imfit} to fit two dimensional Gaussians to these mostly unresolved point sources. A few targets, including NGC 3690, CGCG 448-020, IRAS 17132+5313, VII Zw 031, VV 250, VV 340, and VV 705, showed some extent or multiple components in the D configuration maps. In most of these cases, we tapered the D configuration data to a lower resolution until the morphology became a single point-like source. Then we fit a 2D-Gaussian to this degraded image. NGC 3690 and VV 250 show well separated components that can only be fit using two Gaussians, even in the tapered images. We report the sum of these two components as the integrated flux density. 

The uncertainties that we report sum (in quadrature) the statistical error calculated by \texttt{imfit} with uncertainty in the calibration of the flux scale, which we estimated to be $\sim$ 12\% in \citet{BM15}. For the two faintest galaxies in our sample, UGC 04881 and IRAS 08572+3915, the signal-to-noise ratio of the D configuration data only was not high enough to recover integrated flux densities. For these two systems, we instead report results from the combined data using all configurations, which we tapered until we recovered point-like structures that could be fit using Gaussians.

\subsection{Spectral Indices Involving $\nu = 32.5$~GHz}
\label{sec:alpha}

In addition to our new $32.5$~GHz flux densities, Table \ref{table:tbl-3} reports literature flux densities for our sources at $\nu =1.49$ and $5.95$~GHz. We combine these with the $\nu=32.5$~GHz measurements to calculate the galaxy-integrated spectral index between 1.49 GHz and 5.95 GHz ($\alpha_{1.5-6}$) and between 5.95 GHz and 32.5 GHz ($\alpha_{6-33}$). Here, we define the spectral index, $\alpha$, by $\mathrm{S_{\nu} \propto \nu^{\alpha}}$. Note that because we use the flux density integrated over the whole galaxy, we do not expect the different angular resolutions at different frequencies to affect these calculations.

\begin{figure*}[tbh]
\centering
\includegraphics[width=3.5in]{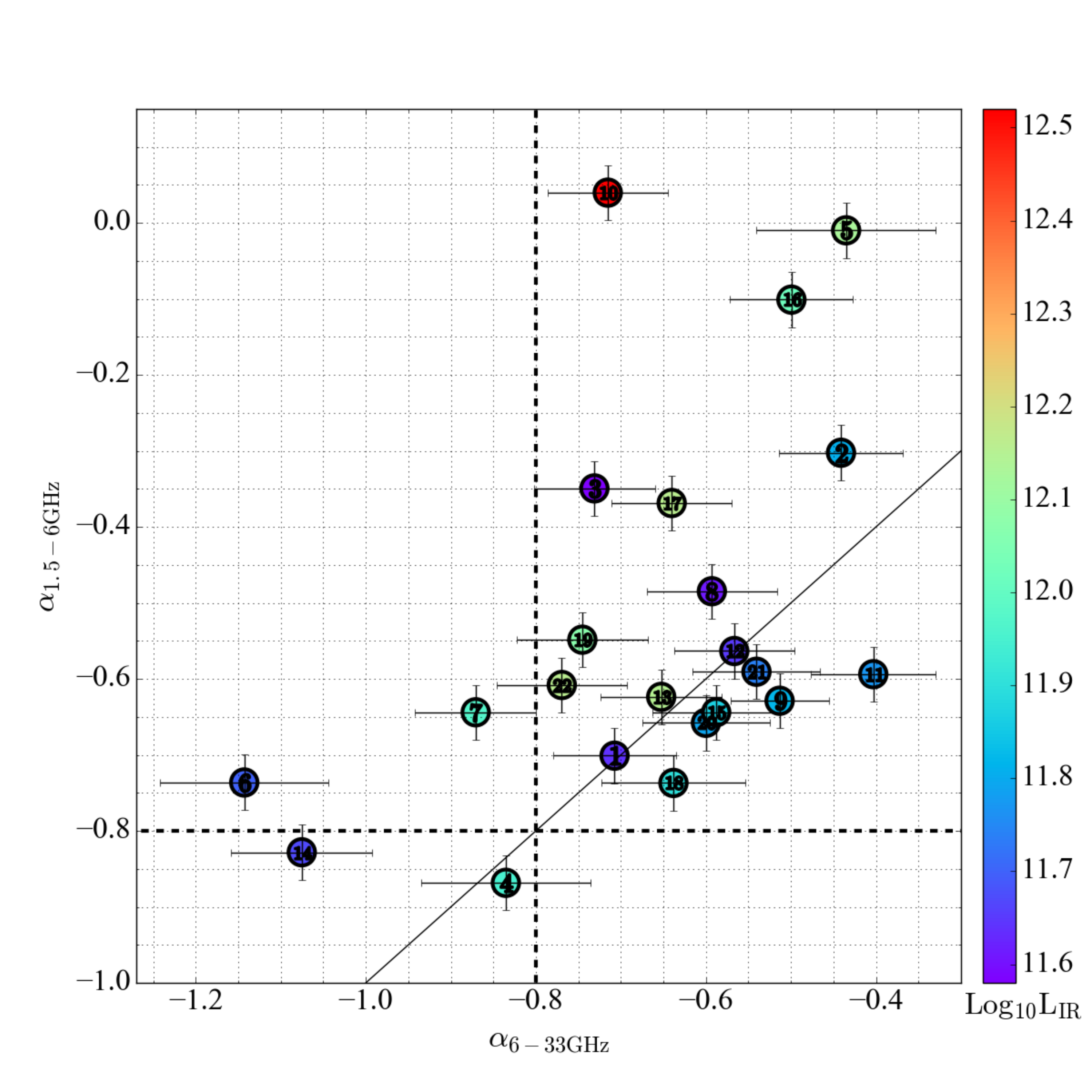}
\includegraphics[width=3.5in]{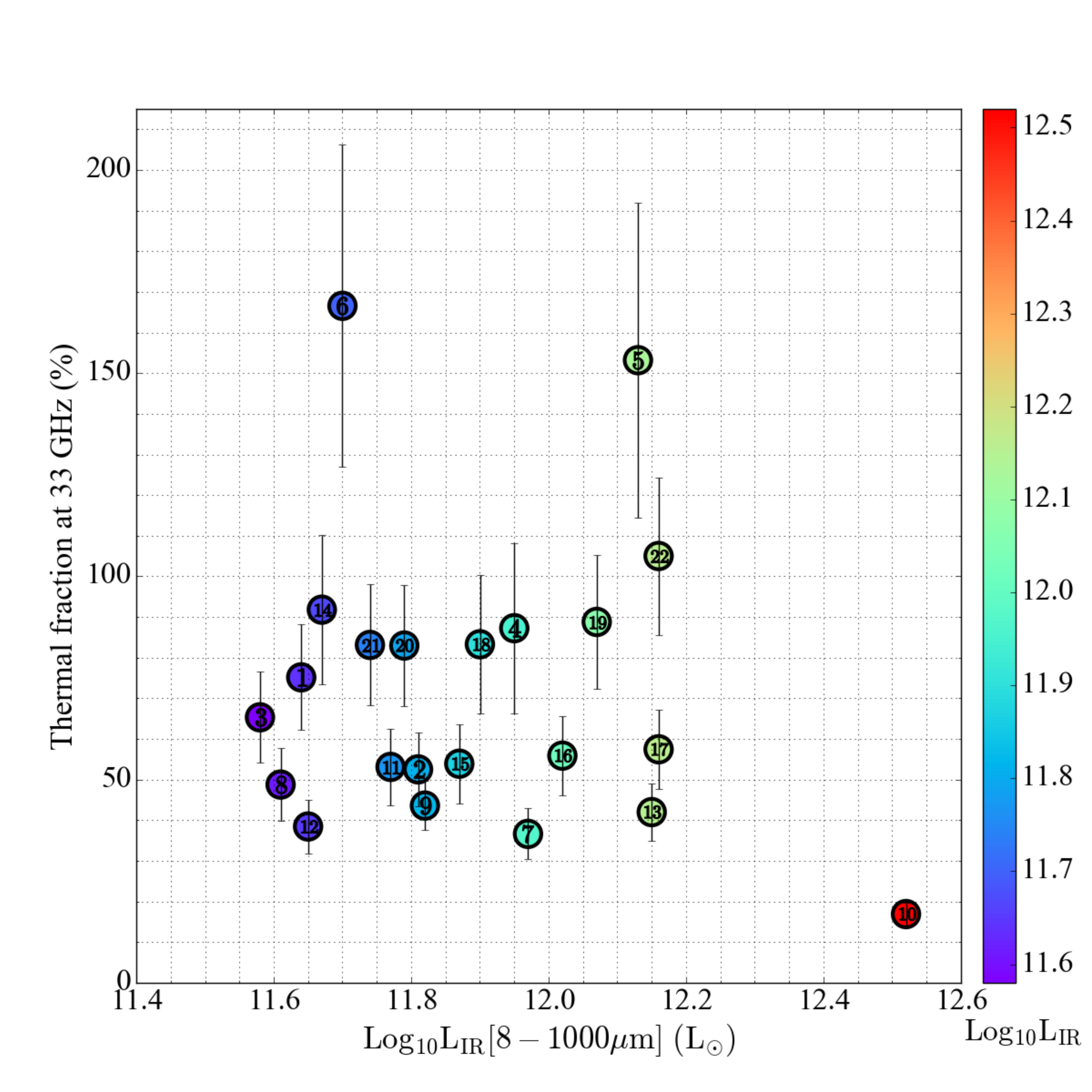}
\includegraphics[width=3.5in]{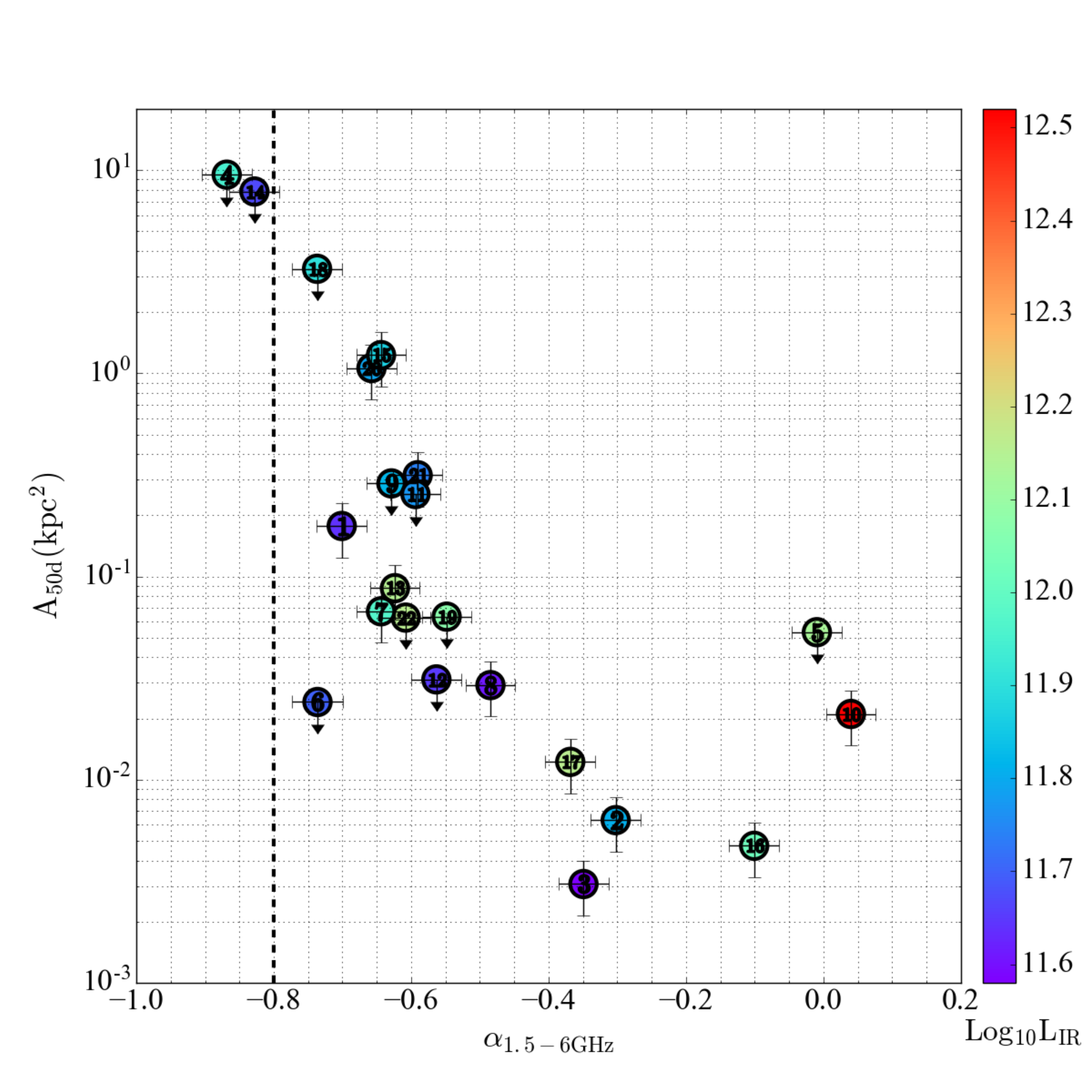}

\caption{Assessment of the nature of the radio emission at 33 GHz. ($Top~Left$) Galaxy-integrated spectral indices, $\mathrm{\alpha_{1.49-6GHz}}$ versus $\alpha_{6-33GHz}$. The solid line shows a slope unity (i.e., a single spectral index across all bands). The dashed lines indicate a typical, optically thin synchrotron emission slope of $-0.8$. We find a median $\mathrm{\alpha_{1.49-6GHz}}$ of $-0.62$ and a median $\mathrm{\alpha_{6-33GHz}}$ of $-0.64$. There is some tendency for the radio SED to become steeper at high frequency (i.e., for points to lie above the line). ($Top~Right$) Predicted thermal fraction at 33 GHz, based on comparing the IR luminosity to the integrated flux density at 33 GHz, as a function of the infrared luminosity. Most of the systems show thermal fractions of $\geq~50\%$, in agreement with SED models \citep{Condon&Yin90,Condon92}. ($Bottom$) Half-light area as a function of $\mathrm{\alpha_{1.49-6GHz}}$. There is a tentative correlation of flatter spectral index in the range 1.5 to 6 GHz for more compact sources. This could be expected given that compact sources are more obscured, and therefore more subject to free-free absorption at low frequencies. In all panels, individual systems are labeled by the ID assigned in Table \ref{table:tbl-1} and color coded by their infrared luminosity.
\label{fig:fig2}}
\end{figure*}

In Figure \ref{fig:fig2}, we show the derived spectral indices. We plot $\alpha_{1.5-6}$ as a function of $\alpha_{6-33}$. Here the solid line shows equal spectral indices for both pairs of bands, which we would expect if a single spectral index holds across the entire radio regime (from 1.5 to 33 GHz). Dashed lines show $\alpha = -0.8$, a typical spectral index for synchrotron emission without any opacity effects \citep[e.g.,][]{Condon92}.

\subsection{Size of the Radio Emission}
\label{sec:sizes}

A main goal of our study is to measure the extent of the radio continuum emission in our targets with the purpose of constraining the size of the energetically dominant region. 

To do this, we analyzed the final images combining data from all the array configurations. These high resolution images are sensitive to the brightest compact cores, but they have lower surface brightness sensitivity than the D configuration data that we used to determine the total flux density. Therefore, they may miss extended, low surface brightness emission. To take this into account, we measure the size of the energetically dominant region from the half-light area (A$_{50}$). This is the area enclosed by the highest intensity isophote that includes half of the total integrated flux density of the system, which we measured from the lower resolution data above and expect to be complete. Note that this approach measures the observed A$_{50}$, which reflects the true size of the source convolved with the synthesized beam of the array.

We require the intensity of the isophote enclosing the half-light area, or C$_{50}$, to be at least $5$ times the rms noise in the image. If C$_{50}$ would be less than 5$\sigma$ in the combined image, we interpret this to indicate an important component of extended, low surface brightness emission. In order to recover this emission, we measure A$_{50}$ for these systems from lower resolution versions of the data that have better surface brightness sensitivity. In these cases, we first tried using \texttt{natural} weighting instead of \texttt{Briggs} (see Section \ref{sec:red}). If we still could not recover half of the light within a S/N$>5$ contour, then we produced progressively lower resolution images by applying larger and larger $u-v$-tapers to the data. We stepped the size of the taper by 0$\,\farcs$2 and used \texttt{Briggs} weighting schemes with \texttt{robust} parameter 0.5 at each step. In this way, we measure A$_{50}$ from the highest resolution image where C$_{50}$ can be reliably measured, i.e., where C$_{50}$ $\geq$ 5$\sigma$.

The following systems showed extended, low surface brightness emission and required $u-v$-tapering: CGCG 436-030, CGCG 448-020, IRAS 21101+5810, IRAS 17132+5313, VV 340a and VV 705. For NGC 3690, the natural weighting approach was sufficient.

Once we identified a reliable half-light contour, C$_{50}$, we calculated A$_{50}$ by multiplying the number of pixels within the C$_{50}$ contour by the pixel area. Figure \ref{fig:fig1} shows the images that were used to measure A$_{50}$ and the C$_{50}$ contour (in red) for each source.

\begin{figure*}[tbh]
\centering
\includegraphics[width=3.0in]{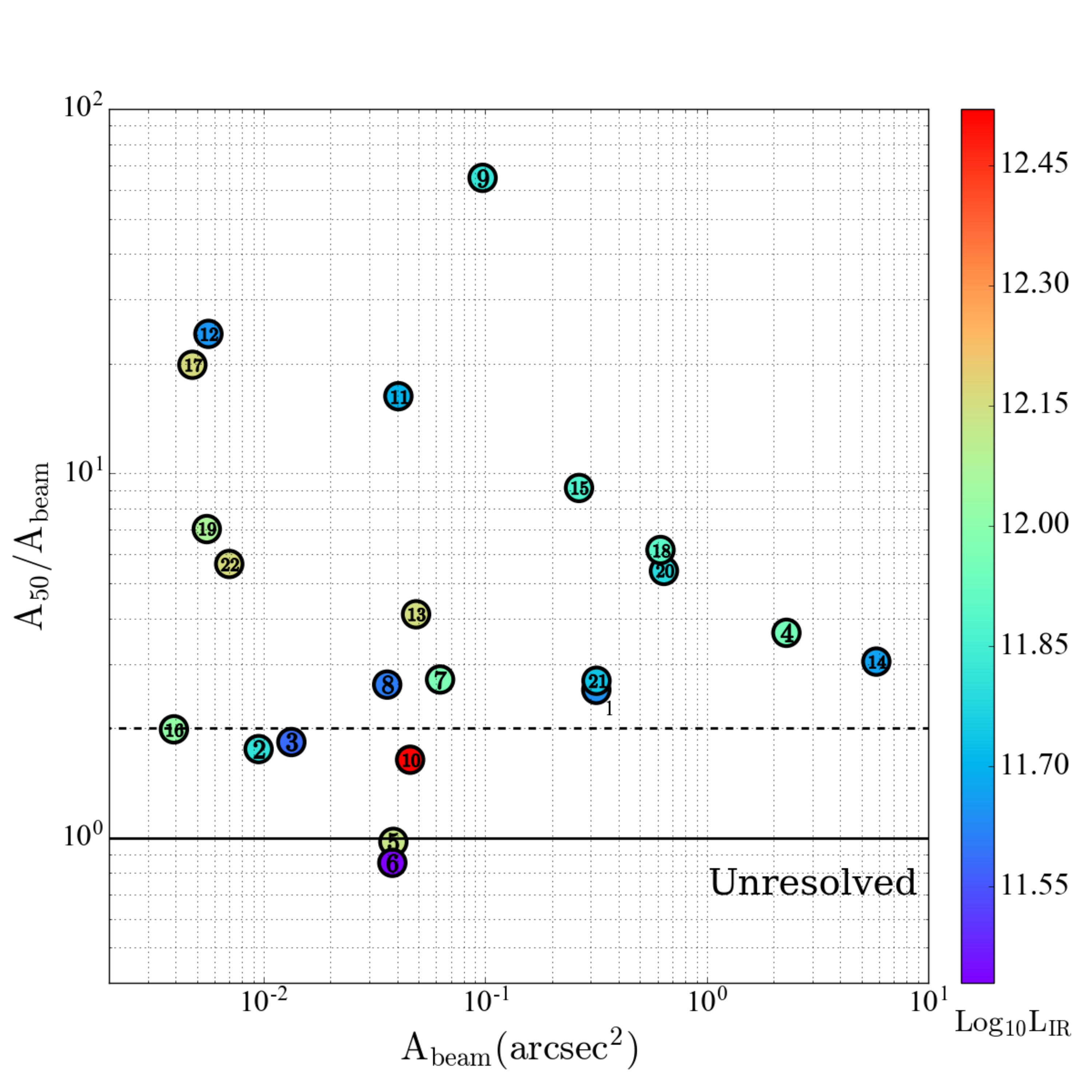}
\includegraphics[width=3.0in]{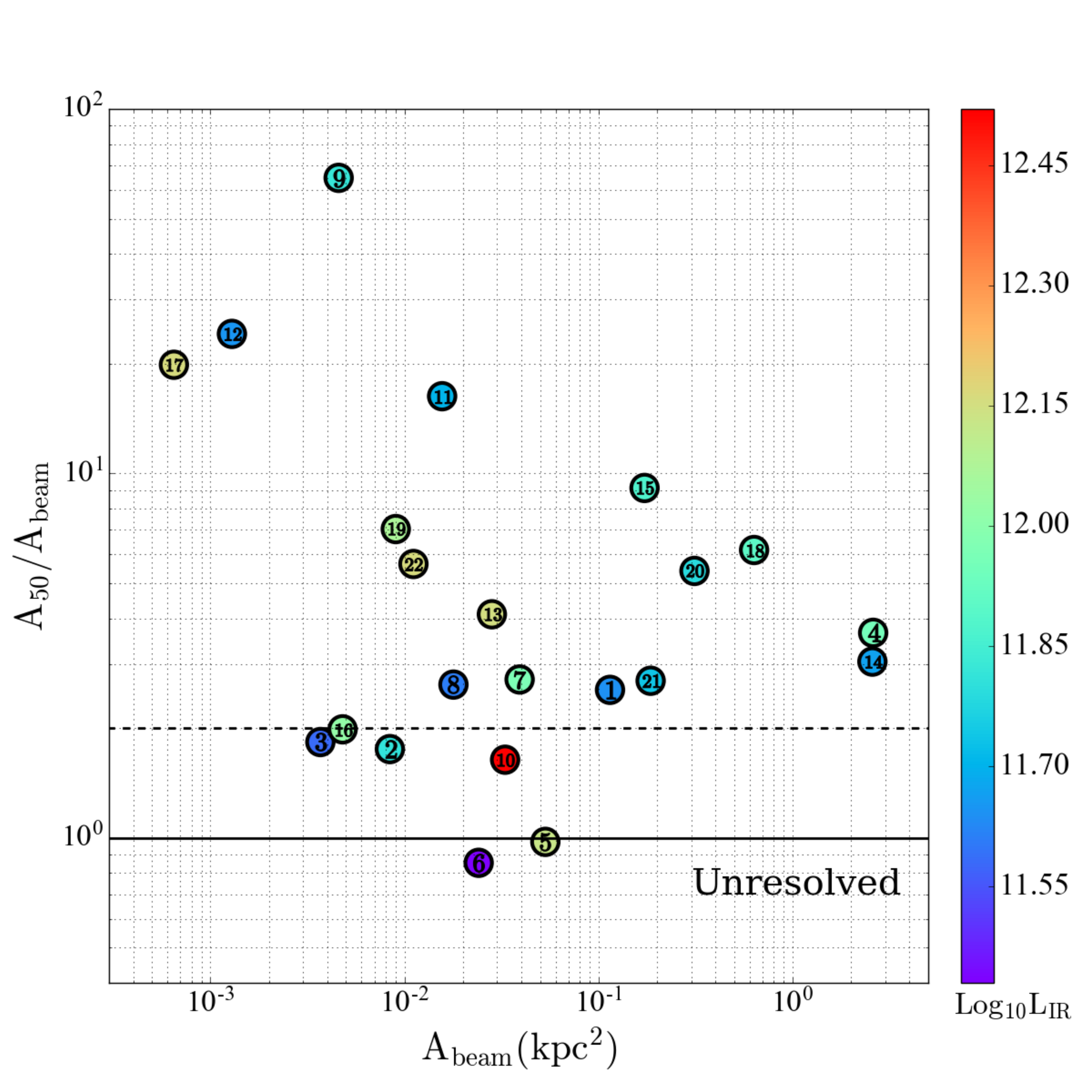}
\includegraphics[width=3.0in]{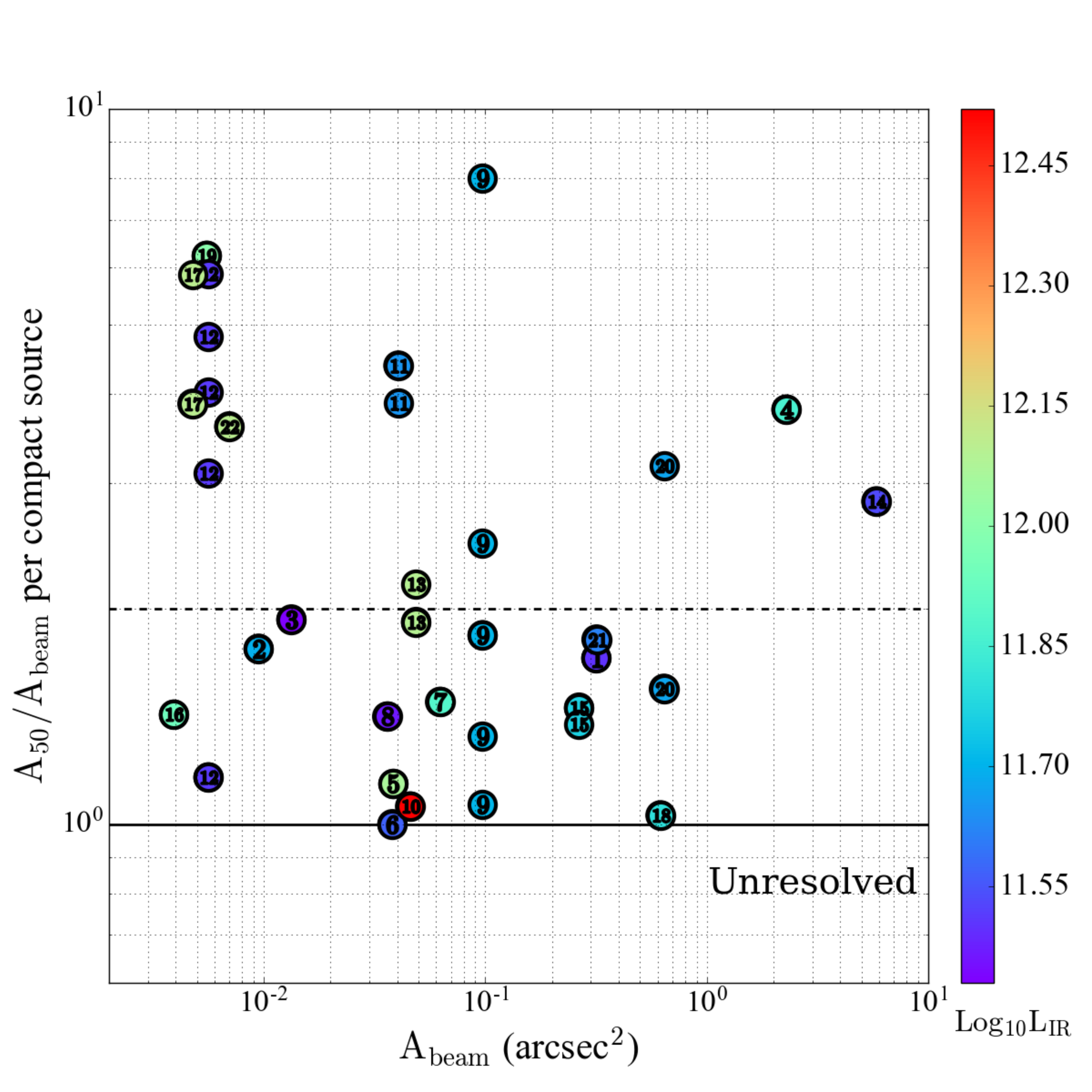}
\includegraphics[width=3.0in]{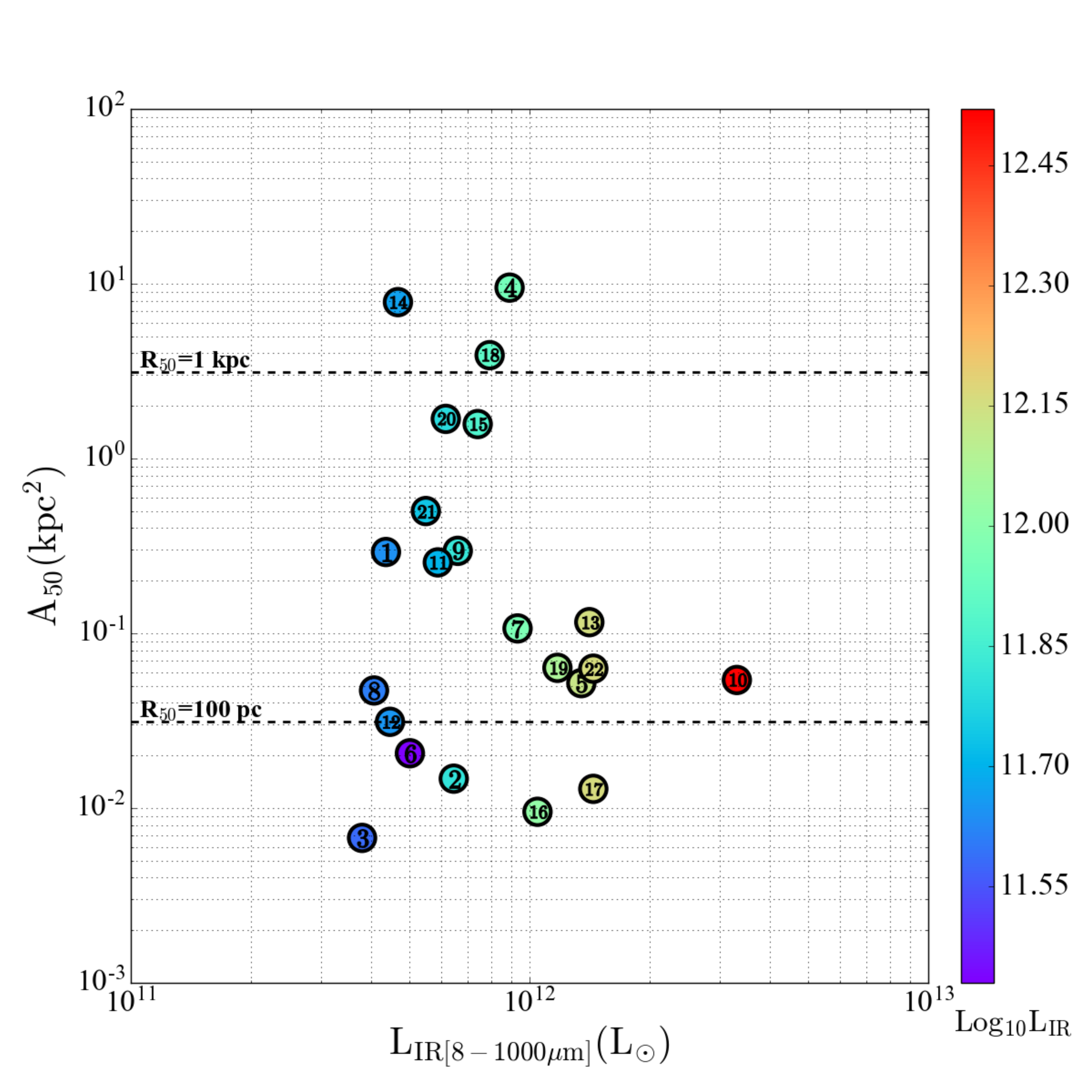}
\caption{Sizes of the 33~GHz emitting regions in our targets. ($Top~left$) Ratio of observed A$_{50}$, the area enclosing 50$\%$ of the total flux density, to beam area, versus beam area in arcsec$^2$. Sources between the solid and dashed line are considered marginally resolved. Above the dashed line sources are resolved, while below the solid line sources are considered unresolved. ($Top~right$) Same as in the previous panel, but now in physical area, kpc$^{2}$. ($Bottom~left$) Same as in the $Top~left$ panel but for compact sources within the observed galaxies obtained from Gaussian fitting (see Section \ref{sec:comp}) ($Bottom~right$) Observed A$_{50}$ versus infrared luminosity of the source, $\mathrm{L_{IR}}$, with source sizes (radii) of 100~pc and 1~kpc marked for reference (see Section \ref{sec:sizes}). The top panels show that we at least marginally resolve all but two of our targets. The bottom right panel shows that the emission often breaks into a collection of compact regions with sizes only a small factor larger than the beam area. The bottom right panel shows a weak tendency for the highest luminosity sources to also be the most compact. In all panels, individual systems are labeled by the ID assigned in Table \ref{table:tbl-1} and color coded by their infrared luminosity.}
\label{fig:fig3}
\end{figure*}

Many of our sources show sizes close to that of the synthesized beam. We show this in Figure \ref{fig:fig3}. There, we plot the ratio of the observed A$_{50}$ to the beam area, $\mathrm{A_{beam}}$, as a function of the beam area in units of arcsec$^{2}$ (top left panel) and kpc$^{2}$ (top right panel). 

The quantity of physical interest is the true size of the 33 GHz emission with the beam deconvolved, $\mathrm{A_{50d}}$. In the top, and bottom left panels of Figure \ref{fig:fig3}, a dashed line indicates a value of $\mathrm{2\times A_{beam}}$, which we consider a practical threshold for the emission to be viewed as resolved. Here $\mathrm{A_{beam}=\frac{\pi \theta_{\rm maj} \theta_{\rm min}}{4}}$ with $\mathrm{\theta_{\rm maj}~and~\theta_{\rm min}}$ the FWHM of the synthesized beam along its major and minor axis. In this definition, $A_{beam}$ refers to the area expected to enclose half the total power in the beam. This definition is consistent with our measured area $A_{50}$, and appropriate for deconvolution.

We treat the sources that show extent larger than the beam but size smaller than $2\times A_{beam}$ as marginally resolved (region between the solid and dashed lines in Figure \ref{fig:fig3}). In these cases, we assume that the intrinsic shape (deconvolved size) of the source follows a Gaussian distribution. We then estimate the deconvolved size of the source by $\mathrm{A_{50}(deconvolved)=A_{50}(observed)-A_{beam} \equiv A_{50d}}$, equivalent to deconvolving the FWHM in quadrature. 

In the top panels of Figure \ref{fig:fig3}, two sources lie below the solid line, indicating an observed size smaller than the beam. These are IRAS F08572+3915 and UGC 04881NE. Although statistical fluctuations could produce this situation, the signal-to-noise of the data appear to be too high for this explanation to hold. The most likely culprit is a calibration issue when combining observations using the different array configurations. We adopt a conservative upper limit of $\mathrm{A_{50d} < A_{beam}}$ for these two systems\footnote{For UGC 04881NE, $\alpha_{6-33}\approx -1.2$ which is unusually steep. We also consider its flux density at 33 GHz as a lower limit.}.

In order to determine the best estimate of A$_{50d}$ for ``resolved'' sources with $\mathrm{A_{50} > 2\times A_{beam}}$, we inspected the shape of the C$_{50}$ contour (red in Figure \ref{fig:fig1}) to determine if the source exhibits a Gaussian shape. If it did, then we apply the same approach used for the marginally resolved sources to each component and summed the results to find the total A$_{50d}$. This tended to be the case when more than one component is present, such as VV 705 and CGCG 448-020.

If C$_{50}$ showed a more complex morphology, our simple Gaussian treatment becomes invalid. In these cases we instead assume that the measured, not deconvolved $A_{50}$ represents an upper limit to the true size. This is true for the following galaxies: IRAS 19542+1110, IRAS F23365+3604, UGC 08387, VII Zw 031, VV 250a and VV 340a. 

For two sources, UGC 04881 and VV250, a second, faint component could be recovered only in the low resolution map used to assess the integrated flux density. In both cases, the individual components are unresolved in this integrated map. Here, we had to lower our conservative limit of 5$\sigma$ in order to recover the half-light area. In these two systems, we measure C$_{50}$ from a contour with $S/N \approx 3$ and treat the size estimate as an upper limit (see Table \ref{table:tbl-4}).

For NGC 3690 and IRAS 17132+5313, one component of the C$_{50}$ contour shows a Gaussian distribution while others show more complex morphology. In both cases, we performed the deconvolution on the Gaussian components. Then we have a partially deconvolved estimate, $\mathrm{A_{50d} < A_{50} (observed)}$, which is still an upper limit because of the un-deconvolved more complex structure. We report values for A$_{50d}$ and C$_{50}$ in Table \ref{table:tbl-4}, along with an equivalent R$_{50d}$ value where $\mathrm{A_{50d}=\pi R_{50d}^{2}}$. We caution, however, that $R_{50d}$ is only a representative number reflective of the upper limit to the area in these cases. 

In Table \ref{table:tbl-4}, we also report the degree of Gaussianity, defined as the ratio between the flux density level of the C$_{50}$ contour and the peak flux density. For a two dimensional Gaussian, this value is 0.5.

\subsection{Compact Sources Decomposition}
\label{sec:comp}

In addition to the integrated flux density and a characteristic size, we measured the contribution of compact sources to the overall flux density of each target using the maps of Figure \ref{fig:fig1}. For our purposes, compact sources are those that clearly belong to the system and show a Gaussian morphology.

\begin{figure*}[tbh]
\centering
\includegraphics[width=3.0in]{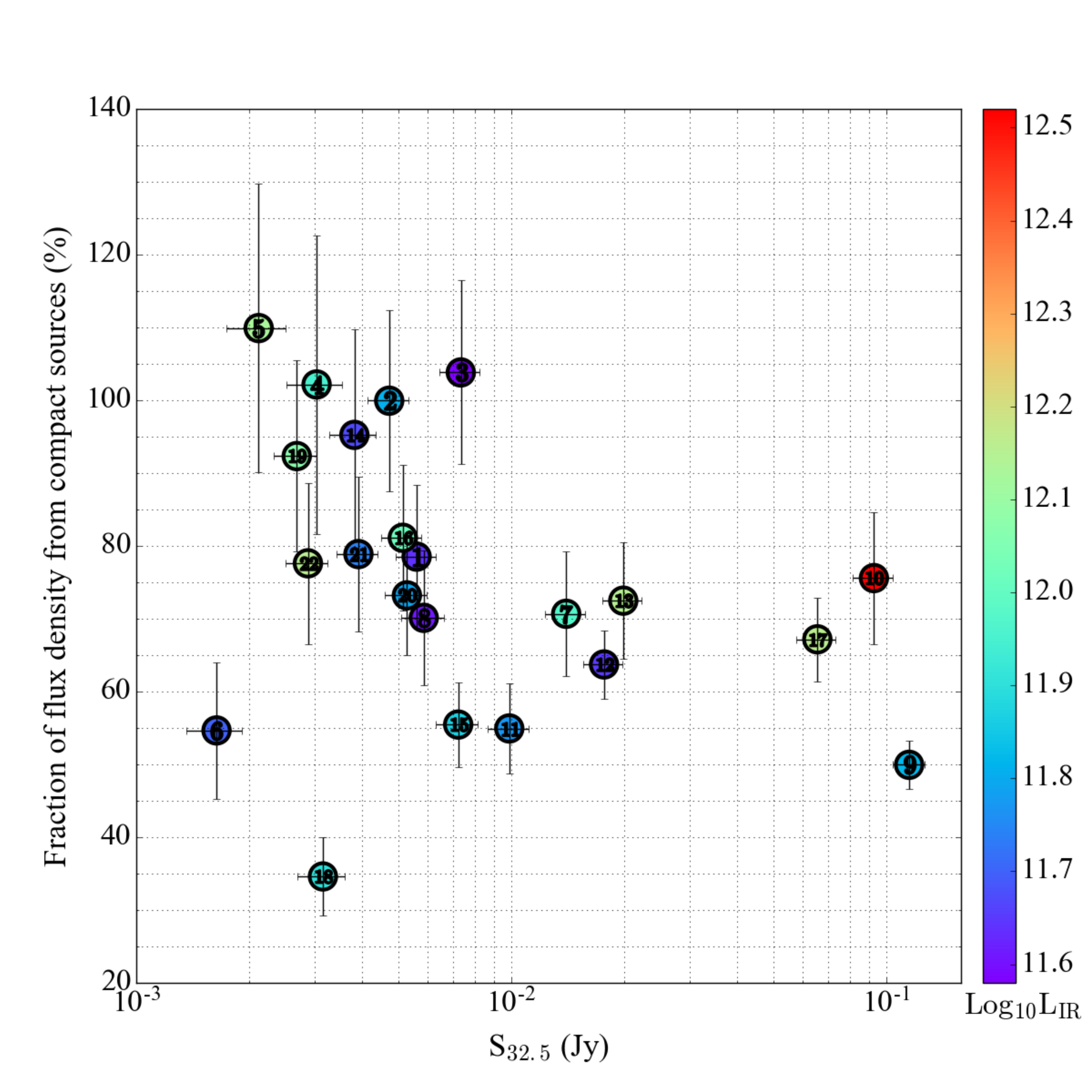}
\includegraphics[width=3.0in]{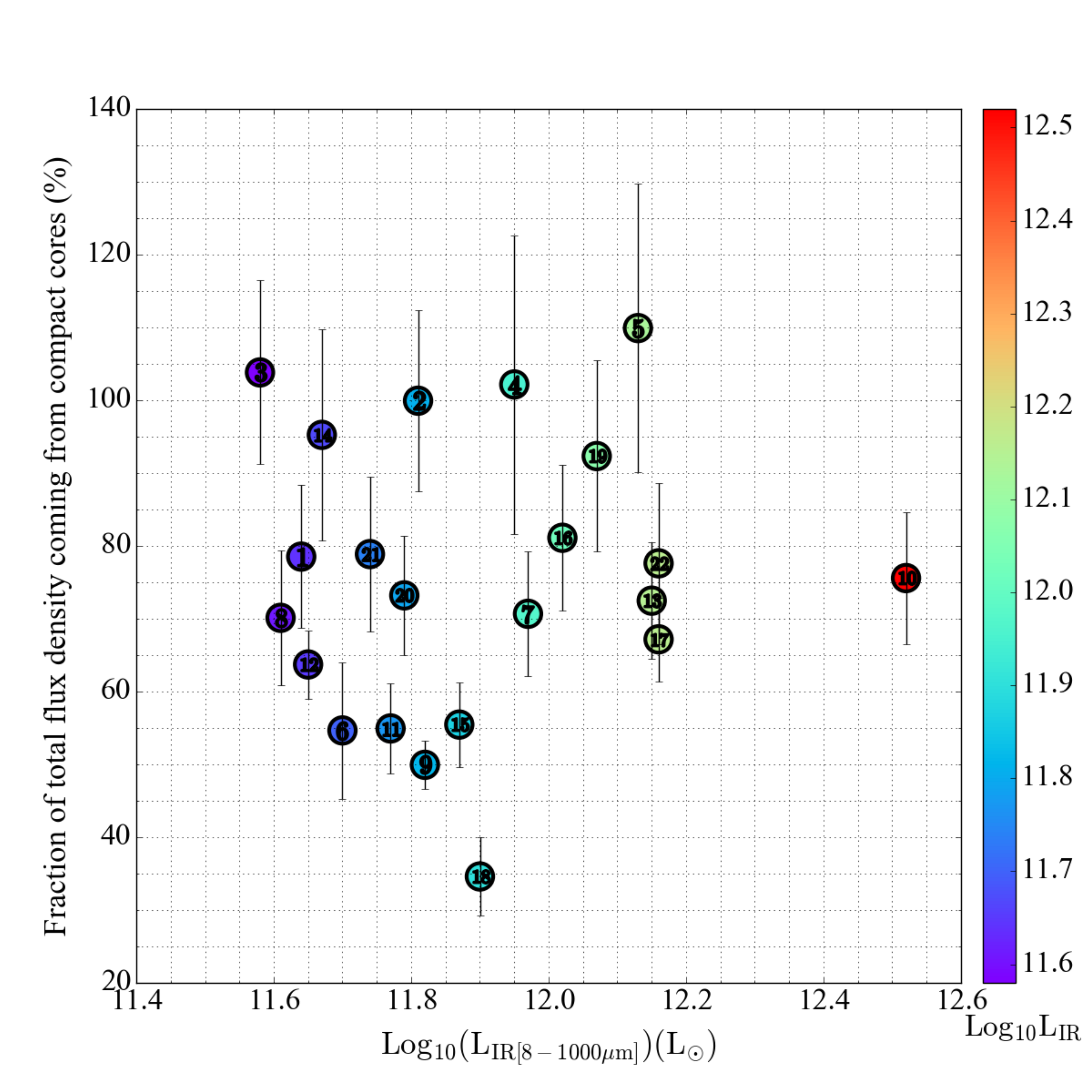}
\includegraphics[width=3.0in]{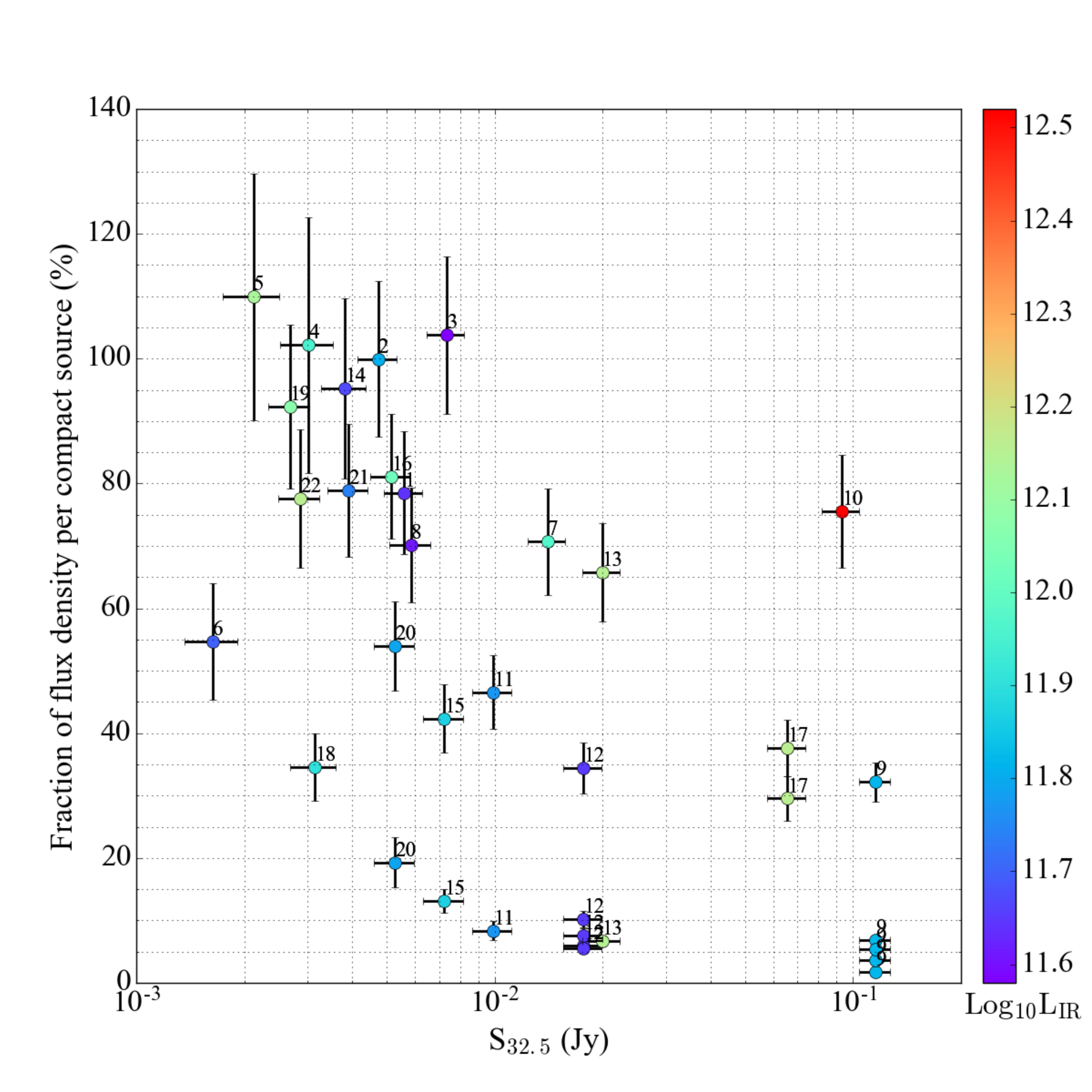}
\caption{Flux density contribution from compact sources. ($Top~left$) Percentage of the 33~GHz flux density arising from compact sources as a function of the total 33~GHz flux density. ($Top~right$) Percentage of the 33~GHz flux density arising from compact sources as a function of the total infrared luminosity. ($Bottom$) Same as $Top~left$ panel, but now plotting each individual compact source as a point. Most of the 33 GHz emission in our sample is concentrated in compact sources instead of extended emission. In all panels, individual systems are labeled by the ID assigned in Table \ref{table:tbl-1} and color coded by their infrared luminosity.
\label{fig:fig4}}
\end{figure*}

For each target, we identify these sources by eye and fit them using \texttt{imfit}, providing estimates of the rms noise and reasonable starting guesses for the sizes, and peak intensity and position. The locations of the fit compact sources appear as white crosses in Figure \ref{fig:fig1}. Their sizes, which are often comparable to the size of the beam, are shown in the bottom left panel of Figure \ref{fig:fig3}. We also calculated the flux density that is originating from all the compact sources in a system, and compared it to the integrated flux density (see top panels in Figure \ref{fig:fig4}). We note that such comparisons may be affected by the different physical resolutions achieved from the observations, however we find no trend relating the fraction of flux in compact sources to beam physical area. In the bottom panel of Figure \ref{fig:fig4} we show instead the contribution of each point source -- especially important when more than one is present -- to the integrated flux density at 33 GHz.

We identified compact sources in each of our targets except the north-east component in IRAS F17132+5313, which shows mostly extended emission. For the cases of the faint components in the systems UGC 04881 and VV250, the Gaussian fit was performed on the low resolution image that was used to obtain the integrated flux density of the system.

A subset of our sources show most of their emission concentrated into a very small area, consistent with a point source producing much of the flux density even at our highest angular resolution. To make the strongest possible measurement of the compactness of these targets, we used our highest resolution images. This is usually the A configuration image ($\sim0\,\farcs1$), except in those cases with B as the longest baseline array configuration observed ($\sim0\,\farcs2$; see Table \ref{table:tbl-3}). 

From this highest resolution image, we measured the flux density detected at S/N$\geq5$, which corresponds well to the total flux density in the compact core of the image. We compared this flux density in the bright core at the highest resolution to the integrated flux density of the system, $f_{A~(or~B)}$. Most of the U/LIRGs in this sample show single bright point sources in the highest resolution image, although a few, including NGC 3690, UGC 08387, Arp 220, and VV 705, show more than one compact core.

We also measured the size of the 33 GHz emission showing significant detection, as set by the 5$\sigma_{A~(or~B)}$\footnote{$\sigma_{A~(or~B)}$ is the rms noise of the A (or B) array configuration image.} contour, at this highest angular resolution image. We report the beam size of the A, or B, array configuration images along with the sizes of the 5$\sigma_{A~(or~B)}$ contour and $f_{A~(or B)}$ of each system in Table \ref{table:tbl-5}. We highlight those sources with most of their emission being contributed by a single bright compact source, being good potential AGN candidates. These include: IRAS F01364-1042, III Zw 035, and IRAS 15250+3609. Arp 220 should also be on this list as it shows $f_{A~(or~B)}>50\%$, however we refer the reader to a more exhaustive discussion on the morphology of its 33 GHz emission presented in \citet{BM15}. There are six other sources with $f_{A~(or~B)}>50\%$, but unfortunately the highest resolution achieved was only $\sim0\,\farcs2$ and the constraint on their compactness is then weaker. However, note that Mrk 231, a known AGN \citep[e.g.,][]{Ulvestad99,Lonsdale03}, belongs to that group.

In Table \ref{table:tbl-5} we also note two systems, VII Zw 031 and VV 340a, with $f_{A~(or B)}<1\%$ indicating most of their emission at $0\,\farcs1$ resolution is filtered out, and then is mostly extended in nature.

\section{Implications of the Radio Sizes}
\label{sec:impl_radio}

From the 33~GHz images, we either measure or strongly constrain the size of the energetically dominant regions in our targets. Radio interferometers are almost unique in their ability to peer through heavy dust extinction while also achieving very high angular resolution. As a result, similar sizes are difficult to obtain at other wavelengths. Here, we assume that the energetically dominant region traced by the radio data has approximately the same size as the region bearing the mass or emitting the light at other wavelengths. This allows the calculation of intensive (per unit area or volume) quantities.

Our method to do this, in general, is to assume that half of the flux at some other wavelength of interest (e.g., 1.4 GHz, IR[8--1000$\mu$m] and CO emission) is enclosed in the $33$~GHz half-light area, $A_{\rm 50,d}$. We then calculate the surface brightness and related parameters (surface density, volume density) implied by this assumption. 

Note that in several cases, we expect optical depth to play a key role (e.g., at 1.4~GHz or in the IR). In this case, the $\tau \approx 1$ photosphere may lie outside the calculated size \citep[e.g., see][]{BM15}. In other cases, our assumption that the radio structure indicates the structure at other wavelengths may break down (e.g., if an AGN contributes significant IR but weak radio emission or if gas traced by CO decouples from star formation). We discuss these cases in the individual sections and report the derived values in Table~\ref{table:tbl-6}.

\subsection{Brightness Temperatures}
\label{sec:Tb}

For a resolved or nearly resolved source, where beam filling is a minor consideration, the brightness temperature, $T_{b}$, offers the prospect to constrain the emission mechanism and opacity of the source \citep[e.g.,][]{Condon91}. At radio frequencies, the brightness temperature, T$_{\rm b}$, follows the Rayleigh-Jeans approximation where

\begin{equation}
\label{eq:Tb}
T_{b} = \Big(\frac{S_{\nu}}{\Omega_{source}}\Big)\frac{c^{2}}{2 k_{B} \nu^{2}}~,
\end{equation}

\noindent with $S_{\nu}$ the flux density at frequency $\nu$ and $\Omega_{source}$ the area of the source. 

\begin{figure*}[tbh]
\centering
\includegraphics[width=3.5in]{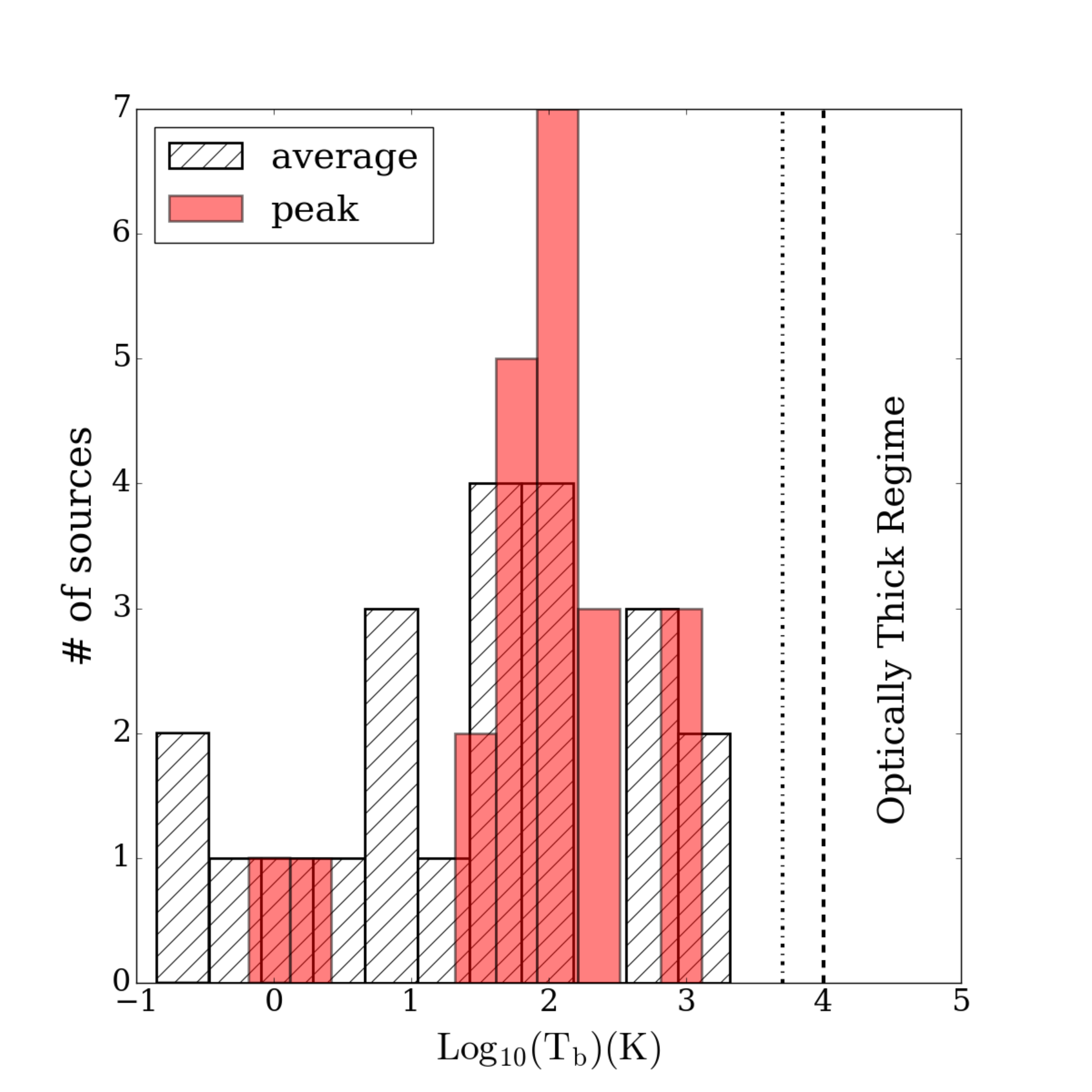}
\caption{Histogram of averaged nuclear T$_{b}$ within A$_{50d}$ and peak T$_{b}$ (see Section \ref{sec:Tb}). The dotted-dashed and dashed vertical lines indicate plausible values for the temperature of {\sc Hii} regions, $\mathrm{T_{e} \sim 5\times10^{3}~K{-}10^{4}}$~K. We expect any source having T$_{\rm b}$ above these limits to have optically thick free-free emission. We measure T$_{b}$ to be below this range, suggesting that our targets are either optically thin at $32.5$~GHz or highly clumped. The images do not appear to resolve into clumpy substructure, and the spectral index also supports an optically thin interpretation.
\label{fig:fig5}}
\end{figure*}

Most of our targets are resolved. Thus an ``averaged nuclear T$_{\rm b}$'' at 32.5 GHz can be derived using $\Omega_{source}$ = A$_{\rm 50d}$ and S$_{\nu}$ = 0.5 $\times$ S$_{32.5}$ (see above for the explanation of the aperture correction). We also calculate T$_{\rm b}$ from the point of highest intensity in the highest resolution image for each target, peak T$_{\rm b}$, where $\Omega_{source} = \Omega_{beam}$ in that case. Figure \ref{fig:fig5} shows histograms of these peak and averaged nuclear T$_{\rm b}$ at $\nu = 32.5$~GHz.

The averaged nuclear $T_{\rm b}$ for our targets is typically a few 10s of Kelvin to a few times 100~K, reaching up to a few thousand Kelvin in the brightest targets. 

For only free-free emission filling the beam, we would expect $T_{\rm b}$ for optically thick emission to approach $T_{\rm e}$ for the {\sc Hii} regions. For physical conditions like those present in our sample, the expected electron temperature, T$_{\rm e} \sim 5000{-}10^{4}$~K \citep{Hummer&Storey87,Condon92}. In metal rich environments, such as the central regions of ULIRGs \citep{Veilleux09}, the cooling is more efficient and T$_{\rm e}$ may tend towards the low end of this range, $\sim 5000$~K \citep[e.g.,][]{Puxley89}, though note that \citet{Ana00} found T$_{\rm e}$ of 7500 K for Arp 220 from integrated measurements of radio recombination lines.

In Figure \ref{fig:fig5} we observe T$_{\rm b}$ does not exceed either 10$^{4}$~K or 5000~K for any galaxy. In theory the unresolved, or marginally resolved, sources could be optically thick and highly clumped at scales much smaller than the beam size. However, both the observed spectral index (which would be positive with $\alpha \sim 2$ for the free-free emission if optically thick) and the relative smoothness of the images argue against such a scenario. Instead, low opacity at $32.5$~GHz appears to be the natural explanation for the $T_{\rm b}$ that we observe.

In Figure \ref{fig:fig5} we observe the peak brightness temperatures do not exceed the likely $T_{\rm e}$. However, T$_{\rm b}$ (peak) should be treated as a lower limit for the unresolved and marginally resolved sources. Are these sources likely to be optically thick? Excluding the case of Mrk 231 since it hosts an AGN, the lower limits for the peak T$_{\rm b}$ go from 20 K up to 690 K, with the unresolved case, UGC 04881, having a temperature of 22 K. In the marginally resolved cases, we can gain insight into the likely size of the source by contrasting the peak and average T$_{\rm b}$. Figure \ref{fig:fig3} shows that for most of these marginally resolved with T$_{b}$(peak) $<$ T$_{b}$(average), we would expect to be able to resolve them with a beam area that is 2 times smaller at most. This would imply a true $T_{\rm b}$ peak $\sim 2$ times larger than what we measure, still not enough for these sources to reach the optically thick regime. In these marginally resolved cases, in particular, the substructure of the emission remains unclear. Our data offer limited insight into whether the data may be structured into smaller optically thick regions beneath the beam.

For the unresolved source, the situation is less clear. With the size unconstrained, the source could be optically thick at 33 GHz and heavily beam diluted. However, we note again that the spectral index that we observe does not appear consistent with optically thick free-free emission. We proceed assuming that we observe optically thin 33 GHz emission for this source.

The flux densities of many of our targets have been measured at $1.4$~GHz (Table \ref{table:tbl-3}), but even in its most extended configuration, the VLA reaches only $\approx 1\arcsec$ resolution at this frequency. Using the measured $1.4$~GHz flux densities, we calculate the averaged nuclear T$_{\rm b}$ at $1.4$~GHz assuming that the $32.5$~GHz sizes also describe the true extent of the $1.4$~GHz emission. These span 10$^{3}$ up to 10$^{6.5}$ K. 

These are high values. Values of T$_{\rm b}>5\times10^{3}$ or 10$^{4}$ K, imply that the emission at 1.49 GHz is mostly synchrotron in nature, because the source function of the free-free emitting ionized gas is a black body at $T \sim 5\times10^3$ - 10$^{4}$~K, as explained above.

Dominant synchrotron emission may be expected at $1.4$~GHz, but the values that we find may in fact be too high for the standard mixture of free-free and synchrotron emission seen in starburst galaxies. Considering such a mixture, \citet{Condon91} suggested a maximum T$_{\rm b}$ for a starburst of $\sim$10$^{4.6}$ K at 1.49 GHz (their Equation 9, using $\mathrm{T_{e}\sim5000~K}$). At least 12 sources in the sample show T$_{\rm b~1.4GHz}>10^{4.6}$ K when we combine the 33~GHz sizes and the 1.49~GHz flux densities (see Table \ref{table:tbl-6}). This could imply that the 1.49~GHz emission from these sources includes a significant AGN contribution. One of those sources, Mrk 231, is well known to be dominated by an AGN, which explains why it has the highest predicted averaged nuclear T$_{\rm b}$ at 1.49 GHz.

Based on this line of argument, for these high brightness $1.49$~GHz sources, we would expect much of the flux density to be confined to an unresolved core in VLA $1.49$~GHz imaging. In Mrk 231, most of the emission is unresolved at 1.49 GHz, however, other sources show resolved emission at 1.49 GHz. In these cases, the 33 GHz sizes, which are small compared to the $\sim 1.5''$ VLA beam, may not be representative of the true $1.49$~GHz emission. Indeed, we might worry that the 33~GHz size will underestimate the size at 1.49 GHz if the system is optically thick at these lower frequencies. In such case, the emission will emerge from a photosphere larger than the emitting (optically thin) region at $33$~GHz and the true brightness temperature at $1.49$~GHz will be lower than our estimate. 

Another alternative is that an extended synchrotron component may contribute to the integrated flux density. This component would have to have a spectrum steep enough that it does not contribute much to the flux density at $33$~GHz, implying substantial variations in the resolved spectral index.

On the other hand, several sources have high $T_{b}$ and remain barely resolved even at 8.44 GHz \citep[see maps in][]{Condon91}: IRAS 08572+3915, IRAS 17132+5313, IRAS 15250+3609 and III Zw 035. These are our best AGN candidates based on $T_{b}$ arguments. Here extra information is needed to determine whether they are powered by an AGN and/or starbursts.  In an upcoming paper, we investigate this possibility by combining the current observations with the lower frequency ($\nu = 4{-}8$~GHz) part of our survey (L. Barcos-Mu\~{n}oz et al. in preparation).

\subsection{Star Formation Rate and IR Surface Density} 
\label{sec:Ssfr}

Infrared luminosity, L$_{\rm IR}[8-1000\mu m]$, and radio emission both trace recent star formation in starburst galaxies. IR luminosity reflects reprocessed light from young stars, while the 33~GHz continuum predominately captures a mix of synchrotron and thermal emission, both of which originate indirectly from young stars.

Considering a mix of synchrotron and thermal emission, \citet{Murphy12} relate the recent star formation rate to the 33~GHz luminosity, $L_{\rm 33GHz}$, via

\begin{equation}
\label{eq-sfrrad}
\begin{split}
\left(\frac{\rm SFR_{\nu}}{M_{\sun}\,{\rm yr^{-1}}}\right) &= 10^{-27}  
\left[2.18 \left(\frac{T_{\rm e}}{10^{4}\,{\rm K}}\right)^{0.45} \left(\frac{\nu}{\rm GHz}\right)^{-0.1}\right. + \\
%&\left.1300 \left(\frac{q_{\rm SNR}}{\rm 0.0116~yr^{-1}}\right) \left(\frac{\nu}{\rm GHz}\right)^{-\alpha^{\rm NT}}\right]^{-1}\\
&\left.15.1 \left(\frac{\nu}{\rm GHz}\right)^{\alpha^{\rm NT}}\right]^{-1} \left(\frac{L_{\nu}}{\rm erg~s^{-1}~Hz^{-1}}\right).  
\end{split}
\end{equation}

\noindent where $T_{e}$ is the electron temperature and $\alpha_{\rm NT}$ is the nonthermal spectral index. \citet{Murphy12} relate the infrared luminosity to the recent star formation rate via

\begin{equation}
\label{eq:sfrir}
\mathrm{\left(SFR_{IR} \over {\rm M_\odot~yr^{-1}} \right) = 3.15\times 10^{-44} \left ( L_{IR}[8-1000\mu m] \over {\rm ergs~s^{-1}} \right)}~.
\end{equation}

\begin{figure*}[tbh]
\centering
\includegraphics[width=3.5in]{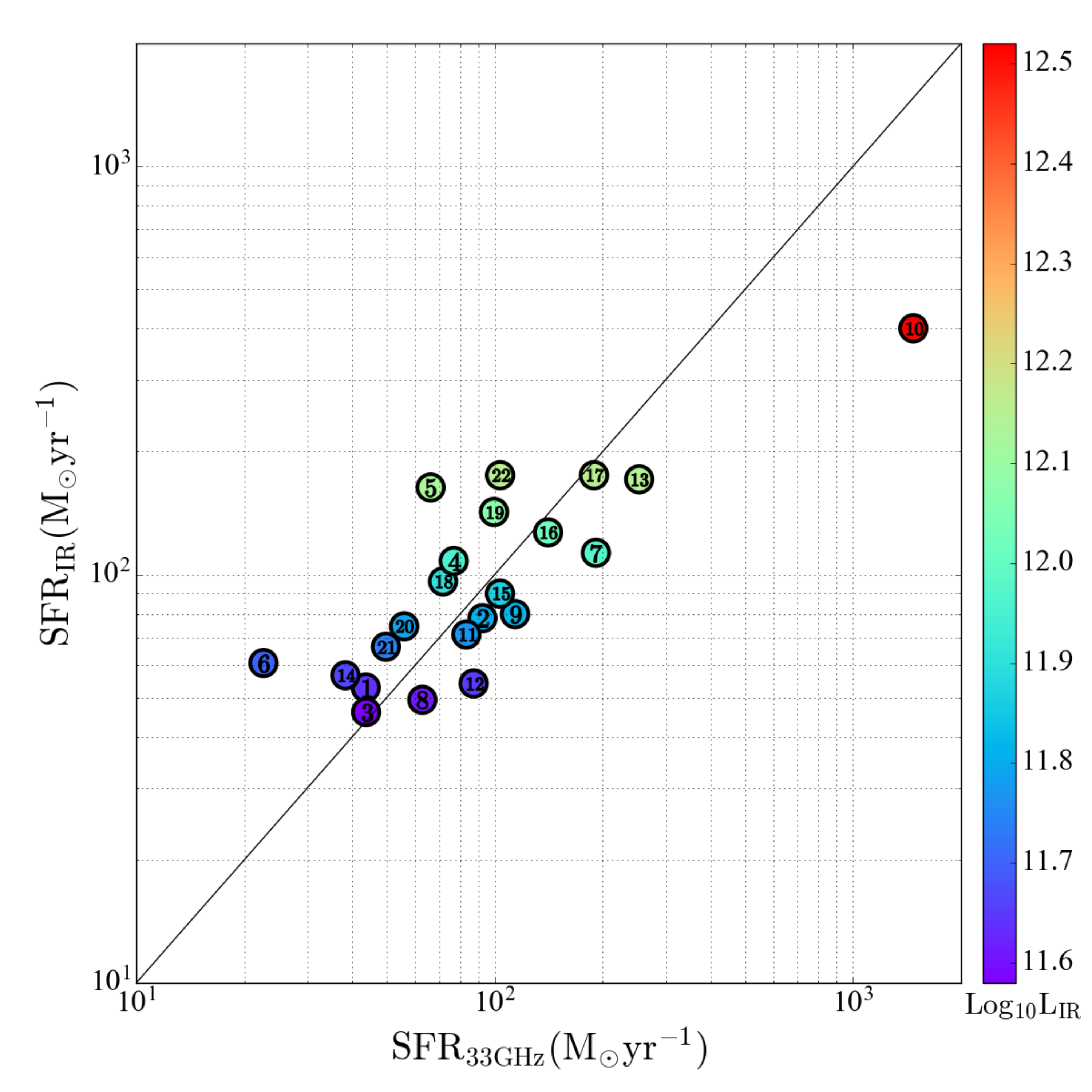}
\includegraphics[width=3.5in]{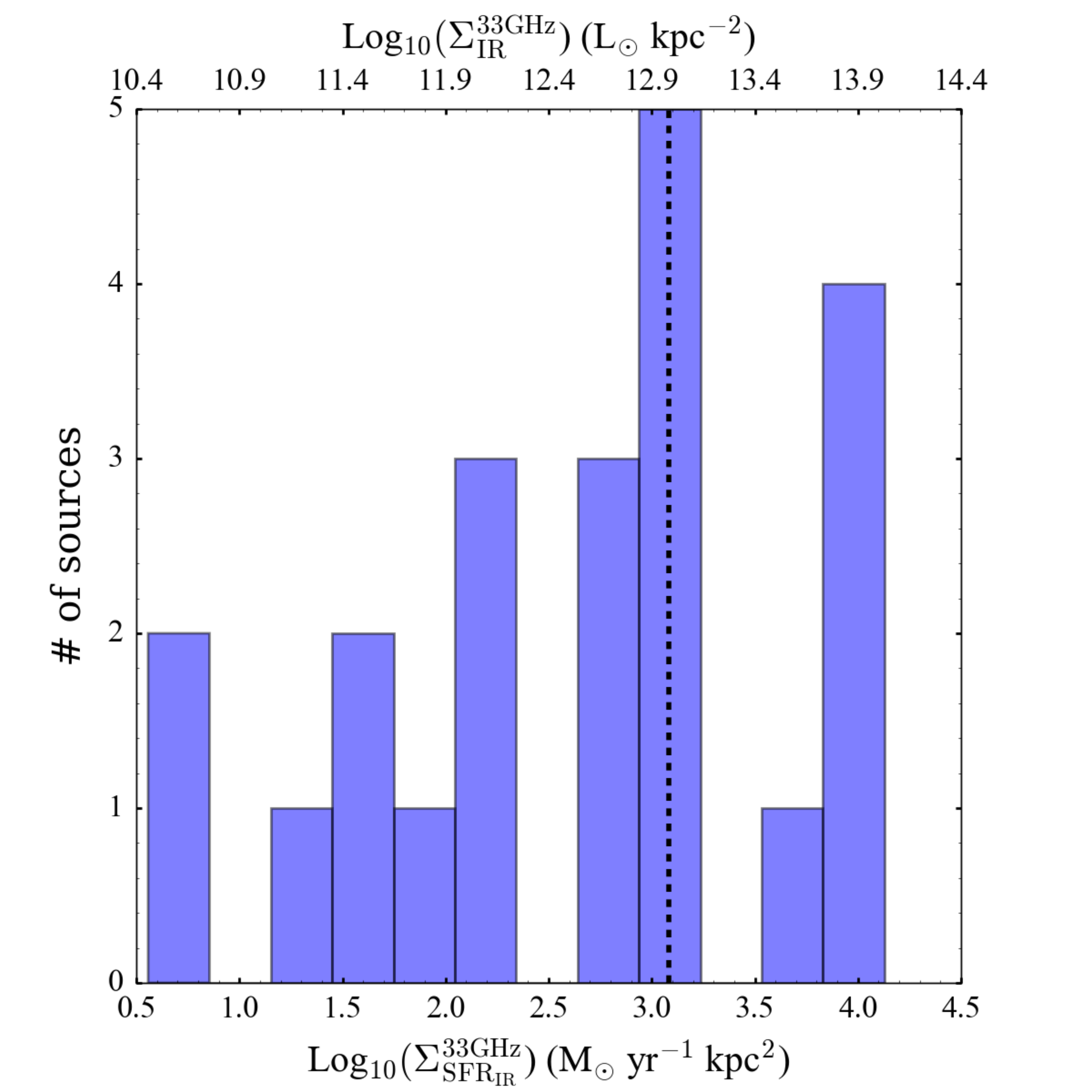}
\caption{$(Left)$ SFR calculated from the IR (see equation \ref{eq:sfrir}) versus SFR calculated from the 33 GHz (see equation \ref{eq-sfrrad}). The solid line shows a one-to-one relation. Individual systems are labeled by the ID assigned in Table \ref{table:tbl-1} and color coded by their infrared luminosity. There is an overall agreement between the two methods, indicating the systems follow a version of radio-IR correlation at 33~GHz and that the calibrations \citet{Murphy12} are consistent with this relation. The outlier with high radio flux is Mrk 231, a known AGN. $(Right)$ Histogram of $\Sigma_{\rm IR}^{\rm 33GHz}$ and the corresponding $\Sigma_{\rm SFR_{IR}}^{\rm 33GHz}$, implied by combining our radio sizes with the infrared luminosity. The dashed line indicates $\mathrm{\Sigma_{IR}^{33GHz}=10^{13}~L_{\odot} kpc^{-2}}$, which is the characteristic Eddington limit set by radiation pressure on dust for optically thick U/LIRGs (see Section \ref{sec:edd}).\label{fig:fig6}}
\end{figure*}

\noindent In the left panel of Figure \ref{fig:fig6} we compare IR-based and 33~GHz based SFRs estimated for each U/LIRG in our sample. Following \citet{Murphy12}, we adopt T$_{\rm e}=10^4$ K and $\alpha_{\rm NT} = -0.8$ at $\nu=32.5$~GHz, but note both as a source of uncertainty. If we use T$_{\rm e}$ = 5000 K, the SFR based on 33 GHz increases by $\sim$37\%.

The left panel of Figure \ref{fig:fig6} shows that these two simple SFR estimates agree in our sample. The strong outlier, source \#10, is Mrk 231. This system is known to be dominated by an AGN that appears to contribute substantially to the 33 GHz emission. The other sources are consistent with a simple radio-infrared correlation that has a normalization in agreement with the \citet{Murphy12} relations.

If the assumption is made that the 33~GHz size, A$_{\rm 50,d}$, reflects the distribution of star formation, we can derive a star formation rate surface density, $\Sigma_{\rm SFR_{IR}^{33GHz}}$. As above, we take $\mathrm{\Sigma^{33GHz}_{SFR_{IR}} = 0.5\times SFR_{IR}/A_{50,d}}$\footnote{In order to obtain values that are comparable to those in the literature, we use SFR = SFR$_{\rm IR}$ to derive $\mathrm{\Sigma^{33GHz}_{SFR_{IR}}}$.}.

The right panel in Figure \ref{fig:fig6} shows our calculated $\Sigma_{\rm SFR_{IR}}^{\rm 33GHz}$. These span from $\mathrm{10^{0.6}}$ up to $\mathrm{10^{4.1}~M_{\odot}~yr^{-1}~kpc^{-2}}$ (right panel, bottom axis, in Figure \ref{fig:fig6}). The high end of this range represents the highest $\Sigma_{\rm SFR}$ found for any galaxy in the local universe. The wide range indicates diverse conditions. Even though we have observed the brightest and closest U/LIRGs, these span about four orders of magnitude in $\Sigma_{\rm SFR_{IR}}^{\rm 33GHz}$.

The IR surface brightness is also of interest. In local U/LIRGs, most of the bolometric luminosity is emitted in the 8$-$1000 $\mu$m range. By assuming that half of L$_{\rm IR}$ is concentrated within A$_{50,d}$, we estimate $\mathrm{\Sigma^{33GHz}_{IR}}$ for this inner region. For our approach from Equation \ref{eq:sfrir}, $\mathrm{\Sigma^{33GHz}_{\rm IR}}$ is identical to $\mathrm{\Sigma^{33GHz}_{\rm SFR_{IR}}}$ within a constant factor. Therefore we show the $\mathrm{\Sigma^{33GHz}_{\rm IR}}$ axis along the top of the right panel of Figure \ref{fig:fig6}. 

The U/LIRGs in this sample have $\mathrm{\Sigma^{33GHz}_{\rm IR}}$ ranging from $\mathrm{10^{10.5}~to~10^{14.1}~L_{\odot}~kpc^{-2}}$. The high end of this range is of particular interest. The dashed vertical line in Figure \ref{fig:fig6} indicates $\mathrm{\Sigma^{33GHz}_{IR}=10^{13}~L_{\odot}~kpc^{-2}}$. This value of $\mathrm{\Sigma^{33GHz}_{\rm IR}}$ has been argued to correspond to the characteristic Eddington limit set by radiation pressure on dust in self-regulated, optically thick disks \citep{Thompson05}. Some sources in our sample show $\mathrm{\Sigma^{33GHz}_{IR} \geq 10^{13}~L_{\odot}~kpc^{-2}}$, indicating they may be Eddington-limited starbursts (see Section \ref{sec:edd} for further discussion).

Note that for systems that are optically thick in the infrared, the $\tau_{\rm IR} \sim 1$ photosphere may be larger than the 33 GHz size. In this case,  the $\mathrm{\Sigma^{33GHz}_{IR}}$ that we calculate would never be observed, even if very high resolution FIR observations were available. This does not mean that this quantity lacks physical meaning, however. These systems are incredibly opaque to UV and optical light, which we expect to be generated in the region of active star formation traced by our 33~GHz data. This will then be quickly reprocessed into IR light, which then scatters out to the photosphere before leaving the system.

In this case, that inner region captured by the 33~GHz emission and $\mathrm{\Sigma^{33GHz}_{\rm IR}}$, and $\mathrm{\Sigma^{33GHz}_{SFR_{IR}}}$ are the quantities directly related to the region of most intense feedback and the immediate sites of star formation.

Are our sources optically thick in the IR? Infrared observations are limited to relatively coarse resolution, so direct size measurements in this range provide only modest constraints. \citet{DS10} and \citet{Lutz16} measured infrared sizes sizes\footnote{We normalized their sizes to the scale we use in this paper (see Table \ref{table:tbl-1}), defined in the same way we define A$_{\rm beam}$.} for the systems in our sample at 13 $\mu$m ({\it Spitzer}) and 70 $\mu$m ({\it Herschel})\footnote{Full list of Herschel images presented in \citet{Chu17}.}. For most systems, the sizes at 13 $\mu$m are larger than those at 70 $\mu$m, and the latter are larger than those we measured at 33 GHz.

A more powerful constraint comes from comparing our measured size to that implied by the measured dust temperature and luminosity. To do this, we consider the emission emitted in the IR, specifically between 8 and 1000 $\mu$m, and the dust temperature found comparing 63 $\mu$m and 158 $\mu$m emission (for more details, see Diaz-Santos et al. submitted). For an optically thick black body of temperature $T_{dust}$,

\begin{equation}
0.5 \times L_{IR}[80-1000\mu m] = 4 \pi R_{50IR}^{2}\sigma T^{4}_{dust}~,
\end{equation}

\noindent where $\mathrm{L_{IR}}$ is shown in Table \ref{table:tbl-1}. $L_{\rm IR}$ and $T_{dust}$ are measured, and this approach allows for the size expected for a photosphere with $T_{\rm dust}$ to produce $L_{\rm IR}$.

\begin{figure*}[tbh]
\centering
\includegraphics[width=3.5in]{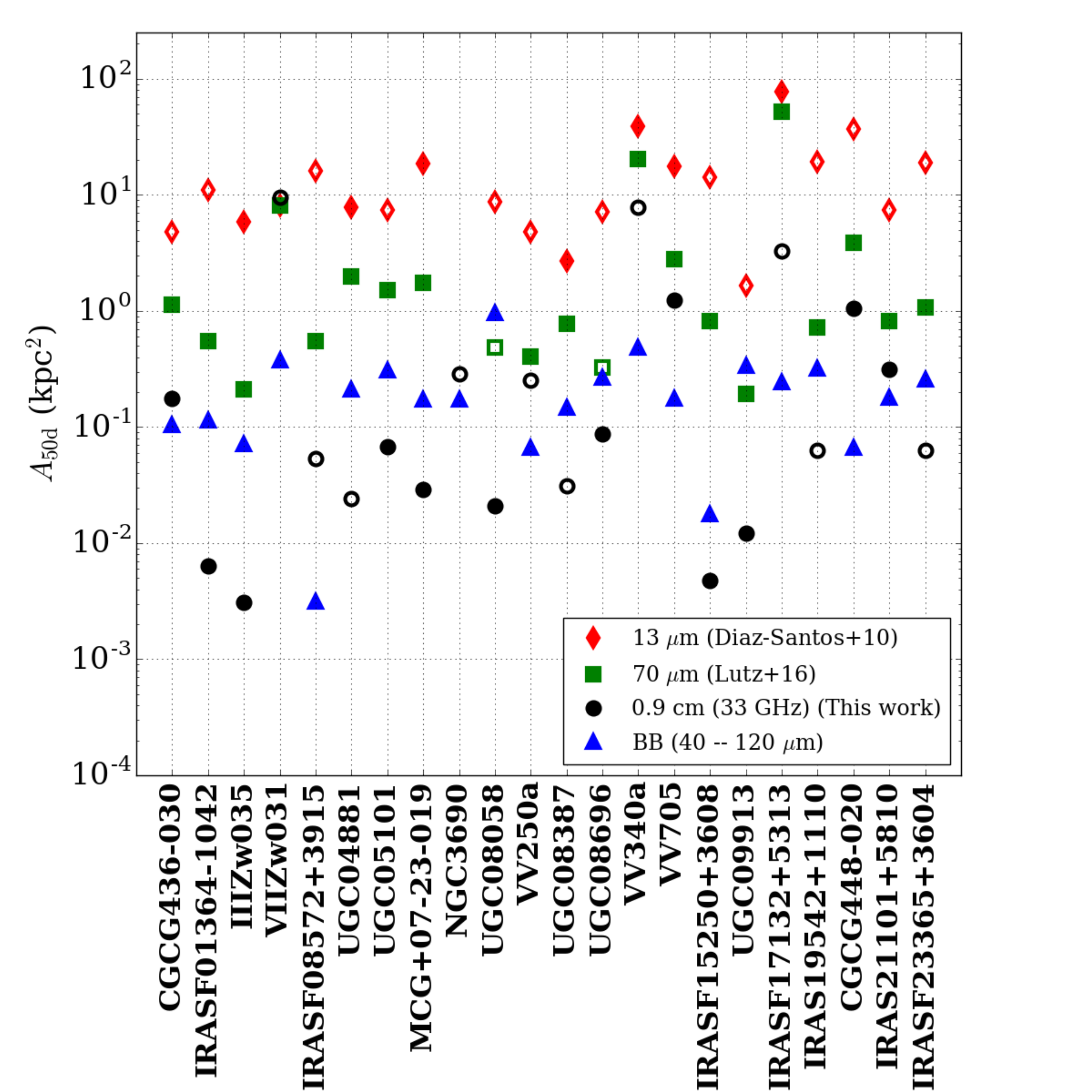}
\includegraphics[width=3.3in]{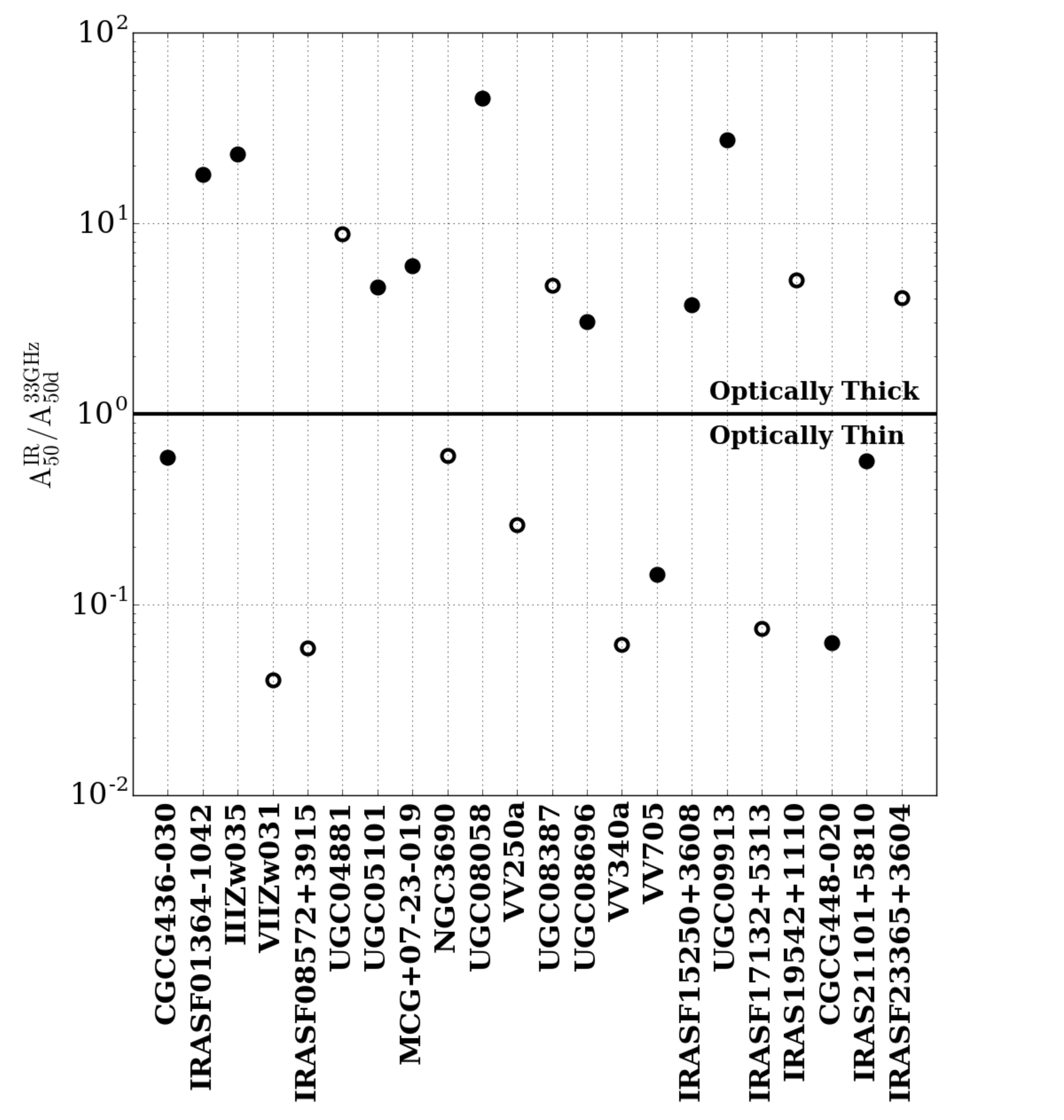}
\caption{{\it(Left)} Sizes measured at 13 $\mu$m (red diamonds), 70 $\mu$m (green squares), and 33 GHz (black circles), and those expected for a black body (blue triangles) with L$_{\rm IR}$ and T$_{\rm dust}$ (see Section \ref{sec:Ssfr} for more details). Open symbols indicate upper limits. {\it (Right)} Ratio of the area expected for a black body and the area measured at 33 GHz. Open symbols indicate lower limits. The solid horizontal line indicate equality among the areas. Sources above (below) this solid line are optically thick (thin) between 8 and 1000 $\mu$m. At least half of the sources in our sample are optically thick at far infrared wavelengths.\label{fig:fig7a}}
\end{figure*}

In Figure \ref{fig:fig7a}, we compare the sizes measured at 13 $\mu$m, 70 $\mu$m, and 33 GHz, and those calculated for a black body (assuming $\mathrm{A^{IR}_{50}=\pi R^{2}_{50IR}}$). We see that at least half of the sources in our sample are optically thick at infrared wavelengths, with our measured 33~GHz size smaller than the blackbody size. Thus IR opacity appears significant in our sample, which might be expected considering we are studying the most obscured systems in the local universe.

\subsection{Gas Surface and Column Density}
\label{sec:Smol}

Our sample consists of gas-rich mergers. In these systems, large masses of gas are funneled to the center, where they become mostly molecular \citep[e.g.,][]{Larson16}. The surface and volume densities of this gas relate closely to its self-gravity and ability to form stars. Again, we assume that the 33~GHz size is characteristic of the system and by combining this with half of the integrated CO (1$-$0) measurements, we estimate these quantities for the sample. 

Both the assumption of the 33~GHz characteristic size and the conversion between CO luminosity and mass (``conversion factor'') introduce uncertainties into the calculation. Our calculation assumes that the gas shares a characteristic size with the star formation traced by the radio. If our targets harbor large amounts of non-star forming gas or the internal relationship between gas and star formation is strongly non-linear, e.g., with stars forming much faster in a subset of very dense gas, the calculation will yield biased results. We do expect the approximation to hold, at least to first order. On larger scales star formation traced by IR and CO emission do track one another approximately one-to-one in major mergers \citep[][]{Daddi10}. More, interferometric CO measurements find that nearly half of the total CO mass is enclosed in the central few kpc in local U/LIRGs \citep[e.g.,][]{D&S98,Wilson08}. 

For a starburst $\mathrm{\alpha_{\rm CO} = 0.8~M_{\odot}~(K~km~s^{-1} pc^{2})^{-1}}$ (including helium) and coexisting gas and radio emission, we infer values for the molecular gas surface density, $\Sigma_{\rm mol}^{\rm 33GHz}$, from $\mathrm{10^{2.5}~to~10^{5.3}~M_{\odot}~pc^{-2}}$. Even the low end of this range corresponds to source-averaged surface densities in excess of many Local Group molecular clouds \citep[e.g.,][]{Bolatto08,Fukui10}. The high end is far in excess of $\sim$~1~g~cm$^{-2}$, which is commonly invoked as an immediate precondition for star formation considering dense substructure inside molecular clouds. Here this gas column density is the average value across the whole energetically dominant area of a galaxy.

These values obviously depend on the mass-to-light ratio adopted to convert CO luminosity to mass. The appropriate conversion factor for starburst galaxies has been a matter of debate, with suggestions ranging from approximately Galactic \citep[e.g.,][]{Pap12,Scoville14} to low \citep[e.g.,][]{D&S98} and highly environment-dependent \citep{Shetty11} values. To see the effect of a higher, Milky Way, $\alpha_{\rm CO}$ one should multiply our nominal surface and volume densities by $\approx 5.4$.

Also note, that this assumption of matched $\Sigma_{\rm gas}$ and $\Sigma_{\rm SFR}$ distributions does not hold for some ULIRGs. For example, for IRAS 13120-5453 the measured starburst size derived from sub-mm continuum is found to be more compact than the emission from dense \citep{Privon17} and more diffuse \citep{Sliwa17} molecular gas tracers. More, recent high resolution observations of the CO emission in Arp 220 \citep{Scoville17} suggest that the gas is distributed in a larger area compared to the star formation area traced by the 33 GHz emission (see Figure \ref{fig:fig1} and \citealt{BM15}). Only $\sim 20\%$ of the total CO emission is coming from the nuclei. At the moment, Arp 220 is uniquely well-studied. These results argue that high resolution interferometric observations of the gas to match our SFR-tracing continuum will yield important information on how the SFR-per-unit gas varies across the system. Lacking such information, we proceed assuming matched gas and SFR. If these ULIRGs represent the general case, the reader may think of our $\Sigma_{\rm gas}$ as an upper limit, with 10s of percent of the material in an extended, comparatively non-star forming disk. This will imply even higher SFR per unit gas mass in the nuclear regions than we calculate below.

One class of models considers the total mass surface density a main driver of the conversion factor, largely via its effect on the line width \citep[e.g.,][]{Shetty11,Narayanan12}. Our measured sizes give us the opportunity to illustrate the effect of such a dependence on derived surface densities. To do this, we use the prescription in \citet[][their equation 31]{Bolatto13} which follows \citet{Shetty11}. Neglecting any metallicity dependence and considering only the regime where $\Sigma > 100$~M$_\odot$~pc$^{-2}$, their prescription is

\begin{equation}
\label{eq:alpha_co}
\left( {\alpha_{\rm CO} \over {{\rm M}_\odot \over [{\rm K~km~s^{-1}~pc^2]}}} \right) \approx 4.35 \left( \frac{\Sigma_{\rm total}}{100~\mathrm{M_\odot~pc}^{-2}} \right)^{-0.5},
\end{equation}

\noindent where $\Sigma_{\rm total}$ is the total mass surface density driving the potential well. We will assume the systems studied here to be gas-dominated in the main CO-emitting region and take $\Sigma_{\rm total} \sim \Sigma_{\rm gas}$. The overall gas mass fraction in local U/LIRGs is closer to $\sim 30\%$ \citep[][]{Larson16}. However we expect the gas to be concentrated relative to the stars, so that we can assume the systems to be locally gas dominated in the emitting region. We assume that in the dense, well shielded central regions of U/LIRGs, the HI content is negligible, and we consider $\Sigma_{\rm gas} \sim \Sigma_{\rm mol}$. 

We calculate the conversion factor from equation \ref{eq:alpha_co} iteratively, because $\Sigma_{\rm mol}$ changes as $\alpha_{\rm CO}$ changes. Numerically iterating, we reach a value of $\alpha_{\rm CO}$ that converges to within 0.1\%. These values go from 0.2 up to 1.65, with a median value of 0.43 for our sample. We report the gas properties derived using this surface-density dependent $\alpha_{\rm CO}$ in brackets, along with $\alpha_{\rm CO}$ for each source, in Table \ref{table:tbl-6}. The effect of applying this correction is to narrow the range of derived gas surface densities, as the high surface density systems have low $\alpha_{\rm CO}$.

The gas surface density values derived here translate to average Hydrogen column densities that range from $\mathrm{10^{22.6}~cm^{-2}~to~10^{25.4}~cm^{-2}}$ when using $\mathrm{\alpha_{CO}=0.8}$, and $\mathrm{10^{22.9}~cm^{-2}~to~10^{24.8}~cm^{-2}}$ when using the surface-density dependent conversion factor. Assuming a Galactic dust-to-gas ratio \citep{Bohlin78}, which may be roughly appropriate \citep{Rupke08,Iono09}, these column densities imply line of sight extinctions of $A_V \sim 22$ to $12,000$~mag, for a starburst conversion factor, and $A_V \sim$ 48 to $3,200$~mag, for a surface-density dependent conversion factor.

\subsection{Gas Volume Density}

The gas volume density, and the corresponding free fall time, are central quantities for many theories of star formation \citep[e.g.,][]{Krumholz12}. We estimate the gas volume density from the measured sizes and the integrated CO luminosities. This requires additional geometric assumptions. We consider the most basic approach and assume that our sources are three dimensional Gaussians. In this case, $\sim 30$\% of the mass exists inside the FWHM of the Gaussian\footnote{This is the correction to obtain the mass inside a sphere of radius R$_{50,d}$ (see Section \ref{sec:sizes}) with a Gaussian mass distribution.}, R$_{50,d}$. 

Adopting this geometry, we find n$_{\rm H_{2}}$ from $\mathrm{10^{0.5}~cm^{-3}~to~10^{4.9}~cm^{-3}}$ for a fixed starburst $\alpha_{\rm CO}$. Using the variable, surface-density dependent $\alpha_{\rm CO}$, we find a narrower range of $\mathrm{10^{0.9}~cm^{-3}~to~10^{4.3}~cm^{-3}}$. The free fall collapse times associated with these densities range from $6{-}100$ ($2{-}190$) Myr with the fixed (variable) $\alpha_{CO}$.

\section{Discussion}
\label{sec:discussion}

The 33 GHz sizes reported in this paper represent the best measurements to date of the energetically dominant regions in this set of bright, nearby U/LIRGs. These sizes, combined with the integrated flux density measurements allow us to study the physical properties of the nuclear regions in the sample. Here, we discuss the implications of these measurements for the nature of the 33 GHz emission, star formation scaling relations, optical depth, and radiation pressure feedback.

\subsection{Nature of the $33$~GHz Radio Emission}
\label{sec:nature}

In models like those of \citet{Condon92} and \citet{Murphy12}, the radio SED reflects a mixture of thermal and nonthermal emission. What powers the emission that we observe from U/LIRGs at 33~GHz? In the \citet{Condon92} model for a starburst galaxy like M82, about 50\% of the total 33 GHz continuum is produced by free-free (``thermal'') emission; for comparison, $<12\%$ of the emission is expected to be produced by free-free emission at 1.5~GHz.

We have several constraints on the nature of the emission mechanism in our targets: the SED shape, the brightness temperature, and the comparison with the SFR implied by the IR. Together, these indicate some 10s of percent contribution of thermal emission to the 33~GHz flux density, with the balance being synchrotron. However, a detailed understanding of the emission mechanism will need to wait for better coverage of the radio SED in these targets (L. Barcos-Mu\~noz et al. in preparation).

{\em Brightness Temperature and Optical Depth:} The brightness temperature of optically thick free-free emission is expected to be $\sim 5\times10^3$~K - 10$^{4}$~K. If the 33~GHz $T_b$ exceeded this value, this would provide evidence that synchrotron dominates the emission. Figure \ref{fig:fig5} shows that the averaged nuclear T$_{\rm b}$ does not exceed this limit. Either the emission is patchy within the beam, or the emission at $33$~GHz is optically thin. Thus, the brightness temperature in the sources allows for a normal mix of emission mechanisms and is consistent with optically thin free-free emission making up a large part (or all) of the observed 33~GHz flux density.

If we neglect filling factor effects and assume that $\approx 50\%$ of the total T$_{\rm b}$ is due to thermal emission, then we can estimate the optical depth of the free-free emission. We derive $\mathrm{\tau_{thermal} \sim T_{b}/T_{e} \leq 0.2}$ for all our sample. This number is still less than 1, therefore optically thin, even if we assume 100\% of the 33 GHz flux density is due to thermal emission.

{\em Spectral Index:} For a mixture of synchrotron (``non-thermal'') emission and optically thin free-free emission, \citet{Condon&Yin90} give the following approximation to the fraction of emission that is thermal,

\begin{equation}
\label{eq:thermal}
\frac{S}{S_{\rm T}} \sim 1+10 \left( \frac{\nu}{\rm GHz} \right)^{0.1+\alpha_{\rm NT}}~.
\end{equation}

\noindent Here $S$ is the total flux density, S$_{\rm T}$ is the flux density from thermal emission, and $\alpha_{\rm NT}$ is the typical non-thermal spectral index $\sim -0.8$. The formula assumes a power-law spectral energy distribution for the non-thermal emission.

We combine equation \ref{eq:thermal} with the S$_{5.95}$ from Table \ref{table:tbl-3} to calculate $S/S_T$ at $5.95$~GHz. Then, knowing that $\mathrm{S_{T} \propto \nu^{-0.1}}$ we predict the spectral index between $\alpha_{6-33}$. Based on this, we expect an average $\alpha_{6-33} = -0.53$. We expect $\alpha_{6-33}$ to approach $\alpha_{\rm NT} = -0.8$ as the thermal fraction decreases to zero, while if the thermal fraction is higher than this estimate, $\alpha_{6-33}$ will be $> -0.53$.

Figure \ref{fig:fig2} shows that 17 out of the 22 systems in our sample have $\alpha_{6-33}<-0.53$, implying that in most of our sample, non-thermal emission is stronger relative to thermal emission than predicted by Equation \ref{eq:thermal}. \citet{BM15} found a similar result comparing 6 and 33 GHz emission in Arp 220. We caution that our assumed $\alpha_{\rm NT}$ affects this result and that we cannot, at present, distinguish between variations in the thermal fraction and $\alpha_{\rm NT}$ from only two frequencies. Indeed, multi-frequency observations, particularly at high frequency, suggest curvature in the radio SED \citep[e.g., see][]{Clemens08,Clemens10,Leroy11,Marvil15} so that the power-law assumption for the non-thermal emission model in Equation \ref{eq:thermal} represents a simplification. Observations that cover a wide band will allow for a more complex treatment for a better disentanglement of the contribution of the two components at these frequencies (Linden et al. in prep).

{\em Spectral Index and Implied Opacity at Lower Frequencies:} Following the same approach, we use Equation \ref{eq:thermal} and the flux at $5.95$~GHz to predict an integrated $\alpha_{1.5-6}$ of $-0.71$. However, less than half of the sample show spectral indices that agree with this predicted value. Most of our targets show shallower spectral indices. This is most likely due to opacity affecting the low frequency emission, especially the observations at 1.49 GHz where free-free absorption is known to play a major role in compact starbursts \citep[see, e.g.,][]{Condon91,Murphy13a}. In fact, in Figure \ref{fig:fig2} we also observe a change in slope as frequency increases for several sources, from shallower to steeper in most cases. Mrk 231 even shows a change from positive $\alpha_{1.5-6}$ to a negative $\alpha_{6-33}$. For a compact starburst this would indicate that $\tau_{\rm thermal}$ becomes one at some frequency between 1.5 and 33 GHz\footnote{This turnover frequency normally occurs at MHz frequencies, when present, and it shifts to higher frequencies for high star forming, very compact systems.}. However, we know Mrk 231 has a very compact core \citep[e.g.,][]{Lonsdale03,Helmboldt07}, which suggest instead the change in slope is most likely due to synchrotron self-absorption at low frequencies. In addition, it is also possible that the flattening in the observed $\alpha_{1.5-6}$ could be caused by ionization and bremsstrahlung losses \citep{Thompson06,Lacki10}, which become important at low frequencies in high density environments such as those found in our sample (see Section \ref{sec:Smol}).

\begin{figure*}[tbh]
\centering
\includegraphics[width=3.5in]{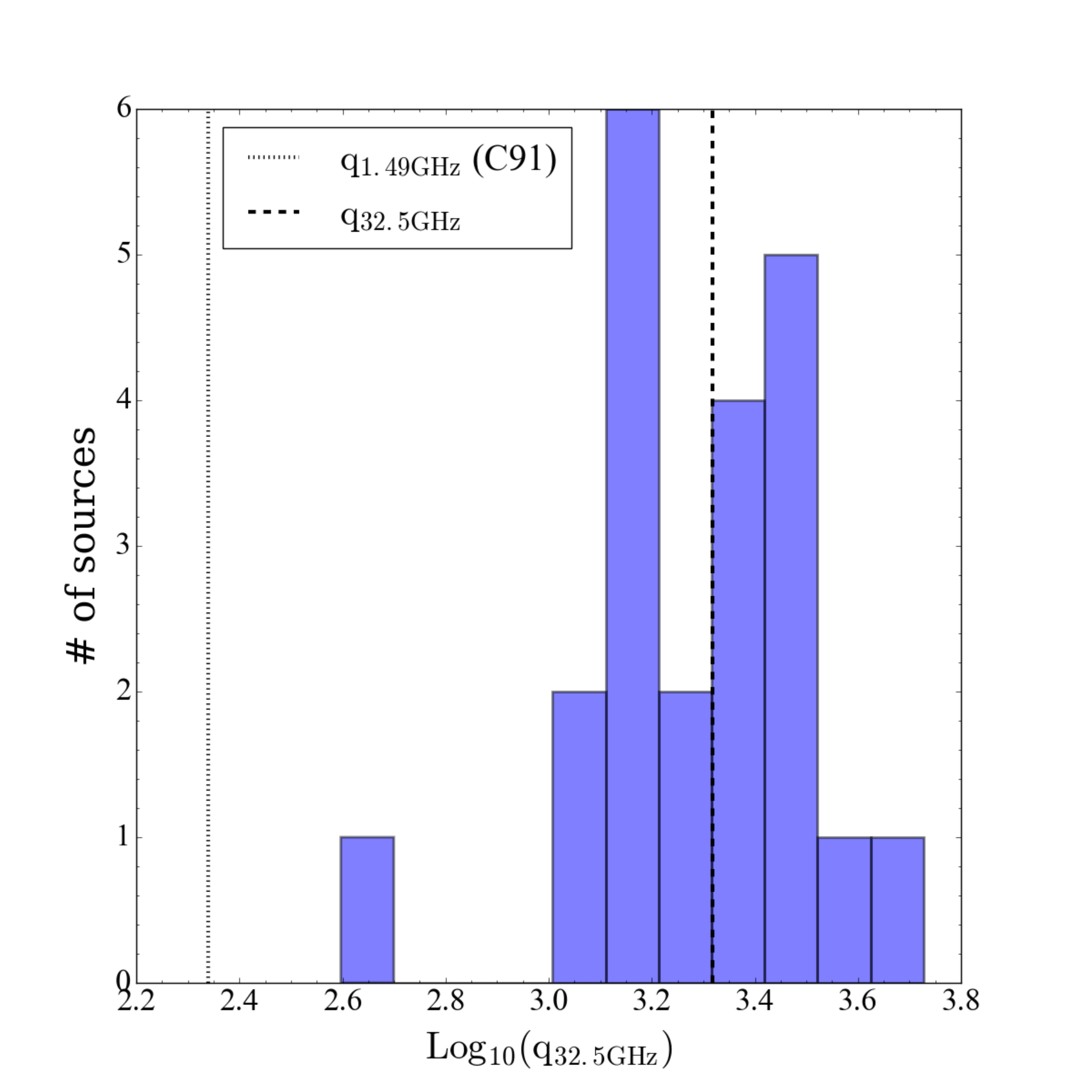}
\caption{Histogram of q$_{\rm 32.5GHz}$ obtained using equation \ref{eq:q}. The dashed line shows the median q$_{\rm 32.5GHz}=3.3$. For comparison, we also show the median value for q$_{\rm 1.49GHz}=2.34$ from \citet{Condon91}. The strongest outlier in the histogram is Mrk 231 (q $\sim$ 2.37), which is a known AGN and does not follow the radio-FIR correlation (see Figure \ref{fig:fig6}).\label{fig:fig7}}
\end{figure*}

Several systems show the opposite trend, exhibiting steep $\alpha_{1.5-6}$ and a shallower $\alpha_{6-33}$. The simplest explanation for these measurements is that these systems have a higher thermal fraction than the other targets. Alternatively, some other source may contribute to the 33~GHz emission, e.g., anomalous dust emission \citep{Draine98,Ali09,Murphy10}. More detailed SED coverage could confirm this interpretation. Another possible explanation includes contribution from thermal dust, which is normally important only at much higher frequencies, $\gtrsim$ 100 GHz. Again, better frequency coverage will play a key role.

In Figure \ref{fig:fig2}, we find a tentative correlation between $\mathrm{\alpha_{1.5-6GHz}}$ and $\mathrm{A_{50,d}}$, showing a shallower spectral index for more compact sources. This trend makes sense if more compact sources are also more opaque. In this case, $1.5$~GHz emission in more opaque systems will be suppressed due to a higher opacity at $1.5$~GHz relative to $6$~GHz.

Integrated spectral indices only give us a partial view of the processes that are powering star formation in our sample. We require more detailed spectral index maps to dissect the distribution of the radio emission. We will report resolved spectral index maps between 6 and 33 GHz in a future paper (Barcos-Mun\~{o}z et al. in prep). These results will be greatly complemented by spectral indices maps between 1.49 and 8.44 GHz reported in \citet{Vardoulaki15} using the \citet{Condon90} and \citet{Condon91} observations. 

{\em Expectations from IR-Based SFRs:} The contrast of the 33 GHz flux density with the total IR emission also sheds some light on the emission mechanism. Inasmuch as the IR tells us about the star formation rate, it also makes a prediction for the expected thermal emission, along with some simplifying assumptions.

We derive the expected free-free emission, S$_{\rm T}$, and then thermal fraction, S$_{\rm T}$/S, at 33 GHz by assuming that all the IR luminosity is due to star formation and that none of the ionizing photons (that will potentially produce free-free emission) are absorbed by dust. Note that if an AGN is present and contributes significantly to the IR luminosity, then the SFR derived by this method will be overestimated \citep[see][for an estimation of the AGN contribution to L$_{IR}$ in local U/LIRGs]{Armus07,Petric11}. We use equation \ref{eq:sfrir} and the thermal SFR from Table 8 in \citet{Murphy12}, which relates SFR and the thermal luminosity, L$^{\rm T}$, by the following equation,

\begin{equation}
\label{eq-sfrt}
\begin{split}
\left(\frac{\rm SFR_{\nu}^{T}}{M_{\sun}\,{\rm yr^{-1}}}\right) &= 4.6\times10^{-28}\\
&\left(\frac{T_{\rm e}}{10^{4}\,{\rm K}}\right)^{-0.45} \left(\frac{\nu}{\rm GHz}\right)^{0.1} \left(\frac{L_{\nu}^{\rm T}}{\rm erg~s^{-1}~Hz^{-1}}\right),  
\end{split}
\end{equation}

\noindent where we assume T$_{\rm e}\sim10^{4}$ K (see Section \ref{sec:Tb} and \ref{sec:Ssfr} for further discussion on this assumption). In this way, we predict the thermal radio emission expected given the IR luminosity. Comparing it to L$_{\rm IR}$, we derive the thermal fractions shown in the top right panel in Figure \ref{fig:fig2}. We see no clear trend, however note that T$_{\rm e}$ is uncertain, and the derived thermal fractions depend on it. Lower values of T$_{\rm e}$, or higher thermal optical depths, imply lower thermal fractions. We also observe that Mrk 231 shows the lowest predicted thermal fraction in our sample. This is expected since it does not follow the radio-IR correlation (see Figure \ref{fig:fig6}), with SFR$_{\rm 33GHz}$ being $\sim$ 4 times higher than SFR$_{\rm IR}$. By comparing the thermal fractions shown in Figure \ref{fig:fig2} with the radio-IR correlation shown in Figure \ref{fig:fig6}, we see that all 11 sources with low thermal fraction (i.e., thermal fractions $< 60\%$) are below the equality line in Figure \ref{fig:fig6}. This is consistent with Equation \ref{eq-sfrrad} underestimating the SFR due to a more dominant non-thermal component (i.e., a plausible shallower $\alpha_{\rm NT}$) than what is assumed for the equation (-0.8).

From our analysis of the spectral index, we expect thermal fractions $\leq$ 50\%. Figure \ref{fig:fig2} shows that, based on the prediction from the IR, most of the sources have thermal fractions $\approx 50$--$100\%$. We expect that this is the combination of three effects. First, even if the IR is all powered by star formation, some of the ionizing photons produced by young stars that could otherwise produce free-free emission will be absorbed dust and thus not produce free-free emission. These should not be counted in our prediction for the thermal emission, and the true thermal fraction would be accordingly smaller. We highlighted a similar situation in Arp 220, where the predicted thermal fraction is $\sim$ 50\% but SED analysis shows it should be closer to 35\% \citep[see][]{BM15}. Second, as noted above, the SED-based estimates remain hampered by the lack of sensitive, wide-band coverage of the spectral energy distribution. As long as the adopted non-thermal spectral index (or SED) remains uncertain, so will do the thermal fractions estimated in this way. Third, if an AGN contributes a substantial amount to the IR emission, then the thermal fraction would be overestimated because the AGN will not contribute to the free-free emission in the same way as star formation.

Two sources, UGC 04881NE and IRAS 08572+3915 show thermal fractions $>$ 100\%, meaning that they have very high ratios of IR to radio emission (see Figure \ref{fig:fig6}). This IR excess has been reported before for IRAS 08572+3915 \citep[see discussion in][]{Yun04}, and this system was already noted as an interesting source in discussion of first results from this survey \citep{Leroy11}. See the Appendix for further discussion on these two sources.

{\em Radio-FIR Correlation at 33~GHz:} As a more observational restatement of the previous result, we derive $q_{33}$, the ratio of FIR flux (between $\sim$ 42 and $\sim$ 122 $\mu$m) to radio flux density at 33~GHz:

\begin{equation}
\label{eq:q}
\mathrm{q_{\rm \nu} = Log_{10}((S_{FIR}/3.75\times10^{12}~Hz)/S_{\nu})}~.
\end{equation}

\noindent Here, $S_{\nu}$ is the flux density at frequency $\nu$ in units of $\mathrm{W~m^{-2}~Hz^{-1}}$, and $\mathrm{S_{\rm FIR}[42-122\mu m] = 1.26\times10^{-14}~(2.58~S_{60\mu m}+ S_{100 \mu m})}$, in units of $\mathrm{W~m^{-2}}$, is the far-infrared flux, with the flux density at 60 and 100 $\mu$m measured in Jy. 

We show a histogram of $q_{32.5GHz}$ in Figure \ref{fig:fig7}. We find a median $q_{33}\approx3.32$ and a dispersion of $0.19$~dex. $q_{33}$ is similar to that found by \citet{Rabidoux14} studying regions in local star forming galaxies, but we find a tighter correlation. Their measured dispersion is $0.1$~dex larger than ours. In fact, the 0.19 dex in dispersion we observed for q$_{33}$ is similar to that found in \citet{Condon91} at 1.49 GHz. The tighter dispersion found for our global measurements compared to the local ones of \citet{Rabidoux14} appears to corroborate the global nature of the IR-radio correlation. Note as well that $q_{32.5GHz}$ does not appear to correlate with $\Sigma_{SFR}$.

\subsection{Physical Conditions at the Heart of Local Major Mergers}

Our size estimates imply that a large part of the star forming activity, and so presumably also the gas that fuels it, is concentrated in areas with half-light radii from 30 pc up to 1.7 kpc\footnote{This omits the upper limits obtained for the faint components in the systems UGC 04881 and VV250, for which we did not derive the physical parameters described in section \ref{sec:impl_radio}.}. Applying these sizes to global quantities using the proper aperture corrections, we estimate $\Sigma_{\rm SFR}$, $\Sigma_{\rm IR}$, N$_{\rm H}$, and $\mathrm{n_{mol}}$.

The resulting values span a wide range, typically $4$~dex. The high end of the range for each property is among the highest average gas, SFR, or luminosity surface density measured for any galaxy. The low end of the range is still high compared to values found in ``normal'' disk galaxies: the lowest density systems have $\mathrm{\Sigma_{mol}^{33GHz} \sim 10^2}$--$10^3$~M$_\odot$~pc$^{-2}$ and $\mathrm{\Sigma_{SFR_{IR}}^{33GHz} \sim 10^0}$--$10^1$~M$_\odot$~yr$^{-1}$~kpc$^{-2}$. These already resemble the highest kpc-resolution values (which come from active galaxy centers) found in \citet{Leroy13} (see bottom panel in Figure \ref{fig:fig9}). Moreover, the gas surface densities in our sample, even the lowest values, resemble those found for individual molecular clouds, but here they extend over the whole energetically dominant region of a galaxy. This implies average interstellar gas pressures that match or exceed those found inside individual clouds. Because of this high pressure, a Milky Way GMC dropped into any of the targets would not remain an isolated, self-gravitating object. Self-gravitating, overpressured clouds in these targets must be more extreme and denser than clouds in normal galaxies, a conjecture born out by observations of nearby starburst galaxies \citep[e.g.,][]{Keto05,Wei12,Leroy15,Johnson15}.

About half (13) of the 22 targets studied here show galaxy-averaged $\mathrm{\Sigma_{SFR_{IR}}^{33GHz} \geq 10^{2.7}~M_{\odot} yr^{-1} kpc^{2}}$. This corresponds to $\geq$ 2 times higher than the $\Sigma_{\rm SFR}$ that would be inferred based on the IR emission from the Orion core \citep{Soifer00}. Several (7) sources show $\mathrm{\Sigma_{SFR_{IR}}^{33GHz} >10^{3}~M_{\odot} yr^{-1} kpc^{-2}}$, corresponding to $\mathrm{\Sigma_{IR}^{33GHz}>10^{13}~L_{\odot} kpc^{-2}}$. This value has been put forward as the characteristic Eddington limit for $\Sigma_{\rm SFR}$ in a radiation pressure-supported, optically thick disk \citep{Scoville03,Thompson05} (see Section \ref{sec:edd} for further discussion).

The high column densities obscure the energetically dominant regions at non-radio wavelengths. Assuming a ``starburst" conversion factor, $13$ U/LIRGs show hydrogen column densities consistent with being Compton-thick, $\mathrm{N_{H}>1.5\times10^{24} cm^{-2}}$ \citep[e.g.,][]{Comastri04}, which would directly affect the ability of X-ray diagnostics to detect the presence of AGN in these systems. As mentioned above, the implied optical extinctions are extreme, 22$-$12,000~mag for our sample assuming a Galactic dust-to-gas ratio. Even infrared wavelengths, at which a normal star-forming galaxy is usually optically thin, will show significant opacity for these dust columns. At 100 $\mu$m, for a mass absorption coefficient of $\mathrm{\kappa_{100} = 31.3~cm^2~g^{-1}}$ \citep{Li&Draine01}, the dust opacity of these targets is $\tau_{100} \sim $0.02$-$12, with those same 13, but one, Compton-thick sources also being optically thick at 100 $\mu$m, i.e., $\tau_{100}>1$.

\subsection{The {\rm [}{\sc Cii}{\rm ]} Deficit}

Several studies have reported a ``deficit'' in the [C II] 158$\mu$m-to-far infrared luminosity (from 40 to 120 $\mu$m) ratio, $L_{\rm [\sc C II]}/L_{\rm FIR}$, in U/LIRGs relative to lower luminosity star-forming galaxies \citep[e.g.,][]{Malhotra01, DS13, Lutz16}. The $L_{\rm [C II]}/L_{\rm FIR}$ decreases with increasing dust temperature, mid-IR opacity, star formation efficiency ($L_{\rm IR} / M_{\rm H_2}$) and infrared surface density (where {\it Spitzer} and {\it Herschel} data are utilized to measure sizes). The deficit arises because the collisional energy required to produce [{\sc Cii}] is suppressed in the compact, dense starburst environments of U/LIRGs, and/or because the infrared luminosity is increased. 

\begin{figure*}[tbh]
\centering
\includegraphics[width=3.5in]{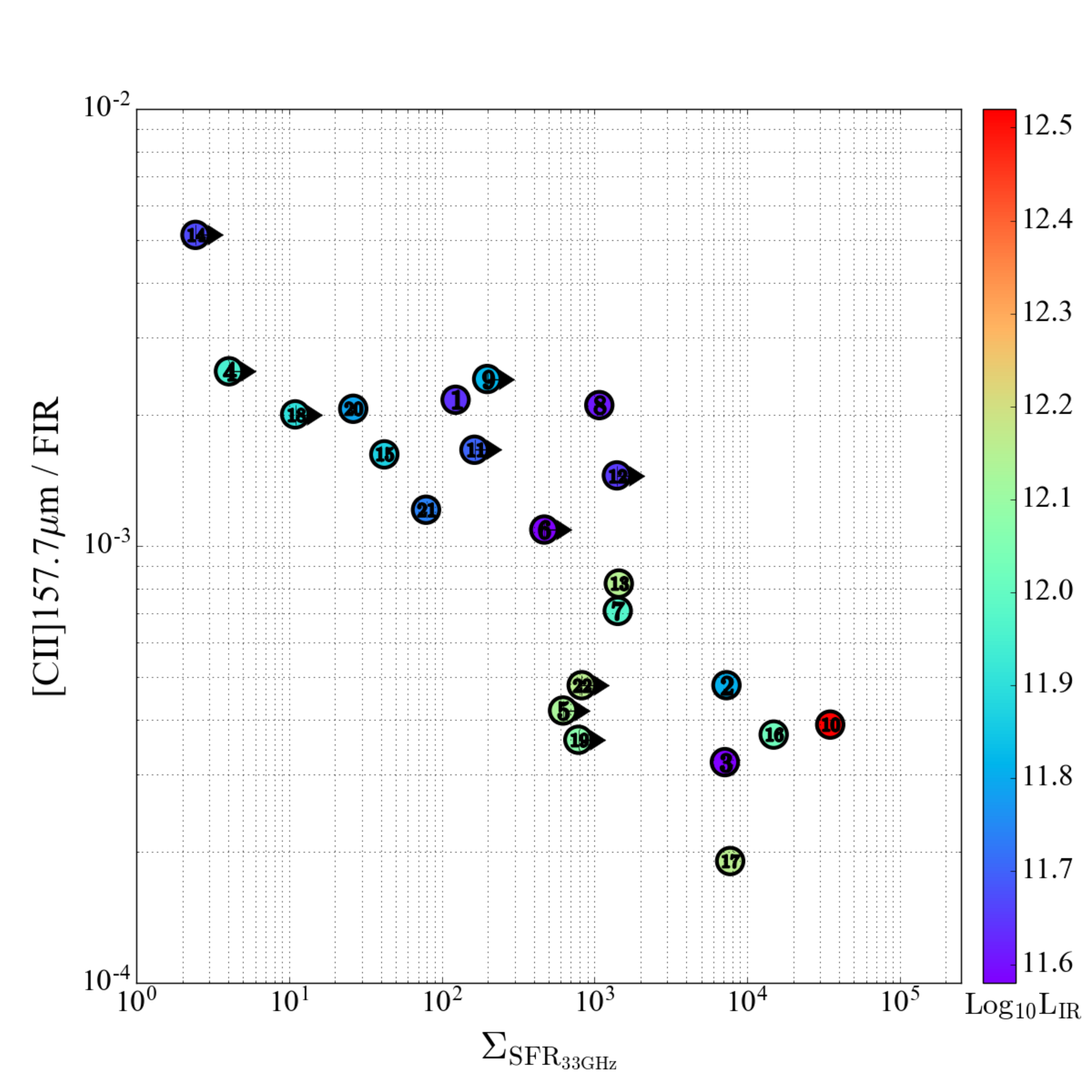}
\includegraphics[width=3.5in]{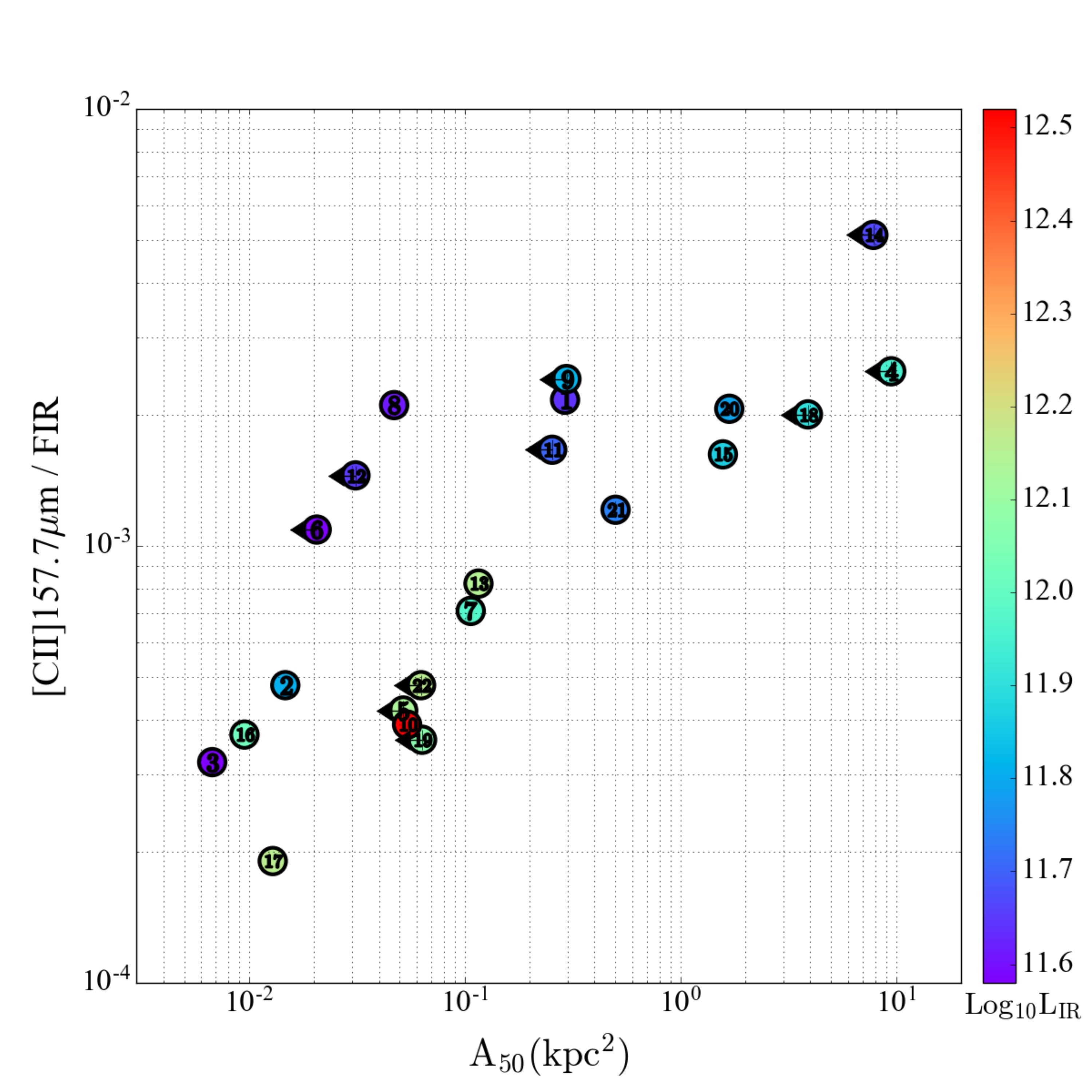}
\includegraphics[width=3.5in]{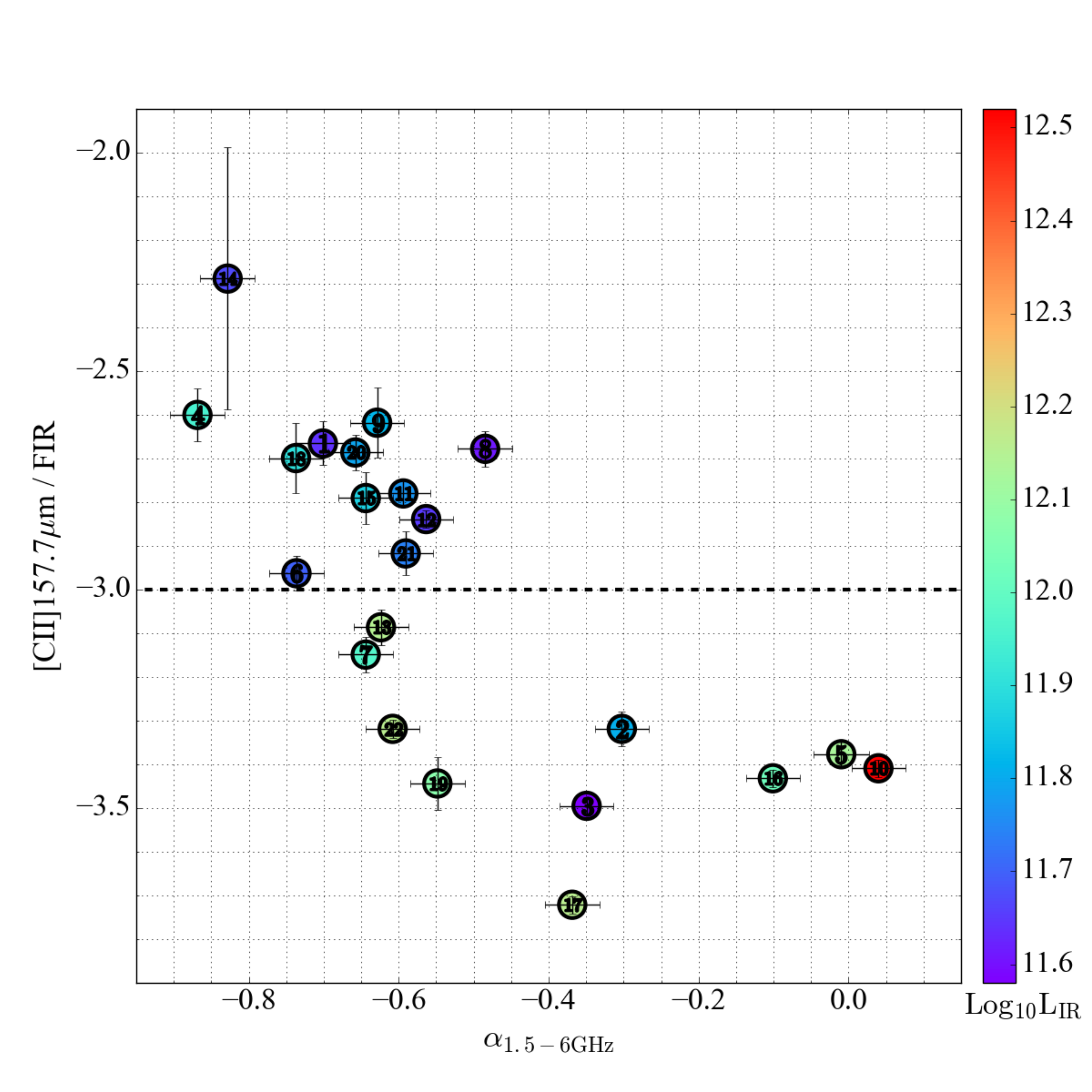}
\caption{($Top~left$) [{\sc Cii}] 158$\mu$m deficit as given by [{\sc Cii}] 158$\mu$m/FIR, where FIR is the far infrared flux density \citep{DS13}, versus $\Sigma_{\rm SFR_{\rm 33 GHz}}$, the surface density of star formation inferred from our $33$~GHz maps. The systems with the highest star formation per unit area show the highest {\sc Cii} deficit (smallest ratio of [{\sc Cii}] flux/FIR). ($Top~right$) [{\sc Cii}]158$\mu$m/FIR as a function of half-light area at 33~GHz. We observe a higher [{\sc Cii}] deficit for more compact objects. ($Bottom$) [{\sc Cii}]158$\mu$m/FIR as a function of $\alpha_{1.5-6GHz}$. The sources with the highest deficit show the flattest spectral index between 1.5 and 6 GHz. This is consistent with the finding that the [{\sc Cii}] deficit is inversely proportional to mid-IR opacity measurements, i.e., low [{\sc Cii}]158$\mu$m/FIR sources are deeply buried. In all panels, individual systems are labeled by the ID assigned in Table \ref{table:tbl-1} and color coded by their infrared luminosity. \label{fig:fig8}}
\end{figure*}

The sizes used to gauge the IR surface brightness in \citet{DS13} come from IR space telescopes, which have much coarser angular resolution than our maps. In Figure \ref{fig:fig9} (top left panel), we plot $L_{\rm [C II]}/L_{\rm FIR}$ from \citet{DS13} as a function of the star formation rate surface density inferred using our sizes, $\Sigma^{\rm 33 GHz}_{\rm SFR_{\rm IR}}$. The plot shows clear, strong anti-correlation between $L_{\rm [C II]}/L_{\rm FIR}$ and $\mathrm{\Sigma^{33GHz}_{SFR_{IR}}}$. The top right panel in Figure \ref{fig:fig8} shows $L_{\rm [C II]}/L_{\rm FIR}$ as a function of $\mathrm{A_{50d}}$. Both plots show that more compact systems with more locally intense star formation show stronger $L_{\rm [C II]}/L_{\rm FIR}$ deficits (lower $L_{\rm [C II]}/L_{\rm FIR}$). This is strong corroboration, using the best size measurements to date, of the correlation found by \citet{DS13} of higher deficit for systems with higher luminosity densities.

The spectral index between 1.5 and 6~GHz may give some indication of the opacity at low frequencies. In the bottom panel of Figure \ref{fig:fig8}, we plot $L_{\rm [C II]}/L_{\rm FIR}$ as a function of this spectral index, $\alpha _{\rm 1.5-6GHz}$. $L_{\rm [C II]}/L_{\rm FIR}$ is lower, and thus the [{\sc Cii}] deficit is larger, for systems with flatter (more nearly $0$) spectral indices. This flattening is believed to be due to increasing opacity \citep[e.g., see][]{Murphy13a}, so the bottom panel of Figure \ref{fig:fig8} shows that the $L_{\rm [C II]}/L_{\rm FIR}$ ratio is lowest for U/LIRGs that are most obscured at radio, as well as infrared, wavelengths.

With the exception of IRAS F08572+3915, the five U/LIRGs (Mrk 231, IRAS 15250+3608, III Zw 035, IRAS F01364-1042, and Arp 220) with the flattest $\alpha _{\rm 1.5-6GHz}$, and among the largest [{\sc Cii}] deficit, also have the lowest estimated thermal fraction at 33 GHz in our sample. These results are broadly consistent with our detailed study of Arp 220 \citep{BM15}, where we presented evidence of suppressed 33 GHz thermal emission and speculated that the suppression is due to the absorption of ionizing UV photons by dust concentrated within the HII regions \citep[see also][]{Luhman03,Fischer14}. Such scenario would also imply a lack of heating of photodissociated regions (PDR) and thus a suppression of the amount of collisional energy available to produce [{\sc Cii}].

\subsection{Implications for Star Formation Scaling Relations}
\label{sec:relations}

\begin{figure*}[tbh]
\centering
\includegraphics[width=3.5in]{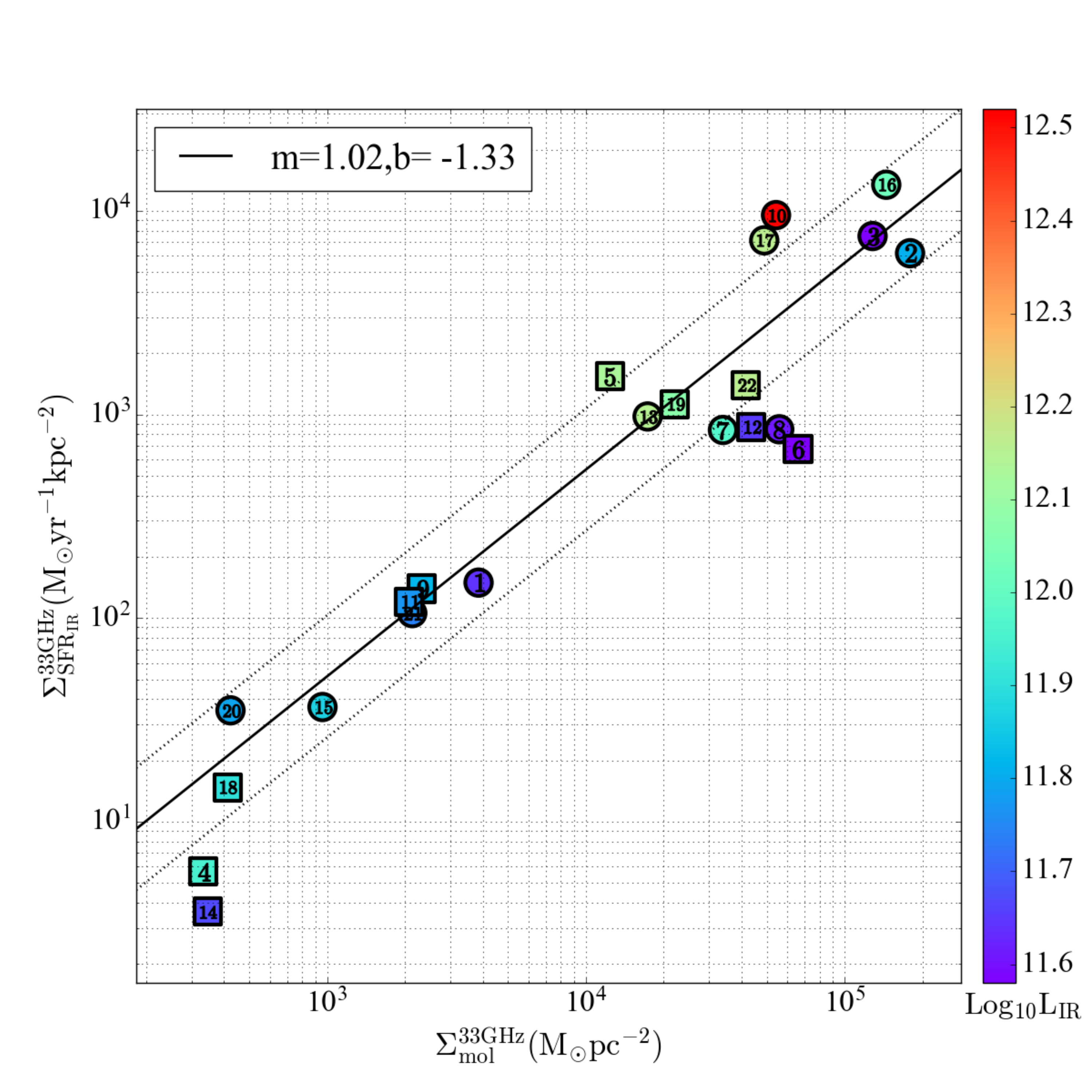}
\includegraphics[width=3.5in]{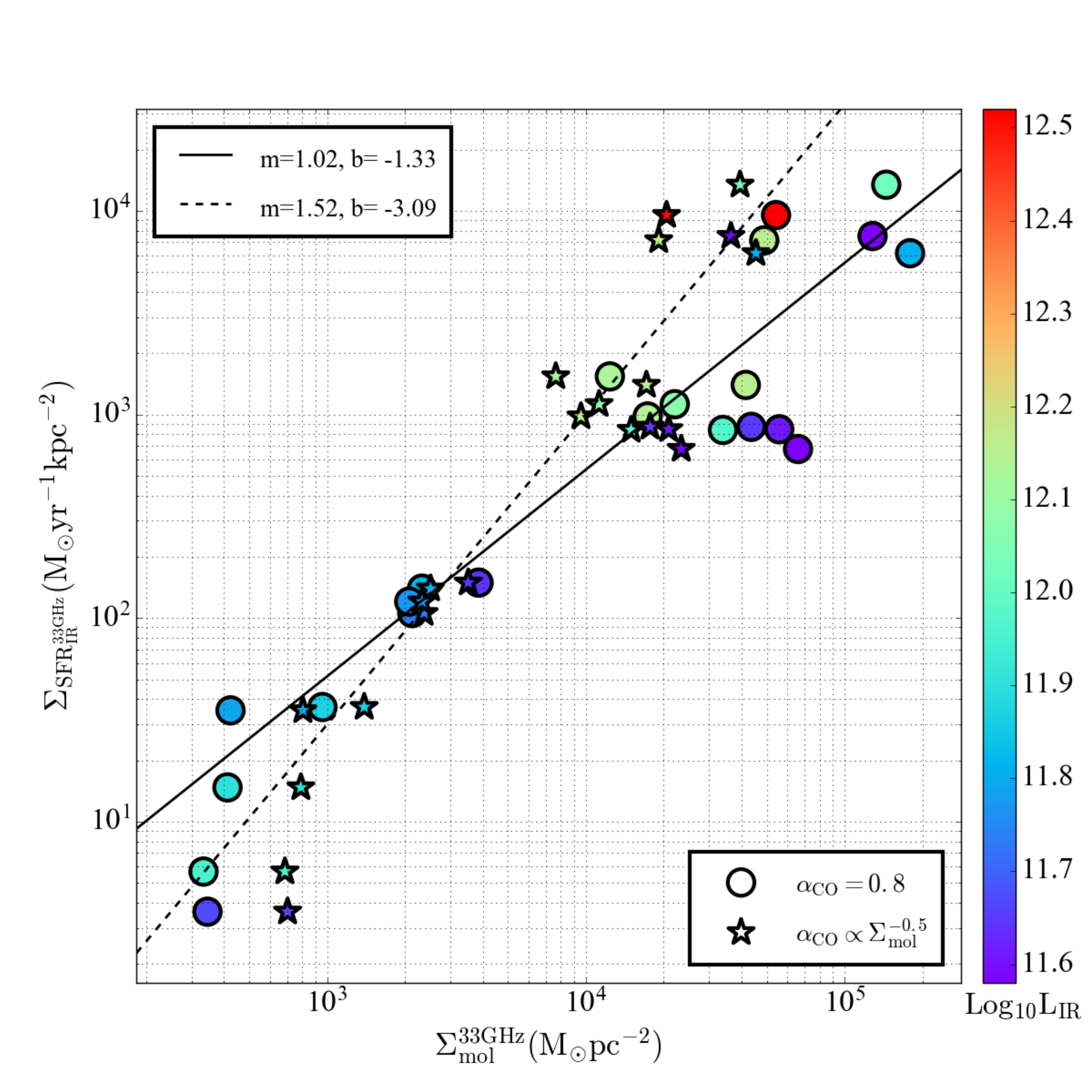}
\includegraphics[width=3.5in]{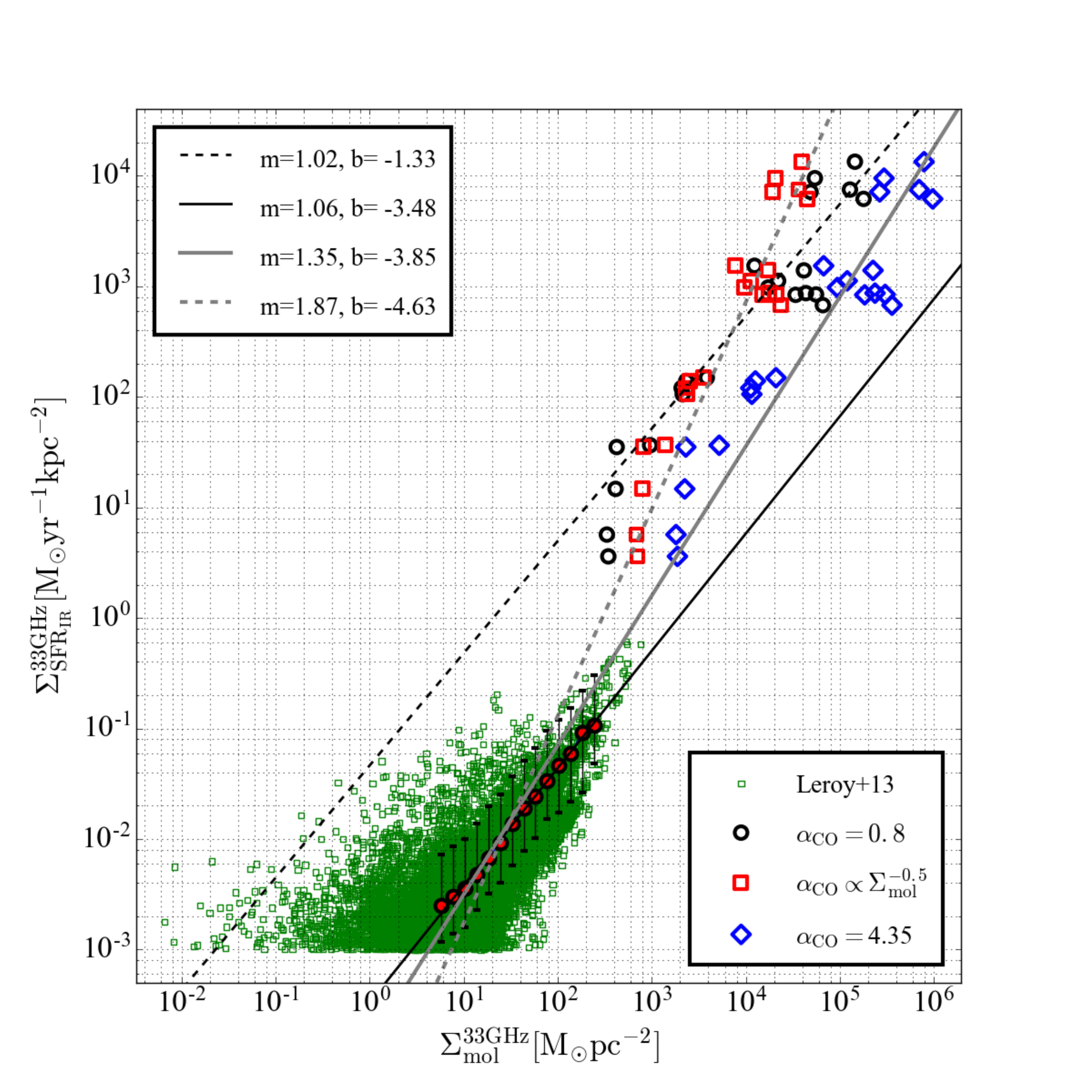}
\caption{Kennicutt-Schmidt (KS) law of star formation within A$_{\rm 50,d}$ (see Section \ref{sec:impl_radio} for details). The SFR was calculated based on the IR luminosity and the molecular gas mass was obtained using an $\alpha_{\rm CO}$ factor. $(Top~left)$ The mass of the gas was calculated using a fixed CO-to-H$_{2}$ conversion factor of 0.8 (typical for ULIRGs). The squares represent lower limits and the solid line is the fit to the data. The dotted lines are separated by 0.3 dex from the fit and the colors represent the infrared luminosity of each system. $(Top~right)$ The circles show the values from the $top~left$ panel, i.e., with the gas obtained using a fixed conversion factor of 0.8, and the solid line shows the fit. The stars show values where the gas was obtained using a conversion factor that varied for each source (see Table \ref{table:tbl-6}) and the dashed line is the fit to the data. $(Bottom)$ Comparison between values from nearby disk galaxies from \citet{Leroy13} (green squares) and from our sample using different values of $\alpha_{\rm CO}$ (other symbols). The dashed line shows the fit to our data using a conversion factor of 0.8, and the solid line shows the fit to the binned data from \citet{Leroy13} (red circles). The best fit to the disk galaxies and U/LIRGs (using a Galactic conversion factor, blue diamonds) is shown by the solid grey line (see Section \ref{sec:relations} for more details). By using a conversion factor that depends on the gas surface density of the source (red squares), we obtain a steeper slope (dashed grey line) compared to that obtained using a fixed conversion factor. This indicates the crucial role that $\alpha_{\rm CO}$ plays when studying the KS law. The nuclear regions in local U/LIRGs occupy the higher end of the star formation law indicating they host more extreme environments in comparison to disk galaxies.\label{fig:fig9}}
\end{figure*}

The observed scaling between star formation rate surface density, $\Sigma_{\rm SFR}$, and gas surface density, $\Sigma_{\rm gas}$, is often used as a main diagnostic of the physics of star formation in galaxies \citep[e.g.,][]{Kennicutt98}. \citet{Kennicutt98} fit a scaling between galaxy-averaged $\Sigma_{\rm SFR}$ and $\Sigma_{\rm gas}$ that describes both normal disk galaxies and starbursts. The starbursts in \citet{Kennicutt98} have high $\Sigma_{\rm SFR}$ and $\Sigma_{\rm gas}$ and include U/LIRGs like those studied here.

The contrast between the normal disks (low $\Sigma_{\rm SFR}$, $\Sigma_{\rm gas}$) and the starburst galaxies (high $\Sigma_{\rm SFR}$, $\Sigma_{\rm gas}$) played a main role in driving the best overall fit of \citet{Kennicutt98}, $\Sigma_{\rm SFR} \sim \Sigma_{\rm gas}^{1.4}$. This contrast depends on the sizes adopted for the starburst galaxies. Changing the size affects both surface densities by the same factor, but because the overall relationship between $\Sigma_{\rm SFR}$ and $\Sigma_{\rm gas}$ is non-linear, the adopted size affects the slope. 

In Figure \ref{fig:fig9} we place each of our targets in the $\Sigma_{\rm SFR_{IR}}^{\rm 33GHz}$-$\Sigma_{\rm gas}^{\rm 33GHz}$ (or $\Sigma_{\rm SFR_{IR}}^{\rm 33GHz}$-$\Sigma_{\rm mol}^{\rm 33GHz}$) plane (see Section \ref{sec:Ssfr} and \ref{sec:Smol} for details on the derivation of $\Sigma_{\rm SFR_{IR}}^{\rm 33GHz}$ and $\Sigma_{\rm mol}^{\rm 33GHz}$). In the top left panel, we show only the U/LIRGs from our sample and adopt a fixed $\alpha_{\rm CO}$ = 0.8~M$_\odot$~pc$^{-2}$~(K~km~s$^{-1}$). These U/LIRGs show high surface densities and an approximately linear relationship. A non-linear least-squares fit\footnote{We used the \texttt{scipy.optimize.curve\_fit} algorithm and a function of the form $\mathrm{Y = slope~X + coefficient}$ to obtain the slope and coefficient, and their standard deviation errors. We excluded sources with upper limits to their sizes.} yields

\begin{equation}
\label{eqn:ks}
\mathrm{log_{10}(\Sigma_{SFR_{IR}}^{33GHz})} = (1.02 \pm 0.10)~\mathrm{log_{10}(\Sigma_{mol}^{33GHz})}-(1.33 \pm 0.47)~.
\end{equation}

\noindent This slope is in good agreement with the results found by \citet{Liu15} for disk galaxies and for U/LIRGs. \citet{Genzel10} also noted that the internal relationship for starburst galaxies was more nearly linear than the relationship using both types of galaxies, giving rise to the idea of ``two sequences'' of star formation. A similar conclusion of ``two sequences'' of star formation is also derived by \citet{Daddi10}, although they obtained a steeper slope ($\sim$1.4) for each type of galaxies that approaches unity within the uncertainty of their measurements. With a slope close to unity, another way to express Equation \ref{eqn:ks} is that for a ``starburst'' conversion factor, we find a typical gas depletion time, $\tau_{\rm dep} \equiv M_{\rm mol}/{\rm SFR}$, of $\tau_{\rm dep} \sim 20$~Myr for the targets studied here. Note that this short timescale would potentially lead to a relatively flat-spectrum radio source inconsistent with the observed FIR/radio correlation (see section \ref{sec:nature}), however the uncertainty in the calculated $\tau_{\rm dep}$ is at least a factor of a few.

In addition to the size, the adopted conversion factor can have a large effect on the results. Because we find an approximately linear relationship within our sample, shifting from one constant $\alpha_{\rm CO}$ to another will not affect the slope. For example, if we use a Galactic $\alpha_{\rm CO}$ = 4.35~M$_\odot$~pc$^{-2}$~(K~km~s$^{-1}$) instead, the coefficient would shift to -2.08$\pm$0.55, raising the depletion time to $\tau_{\rm dep} = 125$~Myr. For comparison, \citet{Leroy13} find a significantly longer $\tau_{\rm dep}$, $\sim 1.6$~Gyr, in the disks of nearby normal galaxies.

Several suggestions posit a continuous variation in $\alpha_{\rm CO}$ that depends on surface density (see Equation \ref{eq:alpha_co}). Adopting such prescription affects the slope of the derived relation. If we adopt the surface density-dependent slope discussed in Section \ref{sec:Smol}, the best fit shifts to

\begin{equation}
\label{eqn:ks2}
\mathrm{log_{10}(\Sigma_{SFR_{IR}}^{33GHz})} = (1.52 \pm 0.16)~\mathrm{log_{10}(\Sigma_{mol}^{33GHz})}-(3.09 \pm 0.66)~.
\end{equation}

\noindent The top right panel in Figure \ref{fig:fig9} shows our data for two cases: a fixed ``starburst'' conversion factor and the mass surface-density dependent value. Internal to the starburst sample, the linearity or non-linearity of the slope depends entirely on the treatment of the conversion factor, and the assumption of the cospatiality between CO and radio emission; the apparent relationship between $\Sigma_{\rm SFR}$ and CO luminosity surface brightness is approximately linear.

As mentioned above, the contrast between normal disk galaxies and starbursts played a large role in determining the \citet{Kennicutt98} fit. The bottom panel of Figure \ref{fig:fig9} explores this contrast. There, we compare our results to those found for kpc-size regions drawn from $30$ nearby disk galaxies by \citet{Leroy13}. Individual regions appear as green squares and the median and scatter in $\Sigma_{\rm SFR}$, in bins of fixed $\Sigma_{\rm mol}$, appear as red points with error bars. Note that, in contrast to \citet{Kennicutt98}, we consider only the molecular gas component of the ISM, and, in the normal galaxies, we consider individual kpc-sized regions. \citet{Kennicutt98} include atomic gas and consider whole-disk averages. We chose our approach to focus on star-forming (molecular) gas in comparable sized regions in order to contrast the ability of gas to form stars in the two types of systems.

Figure \ref{fig:fig9} shows a significant contrast between disks and our starburst sample, even for matched $\alpha_{\rm CO}$ (a similar contrast was seen when comparing $\tau_{\rm dep}$). In that case, $\alpha_{\rm CO}=4.35$~M$_\odot$~pc$^{-2}$~(K~km~s$^{-1}$) for both samples, a fit to our sample and the \citet{Leroy13} bins yield:

%fit using heracles data and Ulirgs using alpha_co=4.35
\begin{equation}
\label{eqn:ks4}
\mathrm{log_{10}(\Sigma_{SFR_{IR}}^{33GHz})} = (1.35 \pm 0.04)~\mathrm{log_{10}(\Sigma_{mol}^{33GHz})}-(3.85 \pm 0.13)~.
\end{equation}

\noindent Meanwhile, adopting the starburst $\alpha_{\rm CO} = 0.8$~M$_\odot$~pc$^{-2}$~(K~km~s$^{-1}$) for our sample only yields:

%fit using heracles data and Ulirgs using alpha_co=0.8
\begin{equation}
\label{eqn:ks3}
\mathrm{log_{10}(\Sigma_{SFR_{IR}}^{33GHz})} = (1.63 \pm 0.07)~\mathrm{log_{10}(\Sigma_{mol}^{33GHz})}-(4.18 \pm 0.22)~.
\end{equation}

\noindent In both cases, the data appear to support the ``two sequences'' idea, at least to some degree, with internal relationships in the two sub-samples that are more nearly linear, and a steep slope when contrasting both populations (but see below). This is particularly the case when we use a starburst conversion factor for our sample.

Adopting $\mathrm{\alpha_{CO} \propto \Sigma_{mol}^{-0.5}}$ (see equation \ref{eq:alpha_co}) we find instead

%fit using heracles data and Ulirgs using alpha_co varying with Sigmal_mol
\begin{equation}
\label{eqn:ks5}
\mathrm{log_{10}(\Sigma_{SFR_{IR}}^{33GHz})} = (1.87 \pm 0.06)~\mathrm{log_{10}(\Sigma_{mol}^{33GHz})}-(4.63 \pm 0.19)~.
\end{equation}

%\noindent This case best aligns the individual points in our sample along a single power law extending from the normal galaxy points.

\noindent In this case we find an even steeper slope when fitting the combined data, from the U/LIRGs studied here and the normal spirals from \citet{Leroy13}, than when we
use a starburst conversion factor for our sample only, and even more so when we fit either sample alone. To some degree, this reinforces the ``two sequences'' view, but with a strong caveat. Our results are consistent with the idea that the depletion time is multi-valued at a fixed gas surface density, but they do {\em not} offer any strong evidence regarding a true bimodality. The data that we use are not complete in any meaningful sense. Therefore, the absence of intermediate $\tau_{\rm dep}$ points near where the two samples would overlap can easily be a selection effect. That is: there may be plenty of parts of galaxies that fill in apparently empty space in Figure \ref{fig:fig9}, our samples are simply not constructed to reveal this. Indeed, \citet{Saintonge11,Huang15,Genzel15} and others have convincingly shown that a continuous range of gas depletion times appear to exist within the population \citep[see also][for further discussion on continuous and bi-modal star formation scaling relations]{Scoville16}.

Our results do strongly reinforce the idea that the disk-starburst contrast is essential to probe the non-linear nature of star formation scaling relations. We also show, following a number of others \cite[e.g., see][]{Bouche07,O&S11} that the adopted conversion factor, in addition to the starburst sizes, plays a large role in the results. We summarize all the different fits to the gas star formation law using the different conversion factors in Table \ref{table:tbl-7}.

{\em Efficiency per Free Fall Time:} A popular class of models posits an approximately fixed fraction of gas converts to stars per gravitational free fall time, $\mathrm{\tau_{ff}^{\rm mol} = \sqrt{3\pi/(32G\rho_{mol})}}$ \citep[e.g.,][]{Krumholz05,Krumholz12}. If we adopt a simple, spherical, with radius R$_{\rm 50,d}$, view of the geometry of the systems, we can estimate $\tau_{\rm ff}^{\rm mol}$. For a three dimensional Gaussian, this implies an aperture correction of $\sim 1/3.4$ for the total gas mass (or SFR) within that volume.

Comparing $\tau_{\rm ff}^{\rm mol}$ to the depletion time of the molecular gas mas, $\mathrm{\tau_{dep}^{mol}}$, we estimate the efficiency of the conversion of the gas mass into stars per free fall time, or $\epsilon_{\rm ff}$ = $\tau_{\rm ff}^{\rm mol}/\tau_{\rm dep}^{\rm mol}$. We find median values for $\mathrm{\tau_{ff}^{mol}}$ of 1.1, 1.5, and 0.5 Myr for ``starburst", surface-density dependent, and Galactic conversion factors. These numbers imply median $\epsilon_{\rm ff}$ of 8\%, 15\%, and 0.6\%. The first two numbers appear high compared to the universal $\sim 1\%$ $\epsilon_{\rm ff}$ assumed in the \citet{Krumholz12} model, and in more agreement with a non-universal star formation efficiency \citep{Semenov16}, but we emphasize the uncertainty in the adopted geometry.

\subsection{Are Local Major Mergers Eddington-Limited Starbursts?}
\label{sec:edd}

\begin{figure*}[tbh]
\centering
\includegraphics[width=3.5in]{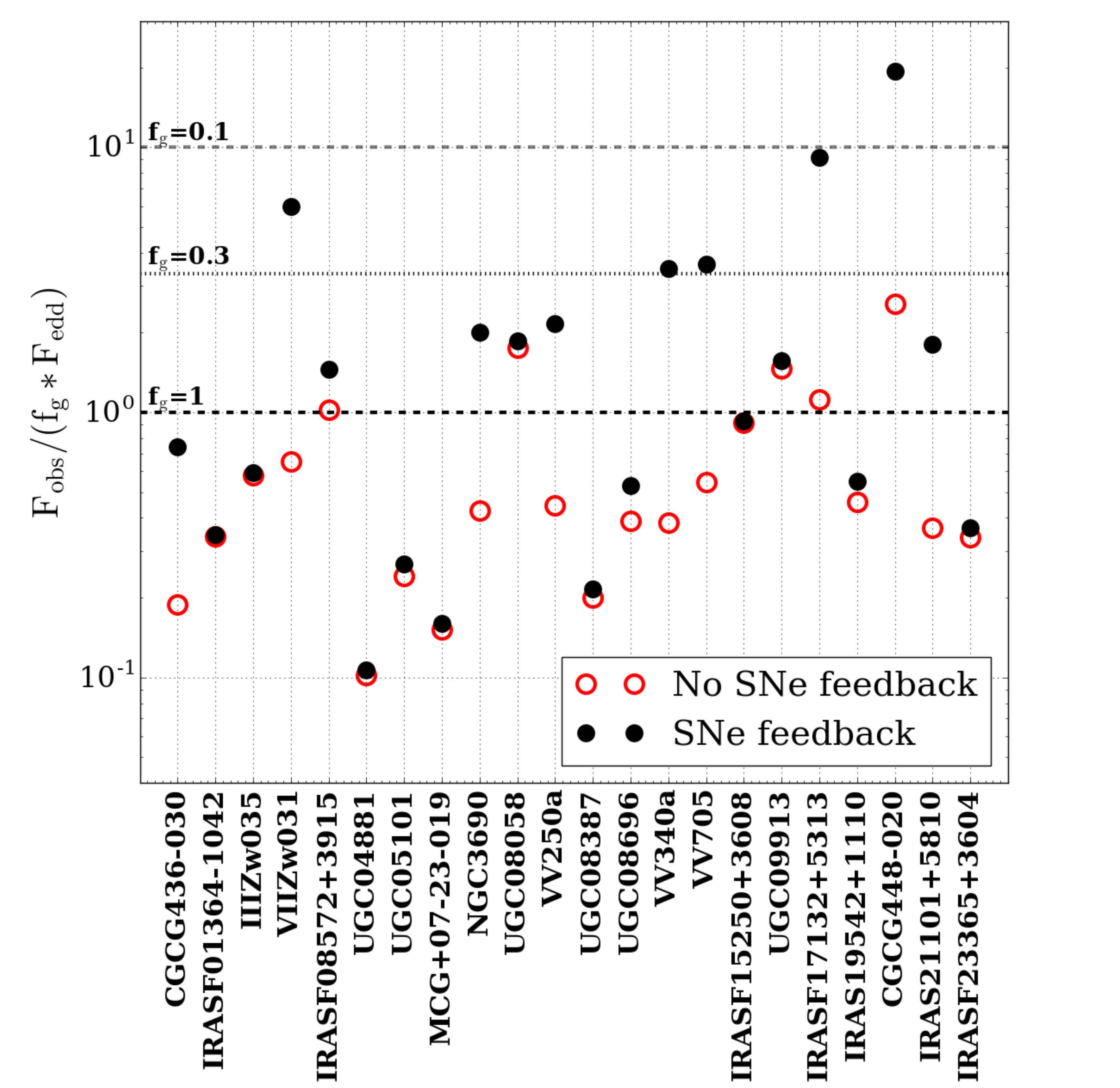}
\caption{Observed infrared flux, F$_{\rm obs}$ (or $\mathrm{\Sigma^{33GHz}_{IR}}$), to predicted Eddington flux times gas fraction, $\mathrm{f_{g}*F_{edd}}$, ratio ($\mathrm{F_{obs}/(f_{g}*F_{edd})}$) for each system in the sample. Solid black circles and open red circles show $\mathrm{F_{obs}/(f_{g}*F_{edd})}$ including and not including SNe feedback, respectively. The different horizontal lines indicate how the Eddington limit varies for f$_{\rm g}$ = 0.1, 0.3, and 1.0. Assuming a central gas fraction of 1 and considering SNe feedback, we find almost half of our sample is super-Eddington.\label{fig:fig10}}
\end{figure*}

The high density of star formation and luminosity in the inner parts of our targets undoubtedly creates strong feedback on the gas. This can suppress or even halt ongoing star formation, and in equilibrium we might expect this feedback to counter-balance the force of gravity, leading to some degree of self-regulation. Radiation pressure on dust has been proposed as the main feedback mechanism for compact, optically thick starbursts \citep{Scoville03,Murray05,Thompson05,A&T11}. Momentum injection by supernova explosions \citep{Thompson05,K&O15} and cosmic ray pressure \citep[e.g.,][]{Socrates08,FG13} also likely play a key role.

The high $\mathrm{\Sigma^{33GHz}_{IR}}$ values derived for our targets and their very dusty nature makes them excellent candidates to be ``Eddington-limited'' starbursts. In such a system, the star formation surface density will increase until it yields a radiation pressure on dust that balances the force of gravitational collapse. Because we expect that the force from radiation pressure must be present, then if a source shows a luminosity surface density above this equilibrium value, then some other assumption in the calculation must break down. This could be the assumption of equilibrium, as the pressure exerted by radiation might temporarily or permanently suppress star formation and/or expel gas from the system in a galactic wind. Alternatively, the source of the luminosity could be something other than star formation. One common inference when this ``maximal starburst'' case is exceeded is that an appreciable part of the luminosity in the system may arise from an AGN. Alternatively, the assumptions about disk structure used to calculate the force of gravity may be wrong. For example, in the models of \citet{Thompson05} the gas fraction and velocity dispersion play a key role.

We have already seen some evidence that this case may apply to our systems. \citet{Thompson05} noted an infrared luminosity surface density of $\mathrm{\Sigma_{IR} \sim 10^{13}~L_\odot~kpc^{-2}}$ as characteristic for dense, optically thick Eddington-limited starbursts. We showed above that a subset of our targets exhibit $\mathrm{\Sigma_{IR}^{33GHz}}$ near, or even above, this limit.

In detail, the exact limiting $\mathrm{\Sigma^{33GHz}_{IR}}$ depends on the detailed structure of the starburst disk, including its size, stellar velocity dispersion ($\sigma$), gas mass fraction (f$_{\rm g}$), dust-to-gas ratio, and the Rosseland mean opacity ($\kappa$) of the system. Thus, the Eddington limit varies from source to source. Taking this in to account, we compare our inferred $\mathrm{\Sigma^{33GHz}_{IR}}$ (or F$_{\rm obs}$) for each target to the predicted Eddington flux. For hydrostatic equilibrium in a disk, the Eddington flux, $F_{\rm edd}$ is:

\begin{equation}
\label{eqn:hidro}
\mathrm{F_{edd}} = \frac{4 \pi G c \Sigma}{\kappa}~,
\end{equation}

\noindent where $\Sigma$ is the surface density of the mass that dominates the gravitational potential involved in the star forming region and $\kappa$ is the effective opacity. 

The effective opacity depends on the characteristics of the system under study. Following \citet{Thompson05} and \citet{A&T11}, for systems that are optically thick to the UV radiation, but optically thin to the re-radiated far infrared emission, $\mathrm{\kappa (thin)\sim\Sigma_{\rm gas}^{-1}}$. For systems that are optically thick to the re-processed far infrared emission, i.e., when $\mathrm{\Sigma_{gas}~\gtrsim~1~g~cm^{-2}}$, $\mathrm{\kappa(thick)\approx\kappa_{o}T^{2}}$, where T is the temperature of the central star forming disk and $\mathrm{\kappa_{o}~\approx2.4\times10^{-4}~cm^{2}~g~K^{-2}}$ \citep{Semenov03}. The transition between regimes is expected to occur when $\mathrm{\Sigma_{gas}~\sim~1~g~cm^{-2}}$. Note that in systems without large IR optical depths, the momentum and turbulence from supernovae is expected to dominate support of the disk, rather than radiation pressure.

For a Milky Way gas-to-dust ratio and $\Sigma = \Sigma_{mol}/f_{g}$ a version of Equation \ref{eqn:hidro} that captures all three possible regimes is

\begin{equation}
\label{eqn:fedd}
{\rm F_{edd}} = \frac{\pi G c \Sigma_{\rm mol}^{2}}{\rm {f_{g}(1 + \tau_{IR} + 10n_{mol}^{-1/7} -exp[-\tau_{UV}])}}~.
\end{equation}

\noindent Here $\mathrm{\Sigma_{mol}=\Sigma_{mol}^{33GHz}}$ is the gas surface density (see Table \ref{table:tbl-6}), and f$_{g}$ is the gas mass fraction in the core of the galaxy. $\tau_{\rm IR}=\kappa_{\rm IR}(\rm T)\Sigma_{\rm mol}/2$ is the infrared optical depth and $\mathrm{\tau_{UV}\sim500-1000(cm^{2}~g^{-1})\times\Sigma_{mol}/2}$ the ultraviolet optical depth. We approximate the contribution to support by supernovae (SNe) as $10n_{\rm mol}^{-1/7}$ (see Appendix of \citealt{FG13}, and \citealt{K&O15}); the numerical prefactor can vary by a factor of several, up to $\sim$30n$_{\rm mol}^{-1/7}$), where $\mathrm{n_{mol}\equiv n_{mol}^{33GHz}}$ is the number density of the gas (see Table \ref{table:tbl-6}). 

In order to derive $\kappa_{\rm IR}(\rm T)$ we assume equation (40) from \citet{Thompson05} describes the relation between $T$, $T_{\rm eff}$, and the vertical IR optical depth. We then solve the implicit equation for $T$ assuming that,

\begin{equation}
\frac{\kappa_{IR}(T)}{\rm (cm^{2}~g^{-1})} = \begin{cases} 2.4\times10^{-4}~T^{2}, & \mbox{if } T\mbox{ $<$ 180 K} \\ 2.4\times10^{-4}~180^{2}\approx7.8, & \mbox{if } T\mbox{ $\geq$ 180 K} \end{cases}
\end{equation}

\noindent where $\mathrm{\Sigma^{33GHz}_{IR} = \sigma_{SB} T_{eff}^{4}}$, and $\sigma_{SB}$ is the Stefan-Boltzmann constant.

In Figure \ref{fig:fig10} we show the resulting $\mathrm{F_{obs}/(f_{g}*F_{edd})}$ estimated for each galaxy. We draw lines at the Eddington limit where $\mathrm{F_{obs} = F_{edd}}$ for different gas fractions, and above which the systems are considered super-Eddington. We include the cases for which we consider SNe feedback (black solid circles) and where we do not (open red circles).

If we assume the gas fraction in the center of the sources is closer to 1 and neglect supernovae feedback, we observe 5 systems that are super-Eddington including Mrk 231 (UGC 08058), Arp 220, and CGCG 448-020. As noted earlier, Mrk 231 is known to host a strong AGN, and if this drives the IR luminosity then this Eddington calculation for a starburst disk does not apply. Arp 220 has mid-IR evidence of energetic AGN -- based on low PAH equivalent widths (= 0.03$-$0.17) and/or high 30-to-15 $\mu$m flux density ratios (= 10$-$20) indicative of very warm, Seyfert-like mid-IR dust emission \citep{Stier13}. Arp 220 is a special case, where the mid-IR diagnostics potentially break due to the high dust opacity of the system. This is also applicable to CGCG 448-020 for which the source dominating the emission at IR and radio wavelengths (north-east component, see Figure \ref{fig:fig1} and the Appendix for more information) is highly obscured.

The sources in our sample are extreme starbursts for which SNe feedback is most likely important, especially in systems that are more extended and warm (T $<$ 180 K). If we include SNe feedback in the calculation of $\mathrm{F_{edd}}$ (solid black circles in Figure \ref{fig:fig10}) we observe that for a gas fraction of 1, 11 out of 22 systems in our sample show super-Eddington values, including the systems mentioned above. Assuming a more conservative gas fraction of 0.3, which is about the system averaged gas mass fraction based on \citet{Larson16}, we find that 5 systems are super-Eddington. Note that CGCG 448-020 is a special case since it shows super-Eddington values independent of the gas fraction, indicating the potential presence of an AGN.

We note that our results highly depend on the adopted gas fraction, and while we might expect some funneling of gas to the center to raise the gas fraction to higher values locally, the best way to further improve our constraints are resolved observations of the disk dynamics, which can yield the total (dynamical) mass, velocity dispersion, and gas mass.

\section{Conclusions}
\label{sec:conclusions}

We present a high resolution imaging survey of 33~GHz continuum emission from local U/LIRGs. Using all four VLA configurations and a bandwidth of $\Delta \nu = 2$~GHz, we achieve very high resolutions of $0\,\farcs07$--$0\,\farcs67$, or 30-720~pc at the distance of these sources, while still retaining sensitivity to emission on large scales. This is the first such survey at such high frequencies (for the VLA). As a result, we improve on the resolution of previous work by \citet{Condon91} and \citet{Condon90} by a factor of $4$. Because of the steep spectral index of galaxies in this range, the improved sensitivity gained from the VLA upgrade was a key element in the survey. 

Using these data, we find:

\begin{enumerate}

\item Most of the 33~GHz emission observed at low resolution arises from sources that appear compact in the highest resolution maps. For the majority of the U/LIRGs studied here, more than 50\% of the integrated flux density at 33 GHz arises from sources with Gaussian-like morphologies at high resolution and extend typically a few times the size of the beam.

\item The 33 GHz emission reflects a mixture of synchrotron and free-free emission. For different approaches, we achieve slightly different results, but within the uncertainties approximately equal fractions of thermal and nonthermal emission could contribute at 33~GHz. To improve on this uncertain number, improved coverage of the radio SED, especially achieving reliable flux densities at many frequencies in the range $15$--$50$~GHz, will be extremely helpful. Unless the emission is highly clumped within the recovered high resolution beam, brightness temperature arguments suggest that all of the observed U/LIRGs are optically thin at 33~GHz.

\item By making use of the 33 GHz size to indicate the active, star-forming region, we provide estimates for the surface densities of gas, star formation, and infrared emission. These quantities are more extreme than those found in typical star-forming galaxies but also vary strongly across the sample, spanning a range of $\approx$ 4~dex. The highest values in our sample are among the highest measured for any galaxies.

\item We also make use of the measured 33 GHz sizes of the sample to estimate their star formation rate surface densities, $\mathrm{\Sigma_{SFR_{33 GHz}}}$. We find that $L_{\rm [C II]} / L_{\rm FIR}$ decreases with increasing $\Sigma_{\rm SFR_{33GHz}}$, increasing opacity (as measured via the flattening of the radio spectral index between 1.5 and 6 GHz), and increasing compactness. These measurement agree with prior studies which used infrared sizes measured at coarser resolution to estimate $\Sigma_{\rm IR}^{\rm 33GHz}$ (or $\Sigma_{\rm SFR_{IR}}^{\rm 33GHz}$). They confirm that the [{\sc Cii}] `deficit' is more pronounced in the most compact and obscured U/LIRGs.

\item We consider the implications for star formation scaling relations from $\Sigma_{\rm SFR_{IR}}^{\rm 33GHz}$ and $\Sigma_{\rm gas}^{\rm 33GHz}$ derived combining the 33 GHz size estimates with unresolved CO (1$-$0) and IR observations. For any single, fixed conversion factor and considering only the U/LIRGs, we find a slope near unity ($\approx 1.02$) relating the two. However, the U/LIRGs studied here contrast with results for normal spiral galaxies from \citep{Leroy13}, and a nonlinear slope is needed to relate the two different populations \citep[consistent with][]{Kennicutt98,Liu15}. 

\item The exact value of the power law index needed to fit both normal disks and these U/LIRGs depends sensitively on the sizes of the U/LIRGs (which we measured), on the assumption that the star formation, traced by 33 GHz, and the molecular gas, as traced by CO, have matched structure, and the prescription for the CO-to-H$_2$ conversion factor (which is highly uncertain). We show results for three common approaches to the conversion factor, and the power law index relating normal disk galaxies to the U/LIRGs studied in this paper varies from $\sim 1.4$ to $\sim 1.9$.

\item The high column densities that we infer imply high opacities outside the $\sim$cm and mm-wave regime. By adopting a ``starburst" conversion factor, the average extinction at optical wavelengths is $A_V \sim 22-12,000$~mag for this sample. $13$ of the observed sources appear X-ray Compton thick, with average $\mathrm{N_{H} > 1.5\times10^{24}~cm^{-2}}$. At IR wavelengths, the opacity is less, $\tau_{100} \sim 0.02-12$, however they are still affected by dust with those same 13 sources, but one, being optically thick at 100 $\mu$m. The combination of the measured sizes at 33 GHz with the $1.5$~GHz flux densities from \citet{Condon90} also indicate that opacity must play a significant role at lower radio frequencies.

\item The targets show high infrared surface brightnesses, with 7 sources having $\mathrm{\Sigma_{IR}^{33GHz}>10^{13} L_{\odot}~kpc^{-2}}$, a characteristic value suggested by \citet{Thompson05} for dusty, radiation-pressure supported starburst galaxies. However, if we consider feedback from supernovae and adopt a nuclear gas fraction of 1, we find 11 out of 22 systems are super-Eddington. This number decreases to 5 if we adopt a gas fraction of 0.3 instead. We note the need for both detailed observations of the inner disk structure and several observational subtleties that should be accounted for in comparing the observed $\Sigma_{\rm IR}^{\rm 33GHz}$ to models.

\end{enumerate}

\acknowledgments
We thank the anonymous referee for providing excellent comments that further improved this paper. We thank Jason Chu for making Herschel/PACS images available for quick visual comparison and further discussions on the mid-IR data. Support for this work was provided by the NSF through the Grote Reber Fellowship Program administered by Associated Universities, Inc./National Radio Astronomy Observatory. L.B-M. was supported by Fulbright, Becas Chile - CONICYT. The work of A.K.L. is supported by the National Science Foundation under Grants No. 1615105 and 1615109. A.S.E., G.C.P. and L.B-M. were supported by NSF grant AST 1109475. G.C.P. was supported by a FONDECYT Postdoctoral Fellowship (No. 3150361). T.D-S. acknowledges support from ALMA-CONICYT project 31130005 and FONDECYT regular project 1151239. This research made use of the NASA/IPAC Extragalactic Database (NED), which is operated by the Jet Propulsion Laboratory, California Institute of Technology, under contract with the National Aeronautics and Space Administration, and NASA's Astrophysics Data System Bibliographic Services. The National Radio Astronomy Observatory is a facility of the National Science Foundation operated under cooperative agreement by Associated Universities, Inc.

\begin{deluxetable}{lccccccc}%[th]
\tabletypesize{\scriptsize}
\tablecaption{Characteristics of the Sample Galaxies\label{table:tbl-1}}
%\tablewidth{0pt}
\tablehead{
\colhead{Galaxy Name} & \colhead{Alternate Name} & \colhead{R.A. (J2000)} & \colhead{Decl. (J2000)}& 
\colhead{$\mathrm{log_{10}(L_{IR})}$} & \colhead{$\mathrm{ D_{L}}$} &\colhead{Scale (kpc/$\arcsec$)}&\colhead{Id. number}\\
\colhead{(1)}&\colhead{(2)}&\colhead{(3)}&\colhead{(4)}&\colhead{(5)}&\colhead{(6)}&\colhead{(7)}&\colhead{8}}
\startdata
CGCG 436-030&MCG +02-04-025&$\mathrm{01^{h}20^{m}02\,\fs722}$ & $\mathrm{+14^{\circ}21'42\,\farcs94}$ & 11.64 & 127 &0.601 & 1\\ 
IRAS F01364-1042 & 2MASXJ01385289-1027113 &$\mathrm{01^{h}38^{m}52\,\fs921}$ & $\mathrm{-10^{\circ}27'11\,\farcs42}$ & 11.81 & 201 &0.942 & 2\\ 
III Zw 035 &  &$\mathrm{01^{h}44^{m}30\,\fs500}$& $\mathrm{+17^{\circ}06'05\,\farcs00}$ & 11.58 & 111 &0.526 & 3\\ 
VII Zw 031 &  & $\mathrm{05^{h}16^{m}46\,\fs096}$ & $\mathrm{+79^{\circ}40'13\,\farcs28}$ & 11.95 & 229 &1.066 & 4\\ 
IRAS 08572+3915 & & $\mathrm{09^{h}00^{m}25\,\fs390}$ & $\mathrm{+39^{\circ}03'54\,\farcs40}$ & 12.13 & 254 &1.176 & 5\\ 
UGC 04881 & Arp 55 & $\mathrm{09^{h}15^{m}55\,\fs100}$ & $\mathrm{+44^{\circ}19'55\,\farcs00}$ & 11.70 & 169 &0.796 & 6\\ 
UGC 05101 &  & $\mathrm{09^{h}35^{m}51\,\fs595}$ & $\mathrm{+61^{\circ}21'11\,\farcs45}$ & 11.97 & 168 &0.792 & 7\\ 
MCG +07-23-019 & Arp 148 & $\mathrm{11^{h}03^{m}53\,\fs200}$ & $\mathrm{+40^{\circ}50'57\,\farcs00}$ & 11.61 & 149 &0.704 & 8\\ 
NGC 3690 & Arp 299 & $\mathrm{11^{h}28^{m}32\,\fs300}$ & $\mathrm{+58^{\circ}33'43\,\farcs00}$ & 11.82 & 45.2 &0.217 & 9\\ 
UGC 08058 & Mrk 231 & $\mathrm{12^{h}56^{m}14\,\fs234}$ & $\mathrm{+56^{\circ}52'25\,\farcs24}$ & 12.52 & 181 &0.849 & 10 \\ 
VV 250 & UGC 08335 NED02& $\mathrm{13^{h}15^{m}34\,\fs980}$ & $\mathrm{+62^{\circ}07'28\,\farcs66}$ & 11.77 & 132 &0.621 & 11\\ 
UGC 08387 & Arp 193, IC 883 & $\mathrm{13^{h}20^{m}35\,\fs300}$ & $\mathrm{+34^{\circ}08'21\,\farcs00}$ & 11.65 & 101 &0.479 & 12\\ 
UGC 08696 & Mrk 273 & $\mathrm{13^{h}44^{m}42\,\fs111}$ & $\mathrm{+55^{\circ}53'12\,\farcs65}$ & 12.15 & 162 &0.761 & 13\\ 
VV 340a & UGC 09618 NED02& $\mathrm{14^{h}57^{m}00\,\fs826}$ & $\mathrm{+24^{\circ}37'04\,\farcs12}$ & 11.67 & 144 &0.665 & 14\\ 
VV 705 & I Zw 107 & $\mathrm{15^{h}18^{m}06\,\fs344}$ & $\mathrm{+42^{\circ}44'36\,\farcs69}$ & 11.87 & 172 &0.807 & 15\\ 
IRAS 15250+3608 & & $\mathrm{15^{h}26^{m}59\,\fs404}$ & $\mathrm{+35^{\circ}58'37\,\farcs53}$ & 12.02 & 238 &1.105 & 16\\ 
UGC 09913 & Arp 220 & $\mathrm{15^{h}34^{m}57\,\fs116}$ & $\mathrm{+23^{\circ}30'11\,\farcs47}$ & 12.16 & 77.2 &0.369 & 17 \\ 
IRAS 17132+5313 & & $\mathrm{17^{h}14^{m}20\,\fs000}$ & $\mathrm{+53^{\circ}10'30\,\farcs00}$ & 11.90 & 217 &1.012 & 18\\ 
IRAS 19542+1110 & &$\mathrm{19^{h}56^{m}35\,\fs440}$ & $\mathrm{+11^{\circ}19'02\,\farcs60}$ & 12.07 & 277 &1.277 & 19\\ 
CGCG 448-020 & II Zw 096 &$\mathrm{20^{h}57^{m}23\,\fs900}$& $\mathrm{+17^{\circ}07'39\,\farcs00}$& 11.79 & 148 &0.698 & 20\\ 
IRAS 21101+5810 & 2MASX J21112926+5823074 &$\mathrm{21^{h}11^{m}30\,\fs400}$ & $\mathrm{+58^{\circ}23'03\,\farcs20}$ & 11.74 & 162 &0.764& 21\\ 
IRAS F23365+3604 & 2MASX J23390127+3621087 & $\mathrm{23^{h}39^{m}01\,\fs273}$ & $\mathrm{+36^{\circ}21'08\,\farcs31}$ & 12.16 & 273 &1.262 & 22
\enddata

\tablecomments{Column 1: Name of the galaxy; Column 2: alternate name; Column 3: Source right ascension (J2000) from NED;
Column 4: Source declination (J2000) from NED; Column 5: total infrared luminosity from 8--1000 $\mu$m in log$_{10}$ Solar units computed from the 
IRAS flux densities from \citet{Sanders03} and following the equation from \citet{S&M96}; Column 6: luminosity distance from NED; Column 7: Scale at Hubble flow distances from NED, corrected for the CMB, used to convert from arcseconds to kpc; Column 8: Number used to identify each system in some of the figures presented in this paper.}
\end{deluxetable}

\begin{deluxetable}{lcc}[th]
\tabletypesize{\scriptsize}
\tablecaption{Summary of the Observations\label{table:tbl-2}}
\tablewidth{0pt}
\tablehead{\colhead{Galaxy Name} & \colhead{Primary Calibrator} & \colhead{Secondary Calibrator Ka band} \\
\colhead{(1)}&\colhead{(2)}&\colhead{(3)}}
\startdata
CGCG 436-030 & 3C 48 & J0117+1418\\ 
IRAS F01364-1042 & 3C 48 & J0141-0928\\ 
III Zw 035 & 3C 48 & J0139+1753\\ 
VII Zw 031 & 3C 48 & J0410+7656\\ 
IRAS 0857+3915 & 3C 286 & J0916+3854\\ 
UGC 04881 & 3C 286 & J0920+4441\\ 
UGC 05101 & 3C 286 & J0921+6215\\ 
MCG +07-23-019 & 3C 286 & J1101+3904\\ 
NGC 3690 & 3C 286 & J1128+5925\\ 
Mrk 231 & 3C 286 & J1302+5748\\ 
VV 250 & 3C 286 & J1302+5748\\ 
UGC 08387 & 3C 286 & J1317+3425\\ 
UGC 08696 & 3C 286 & J1337+5501\\ 
VV 340a & 3C 286 & J1443+2501\\ 
VV 705 & 3C 286 & J1521+4336\\ 
IRAS 15250+3609 & 3C 286 & J1522+3144\\ 
Arp 220 & 3C 286 & J1539+2744\\ 
IRAS 17132+5313 & 3C 286 & J1740+5211\\ 
IRAS 19542+1110 & 3C 48 & J1955+1358\\ 
CGCG 448-020 & 3C 48 & J2051+1743\\ 
IRAS 21101+5810 & 3C 48 & J2123+5500\\ 
IRAS F23365+3604 & 3C 48 & J2330+3348
\enddata

\tablecomments{Column 1: Name of the galaxy; Column 2: Primary calibrator used for the observations; Column 4: Secondary calibrator for Ka Band observations}
\end{deluxetable}

\clearpage
\begin{landscape}
\begin{deluxetable}{lcccccrr}%[th]
\tabletypesize{\scriptsize}
\tablecaption{Summary of Resulting Images\label{table:tbl-3}}
%\tablewidth{0pt}
\tablehead{\colhead{Galaxy Name} & \colhead{R.A. (J2000)} & \colhead{Decl. (J2000)}& \colhead{$\theta_{M}$ $\times$ $\theta_{m}$} & \colhead{$\mathrm{\sigma~(\mu Jy~beam^{-1})}$}  & \colhead{S$_{32.5}$ (mJy)}& \colhead{S$_{5.95}$ (mJy)} & \colhead{S$_{1.49}$ (mJy)}\\
\colhead{(1)} &\colhead{(2)} &\colhead{(3)} &\colhead{(4)} &\colhead{(5)} & \colhead{(6)} & \colhead{(7)}&\colhead{(8)}}
\startdata
CGCG 436-030 & $\mathrm{01^{h}20^{m}02\,\fs628}$ & $\mathrm{+14^{\circ}21'42\,\farcs37}$ & $0\,\farcs109 \times 0\,\farcs090$ & 26.0 & 5.6 $\pm$ 0.7& 18.6 $\pm$ 0.04 & 49.1 $\pm$ 2.5 \\%& 12.58 $\pm$ 0.99\\ 
IRAS F01364-1042 & $\mathrm{01^{h}38^{m}52\,\fs885}$ & $\mathrm{-10^{\circ}27'11\,\farcs54}$ & $0\,\farcs141 \times 0\,\farcs086$ & 31.4 & 4.7 $\pm$ 0.6 & 10.0 $\pm$ 0.04& 15.2 $\pm$ 0.8 \\%&29.59 $\pm$ 3.79\\ 
III Zw 035 & $\mathrm{01^{h}44^{m}30\,\fs536}$ & $\mathrm{+17^{\circ}06'08\,\farcs65}$ & $0\,\farcs145 \times 0\,\farcs117$ & 31.2 & 7.3 $\pm$ 0.9& 25.4 $\pm$ 0.05 & 41.2 $\pm$ 2.1\\%& 7.47 $\pm$ 0.58\\ 
VII Zw 031 & $\mathrm{05^{h}16^{m}46\,\fs028}$ & $\mathrm{+79^{\circ}40'12\,\farcs80}$ & $0\,\farcs797 \times 0\,\farcs566$\tablenotemark{a} & 39.1 & 3.0 $\pm$ 0.5 & 12.5 $\pm$ 0.04 & 41.6 $\pm$ 4.2\tablenotemark{c} \\%& 69.47 $\pm$ 6.57\\ 
IRAS 08572+3915 & $\mathrm{09^{h}00^{m}25\,\fs353}$ & $\mathrm{+39^{\circ}03'54\,\farcs22}$ &  $0\,\farcs254 \times 0\,\farcs193$\tablenotemark{*} & 27.3 & 2.1 $\pm$ 0.4 & 4.44 $\pm$ 0.04 & 4.5 $\pm$ 0.2 \\%& 11.85 $\pm$ 1.79\\ 
UGC 04881 &  &  & $0\,\farcs253 \times 0\,\farcs191$\tablenotemark{*} & 26.6 & 1.6 $\pm$ 0.3\tablenotemark{**} & 11.4 $\pm$ 0.09 & 31.6 $\pm$ 1.6\\% & 67.97 $\pm$ 5.14\\ 
\multicolumn{1}{c}{... NE}& $\mathrm{09^{h}15^{m}55\,\fs513}$ & $\mathrm{+44^{\circ}19'57\,\farcs79}$ & " & " & 0.88 $\pm$ 0.15 & ... & ...\\%&...\\
\multicolumn{1}{c}{... SW}& $\mathrm{09^{h}15^{m}54\,\fs787}$ & $\mathrm{+44^{\circ}19'49\,\farcs83}$ & " & " & 0.76 $\pm$ 0.23 & ... & ... \\%& ...\\
UGC 05101 & $\mathrm{09^{h}35^{m}51\,\fs882}$ & $\mathrm{+61^{\circ}21'10\,\farcs84}$ &$0\,\farcs291 \times 0\,\farcs273$\tablenotemark{*} & 25.7 & 14.0 $\pm$ 1.7 & 61.5 $\pm$ 0.08 & 150.0 $\pm$ 7.5\\% & 40.72 $\pm$ 2.50\\ 
MCG +07-23-019 &  $\mathrm{11^{h}30^{m}54\,\fs018}$ & $\mathrm{+40^{\circ}50'59\,\farcs739}$ & $0\,\farcs228 \times 0\,\farcs202$\tablenotemark{*} & 29.5 & 5.9 $\pm$ 0.8 & 16.0 $\pm$ 0.06 & 31.3 $\pm$ 1.6\\% & 37.00 $\pm$ 2.67\\ 
NGC 3690 &  &  & $0\,\farcs260 \times 0\,\farcs240$\tablenotemark{*} & 25.7 & 115.2 $\pm$ 11.3\tablenotemark{**} & 275.5 $\pm$ 0.34 & 658.0 $\pm$ 32.9\\% & 17.04 $\pm$ 0.83\tablenotemark{**}\\
\multicolumn{1}{c}{... W}& $\mathrm{11^{h}28^{m}30\,\fs851}$ & $\mathrm{+58^{\circ}33'44\,\farcs67}$ &" & " & 39.0 $\pm$ 6.0 & ... & ...\\% & 9.86 $\pm$ 0.66\\
\multicolumn{1}{c}{... E} & $\mathrm{11^{h}28^{m}33\,\fs596}$ & $\mathrm{+58^{\circ}33'48\,\farcs02}$ & " & " & 76.3 $\pm$ 9.5 & ... & ... \\%& 7.19 $\pm$ 0.49\\
UGC 08058 & $\mathrm{12^{h}56^{m}14\,\fs186}$ & $\mathrm{+56^{\circ}52'25\,\farcs29}$ &$0\,\farcs257 \times 0\,\farcs227$\tablenotemark{*} & 25.1 & 92.8 $\pm$ 11.1 & 312.8 $\pm$ 0.2 & 296.0 $\pm$ 14.8 \\%& 26.75 $\pm$ 4.36\\ 
VV 250 &  &  &$0\,\farcs236 \times 0\,\farcs219$\tablenotemark{*} & 24.7 & 9.9 $\pm$ 1.2\tablenotemark{**} & 19.6 $\pm$ 0.05 \\%& 50.4 $\pm$ 2.5 & \\ 
\multicolumn{1}{c}{... a (SE)} & $\mathrm{13^{h}15^{m}34\,\fs890}$ & $\mathrm{+62^{\circ}07'27\,\farcs98}$ &" & " & 8.5 $\pm$ 1.1 & ... & 44.6 $\pm$ 2.2\\% & 9.94 $\pm$ 0.90\\ 
\multicolumn{1}{c}{... b (NW)} & $\mathrm{13^{h}15^{m}30\,\fs359}$ & $\mathrm{+62^{\circ}07'44\,\farcs51}$ &" & " & 1.4 $\pm$ 0.3 & ... & 8.5 $\pm$ 0.4\\% & 12.01 $\pm$ 1.30\\ 
UGC 08387 & $\mathrm{13^{h}20^{m}35\,\fs352}$ & $\mathrm{+34^{\circ}08'21\,\farcs11}$ &$0\,\farcs098 \times 0\,\farcs073$ & 19.2 & 17.7 $\pm$ 2.1 & 46.3 $\pm$ 0.08 & 101.0 $\pm$ 5.1\\%& 34.65 $\pm$ 2.77\\ 
UGC 08696 & $\mathrm{13^{h}44^{m}42\,\fs133}$ & $\mathrm{+55^{\circ}53'13\,\farcs02}$ &$0\,\farcs259 \times 0\,\farcs240$\tablenotemark{*} & 26.2 & 19.9 $\pm$ 2.4 & 60.3 $\pm$ 0.08 & 143.0 $\pm$ 7.2\\%& 28.85 $\pm$ 3.34 \\ 
VV 340a & $\mathrm{14^{h}57^{m}00\,\fs703}$ & $\mathrm{+24^{\circ}37'03\,\farcs69}$ &$0\,\farcs494 \times 0\,\farcs437$\tablenotemark{a} & 22.9 & 3.8 $\pm$ 0.5 & 23.7 $\pm$ 0.12 & 74.6 $\pm$ 3.7\\%& 48.88 $\pm$ 2.89\\ 
VV 705 & $\mathrm{15^{h}18^{m}60\,\fs175}$ & $\mathrm{+42^{\circ}44'44\,\farcs51}$ &$0\,\farcs071 \times 0\,\farcs062$& 17.7 & 7.2 $\pm$ 0.9 & 19.6 $\pm$ 0.05 & 47.8 $\pm$ 2.4\\%& 28.09 $\pm$ 2.20\\
IRAS 15250+3609 & $\mathrm{15^{h}26^{m}59\,\fs440}$ & $\mathrm{+35^{\circ}58'37\,\farcs32}$ & $0\,\farcs075 \times 0\,\farcs067$ & 20.7 & 5.1 $\pm$ 0.6 & 12.0 $\pm$ 0.04 & 13.8 $\pm$ 0.7 \\%& 15.54 $\pm$ 4.38\\ 
Arp 220 & $\mathrm{15^{h}34^{m}57\,\fs260}$ & $\mathrm{+23^{\circ}30'11\,\farcs04}$ &$0\,\farcs087 \times 0\,\farcs069$ & 21.9 & 65.6 $\pm$ 7.9 & 194.5 $\pm$ 0.08 & 324.0 $\pm$ 16.2 \\%& 14.58 $\pm$ 1.22\\ 
IRAS 17132+5313 & $\mathrm{17^{h}14^{m}20\,\fs172}$ & $\mathrm{+53^{\circ}10'29\,\farcs77}$ & $0\,\farcs082 \times 0\,\farcs075$ & 18.5 & 3.1 $\pm$ 0.5 & 9.3 $\pm$ 0.04 & 25.8 $\pm$ 1.3 \\%& 33.70 $\pm$ 6.16\\ 
IRAS 19542+1110 & $\mathrm{19^{h}56^{m}35\,\fs770}$ & $\mathrm{+11^{\circ}19'04\,\farcs98}$ & $0\,\farcs087 \times 0\,\farcs081$ & 21.3 & 2.7 $\pm$ 0.4 & 9.5 $\pm$ 0.04 & 20.3 $\pm$ 2.0\tablenotemark{d} \\%& 22.57\tablenotemark{b}\\ 
CGCG 448-020 & $\mathrm{20^{h}57^{m}24\,\fs229}$ & $\mathrm{+17^{\circ}07'39\,\farcs04}$ & $0\,\farcs090 \times 0\,\farcs079$ & 23.1 & 5.3 $\pm$ 0.7 & 14.6 $\pm$ 0.06 & 36.3 $\pm$ 3.6\tablenotemark{c} \\%& 11.58 $\pm$ 1.81\\ 
IRAS 21101+5810 & $\mathrm{21^{h}11^{m}29\,\fs300}$ & $\mathrm{+58^{\circ}23'08\,\farcs65}$ & $0\,\farcs137 \times 0\,\farcs108$ & 27.1 & 3.9 $\pm$ 0.5 & 9.8 $\pm$ 0.04 &22.2 $\pm$ 2.2\tablenotemark{d}\\%& 16.64 $\pm$ 1.52\\ 
IRAS F23365+3604 & $\mathrm{23^{h}39^{m}01\,\fs259}$ & $\mathrm{+36^{\circ}21'08\,\farcs66}$ & $0\,\farcs098 \times 0\,\farcs091$ & 27.5 & 2.9 $\pm$ 0.4 & 10.6 $\pm$ 0.04 & 24.6 $\pm$ 2.5\tablenotemark{c} \\%&  49.04 $\pm$ 4.29
\enddata

\tablecomments{Column 1: Name of the galaxy; Column 2 and 3: J2000 Right Ascension and declination of the Gaussian fitted to obtain the integrated flux density of the source from the lowest resolution maps; Column 4: Restoring beam size (FWHM major $\times$ minor) obtained combining the different array configurations at 32.5 GHz; Column 5: RMS noise of the final image obtained at 32.5 GHz; Column 6: Integrated flux density at 32.5 GHz; Column 7: Integrated flux density at 5.95 GHz from \citet{Leroy11}; Column 8: Integrated flux density at 1.49 GHz from Condon et al. (1990).} 
\tablenotetext{a}{This is not the highest resolution image we could obtain, but the highest resolution at which emission from the system was not resolved out.}
\tablenotetext{b}{Upper limit}
\tablenotetext{c}{Flux density at 1.425 GHz instead. Value from \citet{Condon96}.}
\tablenotetext{d}{Flux density at 1.4 GHz instead. Value from NVSS \citep{Condon98}.}
\tablenotetext{*}{These systems were not observed with the VLA in A configuration.}
\tablenotetext{**}{Addition of the two components. The uncertainty is the addition in quadrature of the errors of each component}
\end{deluxetable}
\clearpage
\end{landscape}

\begin{deluxetable}{lcccccccc}[th]
\tabletypesize{\scriptsize}
\tablecaption{Summary of 32.5 GHz emission sizes\label{table:tbl-4}}
\tablewidth{0pt}
\tablehead{
\colhead{Galaxy Name} &\colhead{$\theta_{A_{50}M}$ $\times$ $\theta_{A_{50}m}$} & \colhead{C$_{50}$ ($\sigma_{A_{50}}$)}& \colhead{Log$_{10}$(A$_{50d}$) (arcsec$^{2}$)}& R$_{50d}$ (kpc) &\colhead{uplim?} & Gaussianity factor\\
\colhead{(1)} &\colhead{(2)} &\colhead{(3)} &\colhead{(4)} &\colhead{(5)} &\colhead{(6)} &\colhead{(7)}}
\startdata
CGCG 436-030  &$0\,\farcs705 \times 0\,\farcs573$ & 19.9 &-0.309 & 0.238 & no & 0.308\\ 
IRAS F01364-1042  & $0\,\farcs141 \times 0\,\farcs086$ & 40.7 &-2.147 & 0.045 & no & 0.447\\ 
III Zw 035 & $0\,\farcs145 \times 0\,\farcs117$ & 64.5 &-1.956 & 0.031 & no& 0.502 \\ 
VII Zw 031  &$2\,\farcs060 \times 1\,\farcs415$ & 5.7&0.922 & 1.739 & yes & 0.463\\ 
IRAS 08572+3915  &  $0\,\farcs254 \times 0\,\farcs193$& 39.7 & -1.415\tablenotemark{a}& 0.130 & yes & 0.535\\ 
UGC 04881 &  &  &  &  &  & \\
\multicolumn{1}{c}{... NE} & $0\,\farcs253 \times 0\,\farcs191$ & 23.4 & -1.420\tablenotemark{a} & 0.088  & yes & 0.683 \\ 
\multicolumn{1}{c}{... SW} & $5\,\farcs049 \times 3\,\farcs543$ & 3.0 & 1.148\tablenotemark{a} & 1.683 & yes & 0.229\\
UGC 05101  & $0\,\farcs291 \times 0\,\farcs273$ &70.1 &-0.969 & 0.146 &  no & 0.255\\ 
MCG +07-23-019  & $0\,\farcs228 \times 0\,\farcs202$&25.6& -1.230 & 0.096 & no & 0.253\\ 
NGC 3690  &$0\,\farcs379 \times 0\,\farcs327$ &  &0.785\tablenotemark{b} & 0.302 & yes & \\ 
\multicolumn{1}{c}{... W} & " & 5.1 & 0.701 & 0.275  & yes & 0.024\\
\multicolumn{1}{c}{... E} & " & 20.2 & 0.029 & 0.127 & no & 0.030\\ 
UGC 08058  & $0\,\farcs257 \times 0\,\farcs227$ &895.8 & -1.534 & 0.082 & no & 0.345\\
VV 250  && & &  & &\\ 
\multicolumn{1}{c}{... a (SE)} & $0\,\farcs236 \times 0\,\farcs219$  & 5.0 & -0.180 & 0.285 & yes & 0.109 \\ 
\multicolumn{1}{c}{... b (NW)} & $10\,\farcs71 \times 9\,\farcs340$ & 3.2 & 1.895\tablenotemark{a} & 3.106 & yes & 0.058\\
UGC 08387  &$0\,\farcs098 \times 0\,\farcs073$ &7.4& -0.867 & 0.100 & yes & 0.026\\ 
UGC 08696  &$0\,\farcs259 \times 0\,\farcs240$ &56.6& -0.820 & 0.167 & no & 0.205\\
VV 340a  &$2\,\farcs983 \times 2\,\farcs488$ &5.0& 1.249 & 1.581 & yes & 0.491\\ 
VV 705  &$0\,\farcs583 \times 0\,\farcs579$& 5.0&0.276 & 0.625 & no & 0.071\\ 
IRAS 15250+3609 & $0\,\farcs075 \times 0\,\farcs067$ & 52.5&-2.412 & 0.039 & no & 0.366\\
Arp 220 &$0\,\farcs087 \times 0\,\farcs069$ &45.3 &-1.047 & 0.062 & no & 0.132\\ 
IRAS 17132+5313 & $0\,\farcs911 \times 0\,\farcs864$ &5.8& 0.504 & 1.020 & yes & 0.259\\ 
IRAS 19542+1110  & $0\,\farcs087 \times 0\,\farcs081$ &7.6& -1.410 & 0.142 & yes & 0.342\\ 
CGCG 448-020  &$0\,\farcs970 \times 0\,\farcs841$ &6.9& 0.338 & 0.581 & no & 0.141\\ 
IRAS 21101+5810 &$0\,\farcs686 \times 0\,\farcs590$ &11.1& -0.268 & 0.317 & no & 0.293\\ 
IRAS F23365+3604 &$0\,\farcs098 \times 0\,\farcs091$ &6.8& -1.405 & 0.141 & yes & 0.259
\enddata

\tablecomments{Column 1: Name of the galaxy; Column 2: beam size at Ka band (32.5 GHz) of the image used to obtain A$_{50}$ (see Section \ref{sec:sizes}); Column3: Contour level enclosing 50\% of the total flux density of the system in units of $\sigma_{A_{50}}$, the rms of the final image used to measure A$_{50}$; Column 4: Best estimate of the deconvolved area enclosing 50\% of the total emission at 32.5 GHz, A$_{50d}$ (see Section \ref{sec:sizes}); Column 5: Equivalent circular radius of $\mathrm{A_{50d}}$, i.e., assuming $\mathrm{A_{50d} \equiv \pi R_{50d}^{2}}$ ; Column 6: If "yes", the value in Column 4 is an upper limit either because the emission is extended and applying a Gaussian deconvolution was not a good approximation (see Section \ref{sec:sizes}), or because is unresolved; Column 7: Ratio of the flux density of the isophote enclosing 50\% of the total flux density at 32.5 GHz to the peak flux density of the emission. For a Gaussian-like distribution this number is 0.5.}
\tablenotetext{a}{A$_{50}$ for this galaxy is smaller than the beam size, so we adopted the beam area as the best estimate for its size.}
\tablenotetext{b}{A$_{50}$ for this galaxy is the addition of both components.}
\end{deluxetable}

\begin{deluxetable}{cccccc}[th]
\tabletypesize{\scriptsize}
\tablecaption{Analysis of A (or B)\tablenotemark{*} array configuration-only images\label{table:tbl-5}}
\tablewidth{0pt}
\tablehead{\colhead{Galaxy Name} & \colhead{$\theta_{M} \times \theta_{m}$}& \colhead{$\sigma_{A~(or~B)}$ ($\mu$Jy beam$^{-1}$)}&\colhead{$f_{A~(or~B)}$ (\%)} & \colhead{Log$_{10}$(A$_{5\sigma}$) (arcsec$^{2}$)} &\colhead{R$_{5\sigma}$ (kpc)} \\
\colhead{(1)}&\colhead{(2)}&\colhead{(3)}&\colhead{(4)}&\colhead{(5)}&\colhead{(6)}}
\startdata
CGCG 436-030      & 0$\,\farcs072\times0\,\farcs061$ & 36.5 & 16.4& -1.873 & 0.039\\ 
{\bf IRAS F01364-1042}  & 0$\,\farcs101\times0\,\farcs060$ & 44.0 & 59.3& -1.530 & 0.091\\ 
{\bf III Zw 035}   & 0$\,\farcs073\times0\,\farcs062$ & 45.6 & 61.3& -1.476 & 0.054\\ 
VII Zw 031        & 0$\,\farcs119\times0\,\farcs062$ & 24.8 & 0.3&  -3.222 & 0.015\\ 
\underline{IRAS 0857+3915}    & 0$\,\farcs241\times0\,\farcs180$ & 28.8 & 97.5& -0.831 & 0.255\\ 
\underline{UGC 04881}         & 0$\,\farcs247\times0\,\farcs184$ & 27.7 & 84.3& -0.739 & 0.192\\ 
\underline{UGC 05101}         & 0$\,\farcs259\times0\,\farcs240$ & 27.2 & 74.4& -0.112 & 0.393\\ 
\underline{MCG +07-23-019}    & 0$\,\farcs216\times0\,\farcs189$ & 31.5 & 67.8& -0.615 & 0.196\\ 
NGC 3690          & 0$\,\farcs239\times0\,\farcs218$ & 26.6 & 42.3&  0.312 & 0.175\\ 
\underline{UGC 08058}         & 0$\,\farcs227\times0\,\farcs202$ & 27.6 & 80.4& -0.299 & 0.340\\ 
VV 250            & 0$\,\farcs219\times0\,\farcs202$ & 26.5 & 38.2& -0.305 & 0.249\\ 
UGC 08387         & 0$\,\farcs073\times0\,\farcs051$ & 24.4 & 39.8& -1.301 & 0.060\\ 
\underline{UGC 08696}         & 0$\,\farcs210\times0\,\farcs204$ & 28.4 & 65.5& -0.233 & 0.328\\ 
VV 340a           & 0$\,\farcs085\times0\,\farcs065$ & 17.8 & 0.9&  -2.959 & 0.013\\ 
VV 705            & 0$\,\farcs059\times0\,\farcs051$ & 20.9 & 17.2& -1.712 & 0.063\\ 
{\bf IRAS 15250+3609}   & 0$\,\farcs058\times0\,\farcs051$ & 25.9 & 73.2& -1.767 & 0.082\\ 
Arp 220           & 0$\,\farcs066\times0\,\farcs052$ & 24.8 & 66.5& -0.589 & 0.106\\ 
IRAS 17132+5313   & 0$\,\farcs060\times0\,\farcs053$ & 24.5 & 10.5& -2.137 & 0.049\\ 
IRAS 19542+1110   & 0$\,\farcs072\times0\,\farcs063$ & 25.7 & 49.4& -1.472 & 0.132\\ 
CGCG 448-020      & 0$\,\farcs073\times0\,\farcs063$ & 27.8 & 23.1& -1.752 & 0.052\\ 
IRAS 21101+5810   & 0$\,\farcs075\times0\,\farcs052$ & 27.0 & 26.9& -1.785 & 0.055\\ 
IRAS F23365+3604  & 0$\,\farcs069\times0\,\farcs062$ & 37.1 & 33.9& -1.818 & 0.088
\enddata

\tablecomments{Column 1: Galaxy name, with those galaxies that represent good AGN candidates in bold face. Those with a weaker argument to be potential AGNs are underlined (see Section \ref{sec:comp} for more details); Column 2: Beam size of the A (or B) array configuration-only images (major $\times$ minor axis); Column 3: rms of the A (or B) array configuration image; Column 4: Percentage of the total flux density recovered at 33 GHz from the A (or B) array configuration-only image. This A (or B) array configuration-only flux density was obtained adding pixels with emission above 5$\sigma_{A~(or~B)}$; Column 5: Observed area of the 5 $\sigma_{A}$ contour in arcseconds$^{2}$; Column 6: Equivalent radii for Column 5, assuming A$_{5\sigma}$ = $\pi$~R$_{5\sigma}^{2}$ in units of kpc obtained using the scale conversion from Table \ref{table:tbl-1}.}
\tablenotetext{*}{8 systems were not observed with A configuration of the VLA. See Table \ref{table:tbl-3} for reference.}
\end{deluxetable}

\clearpage
\begin{landscape}
\begin{deluxetable}{lcccccccc}[th]
\tabletypesize{\scriptsize}
\tablecaption{Summary of derived average physical parameters based on 32.5 GHz continuum emission sizes\tablenotemark{*}\label{table:tbl-6}}
\tablewidth{0pt}
\tablehead{
\colhead{Galaxy Name} & 
\colhead{$\mathrm{Tb_{33GHz}}$} & 
\colhead{$\mathrm{Tb_{1.4GHz}}$} &
\colhead{$\alpha_{\rm CO}$}&
\colhead{$\mathrm{\Sigma_{mol}^{33GHz}}$\tablenotemark{**}}&
\colhead{N$_{\rm H}$\tablenotemark{**}}&
\colhead{$\mathrm{n_{mol}^{33GHz}}$\tablenotemark{**}}&
\colhead{$\mathrm{\Sigma_{SFR_{IR}}^{33GHz}}$} &
\colhead{$\mathrm{\Sigma_{IR}^{33GHz}}$}\\
\colhead{}&\colhead{(K)}&\colhead{(K)}&
\colhead{}&
\colhead{$\mathrm{(M_{\odot}~pc^{-2})}$}&\colhead{(cm$^{-2}$)}&
\colhead{(cm$^{-3}$)}&\colhead{$\mathrm{(M_{\odot}~yr^{-1}~kpc^{-2})}$}&
\colhead{$\mathrm{(L_{\odot}~kpc^{-2})}$}\\
\colhead{(1)} &\colhead{(2)} &\colhead{(3)} &\colhead{(4)} &\colhead{(5)} &\colhead{(6)} &\colhead{(7)} &\colhead{(8)} &\colhead{(9)}}
\startdata
CGCG 436-030 & 7.47E+00 & 3.12E+04 & 0.73&3.85E+03[3.51E+03] & 4.82E+23[4.40E+23] & 2.89E+02[2.64E+02] & 1.49E+02 & 1.23E+12\\
IRAS F01364-1042 & 4.35E+02 & 6.65E+05 &0.20& 1.79E+05[4.54E+04] & 2.24E+25[5.69E+24] & 7.12E+04[1.80E+04] & 6.19E+03 & 5.11E+13\\ 
III Zw 035 & 4.34E+02 & 1.16E+06 &0.23& 1.28E+05[3.64E+04] & 1.61E+25[4.55E+24] & 7.32E+04[2.08E+04] & 7.51E+03 &6.20E+13 \\ 
VII Zw 031 & 2.37E-01 & 1.55E+03 &1.65& 3.33E+02[6.87E+02] & 4.17E+22[8.60E+22] & 3.41E+00[7.04E+00] & 5.68E+00 & 4.69E+10\\ 
IRAS 08572+3915 & 3.61E+01 & 3.65E+04 &0.49& 1.24E+04[7.66E+03]& 1.55E+24[9.58E+23] & 1.70E+03[1.05E+03] & 1.54E+03 &1.27E+13\\ 
UGC 04881NE\tablenotemark{***} & 2.82E+01 & 2.59E+05 &0.28 &6.61E+04[2.34E+04] & 8.27E+24[2.93E+24] & 1.34E+04[4.75E+03] & 6.76E+02 & 5.58E+12\\
UGC 05101 & 8.55E+01 & 4.35E+05 & 0.35 & 3.39E+04[1.50E+04] & 4.24E+24[1.87E+24] & 4.12E+03[1.82E+03]  & 8.38E+02 & 6.92E+12\\ 
MCG +07-23-019 & 6.50E+01 & 1.66E+05 & 0.30 & 5.59E+04[2.09E+04] & 7.00E+24[2.62E+24] & 1.03E+04[3.86E+03] & 8.45E+02 & 6.98E+12\\ 
NGC 3690\tablenotemark{***} & 1.24E+01 & 3.37E+04 & 0.86 &2.33E+03[2.51E+03] & 2.91E+23[3.14E+23] & 1.37E+02[1.48E+02] & 1.39E+02 &1.15E+12 \\ 
UGC 08058 & 2.08E+03 & 3.16E+06 & 0.30 & 5.44E+04[2.05E+04] & 6.80E+24[2.57E+24] & 1.18E+04[4.47E+03] & 9.52E+03 &7.86E+13 \\
VV 250a\tablenotemark{***} & 9.82E+00 & 2.11E+04 & 0.90 & 2.08E+03[2.33E+03] & 2.60E+23[2.91E+23] & 1.30E+02[1.46E+02] & 1.20E+02 & 9.94E+11 \\ 
UGC 08387 & 8.54E+01 & 2.32E+05 & 0.32 & 4.36E+04[1.77E+04] & 5.46E+24[2.22E+24] & 7.80E+03[3.17E+03] & 8.68E+02 &7.17E+12\\ 
UGC 08696 & 8.63E+01 & 2.95E+05 & 0.44 &1.73E+04[9.58E+03] & 2.17E+24[1.20E+24] & 1.85E+03[1.02E+03] & 9.77E+02 &8.06E+12 \\
VV 340a & 1.41E-01 & 1.31E+03 & 1.63 &3.45E+02[7.04E+02] & 4.32E+22[8.81E+22] & 3.89E+00[7.93E+00] & 3.61E+00 & 2.98E+10\\ 
VV 705 & 2.51E+00 & 7.90E+03 &1.16 & 9.60E+02[1.39E+03] & 1.20E+23[1.74E+23] & 2.74E+01[3.97E+01] & 3.65E+01&3.02E+11\\ 
IRAS 15250+3609 & 8.71E+02 & 1.11E+06 & 0.22 & 1.45E+05[3.95E+04] & 1.81E+25[4.94E+24] & 6.66E+04[1.81E+04] & 1.34E+04 &1.11E+14\\ 
Arp 220 & 4.79E+02 & 1.12E+06 & 0.31 & 4.89E+04[1.91E+04] & 6.12E+24[2.39E+24] & 1.40E+04[5.46E+03] & 7.16E+03 &5.91E+13\\ 
IRAS 17132+5313 & 6.46E-01 & 2.52E+03 & 1.54 & 4.12E+02[7.92E+02] & 5.16E+22[9.92E+22] & 7.20E+00[1.38E+01] & 1.47E+01 &1.21E+11\\
IRAS 19542+1110 & 4.52E+01 & 1.63E+05 & 0.41 & 2.21E+04[1.13E+04] & 2.76E+24[1.41E+24] & 2.77E+03[1.41E+03] & 1.12E+03 & 9.26E+12\\ 
CGCG 448-020 & 1.59E+00 & 5.20E+03 & 1.52& 4.24E+02[8.07E+02] & 5.30E+22[1.01E+23] & 1.30E+01[2.47E+01] & 3.52E+01 & 2.90E+11\\ 
IRAS 21101+5810 & 4.75E+00 & 1.28E+04 & 0.89 & 2.14E+03[2.37E+03] & 2.67E+23[2.97E+23] & 1.20E+02[1.33E+02] & 1.06E+02 & 8.72E+11\\
IRAS F23365+3604 & 4.77E+01 & 1.95E+05 & 0.33 & 4.15E+04[1.71E+04] & 5.20E+24[2.15E+24] & 5.24E+03[2.16E+03] & 1.40E+03 &1.15E+13
\enddata

\tablecomments{Column 1: Name of the galaxy; Column 2: Brightness temperature within A$_{50}$ at 32.5 GHz; Column3: Brightness temperature at 1.49 GHz assuming A$_{50}$ is the true size of the 1.49 GHz emission; Column 4: CO-to-H$_{2}$ conversion factor in units of $\mathrm{M_{\odot}~(K~km~s^{-1}pc^{-2})^{-1}}$ derived for each galaxy using equation \ref{eq:alpha_co}; Column 5: Molecular gas mass surface density; Column 6: Hydrogen gas column density; Column 7: H$_{2}$ particle volume density; Column 8: Star formation rate surface density;Column 9: Infrared surface brightness.}
\tablenotetext{*}{See Section \ref{sec:impl_radio} for details on the derivation of the parameters in each column.}
\tablenotetext{**}{Values in brackets were obtained by using $\alpha_{\rm CO}$ from Column 4, while the other values were obtained by using a fixed ``starbursts" conversion factor of 0.8 $\mathrm{M_{\odot}~(K~km~s^{-1}pc^{-2})^{-1}}$. The integrated CO measurements from which we derived the parameters in these columns will be available in Privon et al. (in preparation).}
\tablenotetext{***}{For simplicity, we include only the brightest components for UGC04881 and VV250, while for NGC 3690 we consider the system as a whole.}
\end{deluxetable}
\clearpage
\end{landscape}

\begin{deluxetable}{cccc}[th]
\tabletypesize{\scriptsize}
\tablecaption{Summary of best non-linear least-squares fit to equation $\mathrm{log_{10}(\Sigma_{SFR}^{33GHz}) = A~log_{10}(\Sigma_{mol}^{33GHz})+B}$ \label{table:tbl-7}}
\tablewidth{0pt}
\tablehead{\colhead{$\alpha_{CO}$} & \colhead{Sample included in fit} & \colhead{A} & \colhead{B} \\
\colhead{$\mathrm{M_\odot~pc^{-2} (K~km~s^{-1})}$}&\colhead{}&\colhead{}&\colhead{}}
\startdata
0.8 & U/LIRGs only (this paper) & 1.02 $\pm$ 0.10 & -1.33 $\pm$ 0.47\\ 
4.35 & U/LIRGs only (this paper) & 1.02 $\pm$ 0.11 & -2.08 $\pm$ 0.55\\ 
$\propto$ $\Sigma_{gas}^{-0.5}$ & U/LIRGs only (this paper) & 1.52 $\pm$ 0.16 & -3.09 $\pm$ 0.66\\ 
4.35 & U/LIRGs (this paper) + kpc-size regions \citep{Leroy13} & 1.35 $\pm$ 0.04 & -3.85 $\pm$ 0.13\\ 
0.8\tablenotemark{a} & U/LIRGs (this paper) + kpc-size regions \citep{Leroy13} & 1.63 $\pm$ 0.07 & -4.18 $\pm$ 0.22\\
$\propto$ $\Sigma_{gas}^{-0.5}$\tablenotemark{a} & U/LIRGs only (this paper) + kpc-size regions \citep{Leroy13}& 1.87 $\pm$ 0.06 & -4.63 $\pm$ 0.19\\
4.35 & kpc-size regions \citep{Leroy13}& 1.06 $\pm$ 0.02 & -3.48 $\pm$ 0.03
\enddata
\tablenotetext{a}{Only applied to the U/LIRGs}
\end{deluxetable}

\appendix
\section{Notes on the sources}

\noindent{\bf CGCG 436-030}: This system has two well separated components (east and west), however we only detected the western component at 33 GHz. 

\noindent{\bf CGCG 448-020}: This is an interacting system showing a complex morphology. It is still not clear whether there are two or more systems interacting. It hosts an off-nuclear starburst (north-east component in Figure \ref{fig:fig1}) which contributes $\sim$ 80\% of the total infrared luminosity of the galaxy at infrared wavelengths (Stierwalt et al. in prep). In the final map (i.e., the one with the highest resolution, not shown in this work), this off-nuclear starburst is still only partially resolved, even at 0$\,\farcs$08, while the more extended component (south-west) is resolved out. 

\noindent{\bf III Zw 035}: This galaxy has the most compact 33 GHz continuum emission in the sample.

\noindent{\bf IRAS 08572+3915}: We only detect the north west (NW) component of this system. The NW component is optically classified as a Seyfert 2 galaxy and is suspected to have a highly obscured AGN\citep[e.g.,][]{Iwasawa11, Rupke13}. The flat spectrum observed in Figure \ref{fig:fig2} suggests this is a flat spectrum AGN, which was also suggested by \citet{Condon91} based on 1.49 and 8.44 GHz continuum observations. The high thermal fraction predicted from the IR luminosity only indicates that the IR emission is mostly dominated by an AGN instead of star formation, and that the 33 GHz emission is dominated by synchrotron instead.

\noindent{\bf IRAS 15250+3608}: This systems is one of the sources emitting at, or close to, the Eddington limit. The optical and mid-IR diagnostics classify this galaxy as a composite source. The fact that it is close to the Eddington limit, agrees with the potential coexistence of an AGN and a strong starburst.

\noindent{\bf IRAS 17132+5313}: This system has two components. The galaxy towards the north east is extended and resolved out in the highest resolution image (0$\,\farcs$08 $\times$ 0$\,\farcs$07). We had to taper the map in order to recover its emission. The galaxy towards the south west is compact and contributes $\sim$ 40\% of the integrated flux density of the system.

\noindent{\bf NGC 3690}: This clearly interacting system consists of multiple components. Two of the components them are associated with the nuclei of the progenitors, NGC 3690E (east) and NGC 3690W (west), while the others are a combination of off nuclear starbursts. The strongest nucleus (NGC 3690E) has been observed with VLBI. At least 30 point sources have been found plus a potential AGN \citep[e.g.,][]{Neff04}.
Due to the proximity of this system and its spatial extent, the 33 GHz emission is resolved at the D configuration resolution ($\sim$ 2"), clearly showing 5 components (see white crosses in map from Figure \ref{fig:fig1}). In order to measure its total flux density, we tapered the D configuration map until the system showed 2 unresolved components (east and west systems). We then proceed as explained in Section \ref{sec:Iflux}, by fitting a Gaussian to each one. 

\noindent{\bf UGC 04881}: This system has two components and its total flux density was recovered by adding the Gaussian fit results of each component separately. The error of this measurement was obtained by adding in quadrature the errors associated to each component (see Section \ref{sec:Iflux}). The D configuration map of this system had low signal-to-noise ratio and the quality was not good enough to recover the total flux density. For this reason, we used the final image (with the different array configurations combined, see Section \ref{sec:red}) tapered such that we recovered a point-like structure for each component. Even though both components contribute about the same to the total flux density observed at 32.5 GHz, the brightest component (north-east) is more compact. The south-west component is resolved out at the highest resolution we can achieve. We measured the size of this component from the image we used to obtain the total flux density ($\mathrm{A_{beam}=20.3~arcsec^{2}}$) and found an upper limit of $\mathrm{A_{50}=19.8~arcsec^{2}}$, i.e., it is unresolved in this coarser map. The brightest component is shown in Figure \ref{fig:fig1}. The flux density of this bright component should also be treated as a lower limit. The potential calibration issues mentioned before could very well be originating the abnormally high thermal fractions observed in Figure \ref{fig:fig2}.

\noindent{\bf UGC 08058}: This is the most powerful IR source in our sample. It is known to host an AGN \cite[e.g.,][]{Lonsdale03,Iwasawa09} and potentially represents the stage before becoming an elliptical galaxy according to the evolutionary model proposed by \citet{Sanders88}.

\noindent{\bf Arp 220}: This is the closest ULIRG in the local universe. This galaxy shows extreme dust opacities and very compact nuclear disks. We present a detailed analysis of the 33 and 6 GHz emission from this galaxy in \citet{BM15}, where we find that the disks are better described by exponential disks, rather than Gaussian. The 33 GHz map reported in \citet{BM15} is slightly different to the one presented here since the imaging procedures differ, however the flux density measured here and the morphology are in agreement with those shown in \citet{BM15}.  

\noindent{\bf VII Zw 031}: At the highest resolution image (0$\,\farcs$8 $\times$ 0$\,\farcs$6, done with natural weighting) the emission was completely resolved out. We had to taper the image heavily in order to recover the emission. This is one of the most extended systems in our sample along with VV 340a.

\noindent{\bf VV 250}: This system has two well separated components, south-east (VV 250a) and north-west (VV 250b). In Figure \ref{fig:fig1}, we only show VV 250a since it contains $\sim$86\% of the total flux density of the system (obtained by adding the flux density of both components). The north-west component is faint with an 11$\sigma$ peak detection. To recover A$_{50}$ for this faint component, we used the tapered D array map ($\sim$ 10" resolution) since we could not recover half of the integrated flux density of this component in higher resolution maps. Even in this low resolution map, we recover A$_{50}$ for $\mathrm{C_{50}=3.2\sigma}$, which is lower than our conservative limit of 5$\sigma$, however we inspected this contour and made sure the emission within it looked real. For the north-west component, $\mathrm{A_{50}=64.8~arcsec^{2}}$ in a map with $\mathrm{A_{beam}=113.4~arcsec^{2}}$, i.e., it is unresolved, and then A$_{50}$ is only an upper limit.

\noindent{\bf VV 340a}: In the final combined image, where we achieved an angular resolution of 0$\,\farcs$5 $\times$ 0$\,\farcs$4 (using natural weighting), the emission from this system was completely resolved out. To recover the extended emission, we had to taper the image heavily. VV 340 has two components, an edge on galaxy to the north (VV 340a), shown in Figure \ref{fig:fig1}, and a face on galaxy to the south (VV 340b). Inconveniently, the pointing of the VLA observation was centered on VV 340b, from which we tentatively detected an off nuclear feature at a $\sim 3 \sigma$ level in our lowest resolution image. The bright edge on galaxy, VV 340a, is clearly detected, although it was hard to perform the Gaussian fit since the source fell close to the edge of the primary beam.

\noindent{\bf VV 705}: This system shows two nuclei in Figure \ref{fig:fig1}, north-west and south-east. In the D configuration map they are indistinguishable.

\end{document}